\documentclass[preprint]{elsarticle}

\usepackage{hyperref}
\usepackage{amsmath}
\usepackage{amsfonts}
\usepackage{mathtools}

\DeclareMathOperator*{\argmin}{argmin}

\begin{document}

\newcommand*{\eref}[2][eq.]{#1~\ref{eq:#2}}
\newcommand{\dd}{\text{d}}
\newcommand{\ii}{\text{i}}
\newcommand{\mat}[1]{\left[#1\right]}
\newcommand{\revL}{\text{\reflectbox{$\mathcal{L}$}}}
\def\tril{\mbox{\begin{picture}(5,5)
		\put(1,0){\line(1,0){5}}
		\put(6,0){\line(-1,1){5}}
		\put(1,0){\line(0,1){5}}
		\end{picture}
	}}
	
      \newcommand{\triu}{\mathbin{\rotatebox[origin=c]{180}{$\tril$}}}

\begin{frontmatter}

\title{Coarse-Grained Modelling Out of Equilibrium}

\author{Tanja Schilling}
\address{Albert Ludwigs Universit\"at Freiburg, Hermann Herder Str.~3, 79104 Freiburg, Germany}

\begin{abstract}
  Active matter, responsive (``smart'') materials and materials under time-depen\-dent load are systems out of thermal equilibrium.
To construct coarse-grained models for such systems, one needs to integrate out a distribution of microstates that evolves in time. This is a challenging task. As a preparation to the topic, we recall equilibrium coarse-graining methods, both theoretical and numerical, such as projection operator formalisms, united atom simulations, numerical reconstruction of memory and Markov State Modelling. Then we review recent developments in theoretical approaches to the non-equilibrium coarse-graining problem, in particular, time-dependent projection operator formalisms, dynamic density functional theory and power functional theory, as well as numerical schemes to contruct explicitly time-dependent memory kernels. 
\end{abstract}

\begin{keyword}
Multiscale Modelling, Non-Equilibrium Statistical Physics, Computer Simulation, Materials Modelling, Scale-Bridging, Coarse-Graining
\end{keyword}

\end{frontmatter}


\tableofcontents

\section{Introduction}
In physics, we hardly ever describe a system in terms 
of all of its microscopic degrees of freedom (with a notable
exception of theoretical elementary particle physics).
Physicists rely on effective coarse-grained
models. Often the use of such models is justified, because the processes of
interest occur on time-, length\nobreakdash-, and energy-scales that can be clearly
separated from the microscopic scales. However, in principle we should be able to derive each
effective model rigorously by integrating out irrelevant
degrees of freedom from the underlying microscopic dynamics. To integrate out degrees of freedom systematically in order to derive an effective model is the task of {\it coarse-graining}.

In addition to the fundamental desire to derive models rigorously, we frequently encounter phenomena in which scale separation is not given and the practical use of models that work on just one scale is limited. Then we resort to {\it multiscale modelling}.
 The concept of multiscale modelling is applied in many branches
 of engineering, physics and the life sciences\cite{Berendsen07,fish2013,hoekstra2014,Weinan11,Attinger04,peter2009}. Already a brief glance at the range of textbooks which cover the topic reveals that the term multiscale modelling is not uniquely defined and that it is used for a wide range of different numerical methods \cite{hoekstra2014}. Inevitably, we need to focus this review on one specific interpretation, select certain topics and leave out others, although they deserve to be reviewed. This review presents the author's personal selection of topics. 

 We will discuss modelling and coarse-graining in the context of computational physics with applications in biophysics and materials science. We will presume that a researcher starts out from a microscopic description of a given system in terms of Hamilton's equations of motion, and that they attempt to obtain a prediction regarding the system's evolution in terms of some effective degrees of freedom. If the reader is interested rather in the mathematical perspective on handling data that is sampled at different scales, we refer them to standard texts such as e.g.~ref.~\cite{ferreira2007}. Also, we will only briefly touch upon the topics of coarse-graining quantum mechanical systems and of continuum models in mechanical engineering and the geosciences. After a brief reminder of Brownian motion as a typical coarse-graining problem, we will review theoretical approaches to coarse-graining under equilibrium and steady state conditions in sec.~\ref{sec:DFT}-\ref{sec:Stochastic}. Sec.~\ref{sec:nlGLE} is more detailed than the other sections, because we have included a derivation of the nonlinear generalized Langevin equation, which we have not seen anywhere in the literature. In sec.~\ref{sec:Numerical} we give a brief review of numerical methods used mostly in coarse-grained simulations of soft materials and biomolecules. Sec.~\ref{sec:Relaxation} covers theoretical approaches to relaxation into equilibrium, such as dynamic density functional theory and projection operator formalisms. In sec.~\ref{sec:timeDependence} we discuss the full non-equilibrium problem of propagators and observables with explicit time-dependence. We review power functional theory, time-dependent projection operator formalisms, a numerical method to construct non-stationary memory kernels and a non-equilibrium Markov State modelling scheme.  

 \subsection{Brownian Motion}
\label{sec:Brownian}
 As an introductory example of coarse-grained dynamics, let us recall Brownian motion. Consider a particle of mesoscopic size suspended in a fluid of many particles of molecular size. Our task is to predict the position of the ``large'' particle as a function of time, i.e.~we intend to integrate out the degrees of freedom of the small particles, because we are not interested in the details of their motion. Two well-known answers to this problem were given in the beginning of the 20th century, one by Einstein \cite{Einstein1905} and Smoluchowski \cite{Smoluchowski1906} and one by Langevin \cite{Langevin1908}. Neither of them derived the equation of motion rigorously from the underlying microscopic physics; they introduced stochastic approaches instead. Einstein argued that within a small time interval due to collisions with many of the solvent particles, the position of the large particle changes by a value that is effectively random. We recall his line of arguments and consider the one-dimensional case for simplicity: If we draw the displacement of the large particle, $\Delta$, from a probability distribution $\phi(\Delta)$ then the probability $f(x,t+\tau)$ to find the particle at position $x$ at a time $t+\tau$ is given by
 \[
f(x,t+\tau) = \int_{-\infty}^{\infty} d\Delta \; f(x - \Delta,t) \; \phi(\Delta) \quad ,
   \]
where $\tau$ is the duration of the small time-interval, during which the particle was subjected to the collisions that produced the displacement.
   We expand $f(x,t)$ to second order in $x$  and  to first order in $t$, impose the symmetry condition $\phi(\Delta)=\phi(-\Delta)$, and obtain an equation of motion for the probability distribution
   \[
\frac{\partial f(x,t)}{\partial t} \simeq \frac{1}{\tau}\frac{\partial^2 f(x,t)}{\partial x^2}
\int_{-\infty}^{\infty} d\Delta \; \phi(\Delta) \; \frac{\Delta^2}{2} \quad .
\]
If the integral exists, we can define $D \coloneqq  \int_{-\infty}^{\infty} d\Delta \; \phi(\Delta) \; \frac{\Delta^2}{2\tau}$. Then we obtain a diffusion equation for $f(x,t)$
\begin{equation}
  \label{eq:diffusion}
\frac{\partial f(x,t)}{\partial t} = D \frac{\partial^2 f(x,t)}{\partial x^2}
 \quad .
\end{equation}
Further, using thermodynamic arguments, Einstein related the diffusion constant $D$ to macroscopic properties of the fluid
\[
  D \gamma = k_BT \quad ,
\]
where $\gamma$ is the Stokes friction coefficient of the large particle in the fluid of the small particles, $k_B$ is Boltzmann's constant and $T$ is the temperature.

The fundamental idea of Einstein's approach, as well as of the approach by Smoluchowski, was to treat the position of the large particle as a stochastic variable. Given that the equations of motion of all microscopic degrees of freedom in the system are deterministic, this idea contained a leap of faith that needed to be justified by a more detailed analysis after their work (we will discuss this justification in sec.~\ref{sec:Zwanzig} and sec.~\ref{sec:Mori}.) 
 
A few years later, Langevin took a route via Newtonian mechanics and introduced a stochastic description of the velocity of the Brownian particle \cite{Langevin1908}. He started out from Newton's equation of motion for a particle of mass $m$ subject to an external force $\xi(t)$ embedded in a Stokesian fluid with a friction constant $\gamma$
\begin{equation}
  \label{eq:simpleLE}
  m\frac{d v(t)}{d t} = -\gamma v(t) + \xi(t) \quad ,
  \end{equation}
  where again, for simplicity we consider the one-dimensional case.
  Langevin argued that the collisions with the solvent particles produce a stochastic force, i.e.~he treated $\xi(t)$ as a stochastic process -- and overlooked the fact that the interpretation of the term $\frac{d v}{d t}$ as a derivative is then incorrect\footnote{If this aspect is new to the reader, we suggest to consult the didactic article on the Langevin equation by Gillespie \cite{gillespie1996}.}.

  As the forces excerted by the embedding fluid have to be symmetrical, the stochastic force needs to fulfill 
  \[
    \langle \xi(t) \rangle = 0 \quad ,
  \]
  where the angle brackets indicate the average over the distribution from which $\xi(t)$ is drawn. Further, Langevin argued that $\xi(t)$ has a vanishingly short autocorrelation time
   \[
    \langle \xi(t) \xi(t^\prime)\rangle = c \delta(t-t^\prime) \quad ,
  \]
  because the motion of the fluid molecules is much faster than the motion of the large particle. ($c$ is a constant, which we will speficy below.)
 If we solve \eref{simpleLE} for the velocity $v(t)$, compute the corresponding position $x(t)$ and then determine the mean-squared displacement of the particle, we obtain
\[
  \langle x(t)^2\rangle = \frac{c}{\gamma^2}t \quad .
\]
The diffusion equation in Einstein's approach, \eref{diffusion}, also produces a mean squared displacement that is linear in time  
\[
 \langle x(t)^2\rangle = 2Dt \quad .
\]
Thus we recognize that the two approaches are equivalent on the level of the ensemble average and we can identify the constant $c$
\begin{equation}
  \label{eq:simpleFDT}
  c = 2D\gamma^2 = 2\gamma k_BT \quad .
  \end{equation}
\eref[Eq.]{simpleFDT} relates a transport coeffcient, a friction coefficient and the thermal energy $k_BT$. This is an example of a fluctuation-dissipation relation.
  
Despite its lack of mathematical rigour, the approach by Langevin was picked up rapidly in various disciplines of physics. Today it is widely used to develop coarse-grained models. We will encounter versions of the Langevin equation in several chapters in this review. 

\section{Equilibrium and Non-Equilibrium Steady State}
\label{sec:Equilibrium}
In the introductory example of Brownian motion we recalled a coarse-grained model that has been brought forward by an educated guess about 100 years ago. At the time, the model was not derived from the equation of motion of the underlying microscopic system. In this chapter we recall methods to derive coarse-grained models, and we show, in particular, how \eref{diffusion} and \eref{simpleLE} can be justified. At first, we restrict the discussion to systems under steady state conditions, i.e.~to systems in which the microscopic density of states does not depend on time. Coarse-graining over a time-dependent density of states will be discussed in sec.~\ref{sec:Noneq}. We also focus the discussion on classical systems, but most of the theoretical considerations we present have either been originally developed for quantum mechanical systems or can be applied to them in the same form as shown here (see ref.~\cite{nakajima1958,zwanzig1960,mori:1965,Zwanzig01} for projection operator formalisms, ref.~\cite{grabert:1982, vrugt2019} for time-dependent projection operators and ref.~\cite{hohenberg1964,kohn1965,parr1989} for density functional theory). 

\subsection{The Structure of Liquids}
\label{sec:DFT}
To derive a coarse-grained model from the underlying microscopic equation of motion, we start out with Hamiltonian mechanics.    
Consider a three-dimensional system of $N$ identical, classical particles of mass $m$ with positions $\vec{r}^N=\{\vec{r}_1,\ldots,\vec{r}_N\}$ and momenta $\vec{p}^N=\{\vec{p}_1,\ldots,\vec{p}_N\}$ which is governed by a Hamiltonian $\mathcal{H}(\vec{p}^N,\vec{r}^N)$.\footnote{We use arrows to indicate vectors in $\mathbb{R}^3$ and boldface symbols to indicate elements of other linear spaces with the exception of some elements of function spaces.} Let the Hamiltonian consist of a potential for pairwise interactions $V_{\rm int}(\vec{r}_i, \vec{r}_j)$, an external potential $V_{\rm ext}(\vec{r}_i)$ and the kinetic energy $\frac{1}{2m}\vec{p}_i\cdot\vec{p}_i$. If we prepare an ensemble of copies of this system with a phase space probability density $\rho_N(\vec{p}^N,\vec{r}^N,t=0)$, the ensemble will evolve according to the Liouville equation
\begin{equation}
  \label{eq:Liouville}
\frac{\partial}{\partial t} \rho_N(\vec{p}^N,\vec{r}^N,t)= \left\{\mathcal{H},\rho_N\right\} =:  -\ii \mathcal{L}\rho_N \quad ,
  \end{equation}
  where the curly brackets are Poisson brackets and we have defined the Liouvillian $\ii \mathcal{L}$. This equation is formally solved by
  \[
     \rho_N(\vec{p}^N,\vec{r}^N,t) = e^{-\ii \mathcal{L} t} \rho_N(\vec{p}^N,\vec{r}^N,0) \quad .
    \]
    Now we tackle the coarse-graining problem and formally integrate out the degrees of freedom of all particles except a subset of $n$ particles
    \[
\rho_n(\vec{p}^n,\vec{r}^n,t) = \frac{N!}{(N-n)!} \int  d\vec{r}^{N-n}\;d\vec{p}^{N-n}\; \rho_N(\vec{p}^N,\vec{r}^N,t) \quad .
\]
We apply the Liouvillian $\ii\mathcal{L}$ to $\rho_n(\vec{p}^n,\vec{r}^n,t)$ in order to propagate the coarse-grained $n$-particle system.
After a few transformations, which are discussed in detail e.g.~in ref.~\cite{Hansen1990}, this leads us to the Bogolyubov-Born-Green-Kirkwood-Yvon equation \cite{bogoliubov1946,born1946,kirkwood1946,yvon1935}
\begin{multline}
    \label{eq:BBGKY}
    \left( \frac{\partial}{\partial t} + \sum_{i=1}^{n} \frac{ \vec{p}_i}{m} \cdot \frac{\partial}{\partial \vec{r}_i} - \sum_{i=1}^{n}\left(\frac{\partial V_{\rm ext}(\vec{r}_i)}{\partial \vec{r}_i} + \sum_{j=1}^{n} \frac{ \partial V_{\rm int}(\vec{r}_i, \vec{r}_j)}{\partial  \vec{r}_i} \right) \cdot \frac{\partial}{\partial \vec{p}_i} \right) \rho_n \\ = \sum_{i=1}^{n} \int  \frac{\partial V_{\rm int}(\vec{r}_i, \vec{r}_{n+1})}{\partial  \vec{r}_i} \cdot \frac{\partial \rho_{n+1}}{\partial \vec{p}_i} \; d\vec{r}_{n+1} \; d\vec{p}_{n+1} \quad .
    \end{multline}  
    \eref[Eq.]{BBGKY} is an exact expression which couples the evolution of the $n$-particle density to the $(n+1)$-particle density. In principle, it would allow us to systematically coarse-grain the full $N$-particle equation of motion into an equation of motion for the $n$-particle system. In practice, however, this cannot be done unless the interaction potential is very simple. Instead, \eref{BBGKY} is used as a starting point for the construction of approximative models.
    Typical approximations either preserve the non-equilibrium nature of the equations of motion, but considerably simplify the higher order densities, or they preserve more of the complexity of the interactions, but consider stationary densities. The first type of approximation leads us to the Boltzmann equation and the Vlasov equation for dilute gases, which we discuss in sec.~\ref{sec:kinetic}. The second type leads us to density functional theory (DFT) and liquid structure theory.

  The basic idea of DFT is to express the condition for the thermal equilibrium structure in terms of a variational principle. The grand potential $\Omega$ as a functional of the configurational one-particle density $\rho_1(\vec{r})$, is mimimized  by the equilibrium structure $\rho^{\text{EQ}}_1(\vec{r})$
  \begin{equation}
    \label{eq:DFT}
\left. \frac{\delta\Omega[\rho_1]}{\delta \rho_1(\vec{r})}\right\vert_{\rho_1(\vec{r}) = \rho^{\text{EQ}}_1(\vec{r})} = 0 \quad .
\end{equation}
  Note that we have removed the dependence on the momenta from $\rho_1(\vec{p},\vec{r})$, because in thermal equilibrium the momenta produce only a trivial prefactor. In the canonical ensemble $\rho_1(\vec{p},\vec{r})$ is related to $\rho_1(\vec{r})$ by
  \[
    \rho_1(\vec{p},\vec{r}) = \left(\frac{\beta}{2\pi m}\right)^{-\frac{3}{2}}\exp{\left(-\beta\frac{\vec{p}\cdot\vec{p}}{2m}\right)}\;\rho_1(\vec{r}) \quad ,
  \]
   where $\beta\coloneqq 1/k_BT$. Throughout this article, we will use the notation $\rho_m(\vec{r}^m)$ for the configurational part of $\rho_m(\vec{p}^m,\vec{r}^m)$, implying that the trivial contribution of the momenta has been factored out. Once we know the one-particle density, for which the condition \eref{DFT} is fulfilled, the $m$-particle densities $\rho_m(\vec{r}^m)$ can be obtained as functional derivatives of the grand potential with respect to the corresponding intrinsic chemical potential.

  The condition on the grand potential, \eref{DFT}, can be written in terms of the excess free energy functional $\mathcal{F}_{\rm exc}\left[\rho_1 \right]$ (``excess'' here means excess over the free energy of the ideal gas), the external potential $V_{\rm ext}$ and the chemical potential $\mu$ as
  \[
k_BT \ln{\left(\left(\frac{h^2\beta}{2\pi m}\right)^{\frac{3}{2}}\rho_1(\vec{r})\right)}  + V_{\rm ext}(\vec{r}) - \mu + \frac{\delta \mathcal{F}_{\rm exc}\left[\rho_1 \right]}{\delta \rho_1(\vec{r})} = 0\quad .
\]
The first term is the contribution of the free energy of the ideal gas to the grand potential, the second term is provided by the definition of the microscopic problem, the third term is specified as a thermodynamic parameter, but, in general the last term, the excess free energy functional, is not known. It is the task of DFT to develop suitable, approximative excess free energy functionals to predict the properties of specific classes of materials.

When modeling materials, the two-particle density $\rho_2(\vec{r}_1,\vec{r}_2)$ is of particular interest, because many material properties can be derived from it and it is closely related to the structure factor, which can be measured in scattering experiments. Thus  on the level of $\rho_2$, models can be easily validated experimentally. Typical approaches to the liquid structure problem take a route via the Ornstein-Zernike relation \cite{ornstein1914}. We define the {\it pair correlation function} 
    \[
h(\vec{r}_1,\vec{r}_2) \coloneqq \frac{\rho_2(\vec{r}_1,\vec{r}_2)}{\rho_1(\vec{r}_1)\;\rho_1(\vec{r}_2)} - 1
\]
and the {\it direct correlation function}
\[
c(\vec{r}_1,\vec{r}_2) \coloneqq -\beta \frac{\delta^2\mathcal{F}_{\rm exc}[\rho_1]}{\delta\rho_1(\vec{r}_1)\delta\rho_1(\vec{r}_2)} \quad . 
\]
The Ornstein-Zernike relation decomposes $h(\vec{r}_1,\vec{r}_2)$ into
\begin{equation}
  \label{eq:OZ}
h(\vec{r}_1,\vec{r}_2) = c(\vec{r}_1,\vec{r}_2) + \int\dd \vec{r}_3\; \rho_1(\vec{r}_3)c(\vec{r}_1,\vec{r}_3)h(\vec{r}_3,\vec{r}_2) \quad .
\end{equation}
If the system is homogeneous, $\rho_1(\vec{r}_3)$ is a constant. Then \eref{OZ} is a Fredholm integral equation of the second kind \cite{tricomi1985}, which could be solved by the resolvent formalism, if we knew $c(\vec{r}_1,\vec{r}_2)$. However, we do not know $c(\vec{r}_1,\vec{r}_2)$, unless we know $\mathcal{F}_{\rm exc}$. One task of liquid structure theory is to propose approximative closure relations for the Ornstein-Zernike relation, i.e.~additional relations between $h(\vec{r}_1,\vec{r}_2)$ and  $c(\vec{r}_1,\vec{r}_2)$, which allow to solve \eref{OZ} in order to construct models on the level of $\rho_2$.

We do not extend the discussion on DFT and liquid structure theory here, because these are equilibrium theories and our review focuses on theories and models for systems out of equilibrium. This subsection is intended to serve as an introduction to the non-equilibrium extensions of DFT (dynamic DFT and power functional theory) which we discuss in sections \ref{sec:DDFT} and \ref{sec:PowerFunctionals}. If the reader is interested in an introduction to classical DFT and liquid structure theory, the book by Hansen and McDonald provides a good basis \cite{Hansen1990}.

  \subsection{Projection Operator Formalism I: Zwanzig's Projection Operator}
  \label{sec:Zwanzig}
  The Liouville equation, \eref{Liouville}, can also be used as a starting point for a rather different kind of coarse-graining strategy. Assume that we are interested in a more general kind of averaged variable, not just a position or momentum of a particle, but any set of phase space observables $\{\mathbb{A}_1(\vec{p}^N,\vec{r}^N), \ldots \mathbb{A}_k(\vec{p}^N,\vec{r}^N)\}$, which we will write for short as $\mathbf{A}(\vec{p}^N,\vec{r}^N)$ or $\mathbf{A}(\Gamma)$ with $\Gamma \coloneqq  (\vec{p}^N,\vec{r}^N)$. To give a few examples, these quantities could be the order parameters of a phase transition, the centers of mass of certain groups of atoms in a macromolecule, or a set of parameters characterizing the roughness of a moving interface.
  
  In the 1960s, Zwanzig proposed to derive an equation of motion for such a set of observables by means of a projection operator formalism \cite{zwanzig:1961,Zwanzig01}. We recall the basic steps of his derivation: Consider an ensemble of copies of the system of interest, initialized such that the observables
  have the same values $\mathbf{a}_0$ in each copy of the system.  If we propagate each copy for a certain duration of time $t$, trajectories which start from different phase space points $\Gamma$ and $\Gamma^\prime$ will in general produce different values of the observables at time $t$, even though they were initialized with coinciding values. It is interesting to study the probability $g(\mathbf{a},t)$ of finding a given set of values $\mathbf{a}$ at time $t$.

  Zwanzig proposed to split the equation of motion of any other observable $f(\Gamma)$ into a contribution that depends on the microscopic degrees of freedom via $\mathbf{A}$, $f_1(\mathbf{A}(\Gamma))$, and a remaining part $f_2(\Gamma) = f(\Gamma) - f_1(\mathbf{A}(\Gamma))$. If the set of observables $\mathbf{A}$ characterizes the macroscopic properties of the system well, then this separation should produce a useful effective description of the dynamics. In Zwanzig's work the distinction between $f_1$ and $f_2$ is therefore motivated as a separation between contributions from ``relevant'' and ``irrelevant'' degrees of freedom. To associate ``relevance'' with certain observables obviously requires knowledge or at least intuition about the system. This point seems to have caused some misinterpretations of Zwanzig's work. Let us stress that Zwanzig's derivation  is exact, irrespective of whether the observables $\mathbf{A}$ are ``relevant'' or ``irrelevant'' to the effective model of a specific problem. (The same holds for the derivation by Mori presented in sec.~\ref{sec:Mori}.) Only the approximations made after the derivation in order to simplify the equations require this kind of knowledge. 

  In his work from 1961 \cite{zwanzig:1961}, Zwanzig focused on an equation of motion for the probability $g(\mathbf{a},t)$ of finding specific values $\mathbf{a}$ of the observables $\mathbf{A}(\Gamma)$
     \[
g(\mathbf{a},t) \coloneqq \int\; d \Gamma\; \delta (\mathbf{A}(\Gamma)- \mathbf{a}) \rho_N(\Gamma,t)
      \]
  To obtain the equation of motion, Zwanzig introduced a projection operator
  \begin{equation}
    \label{eq:DefPZ}
\mathcal{P}^{\rm Z} \mathbb{X}(\Gamma) \coloneqq \frac{\int\;  d \Gamma^\prime \; \delta (\mathbf{A}(\Gamma^\prime)- \mathbf{A}(\Gamma)) \mathbb{X}(\Gamma^\prime)}{\int\; d \Gamma^\prime\; \delta (\mathbf{A}(\Gamma^\prime)- \mathbf{A}(\Gamma))} \quad .
    \end{equation}
    $\mathcal{P}^{\rm Z}$ acts on an observable $\mathbb{X}(\Gamma)$ in such a way that it averages $\mathbb{X}$ over all phase space points $\Gamma^\prime$ for which  $\mathbf{A}(\Gamma^\prime)=\mathbf{A}(\Gamma)$. With this definition we can express $g(\mathbf{a},t)$ 
      as
      \[
        g(\mathbf{a},t) = W(\mathbf{a}) \left(\mathcal{P}^{\rm Z}\rho_N(\Gamma,t)\right)|_{\mathbf{A}(\Gamma)=\mathbf{a}}
      \]
      where
      \[
          W(\mathbf{a})\coloneqq \int\; d \Gamma\; \delta (\mathbf{A}(\Gamma)-\mathbf{a}) \quad .
        \]
        We define $\mathcal{Q}^{\rm Z}\circ\coloneqq (1-\mathcal{P}^{\rm Z})\circ$ and apply $\mathcal{P}^{\rm Z}$ and $\mathcal{Q}^{\rm Z}$ to the Liouville equation, \eref{Liouville}. Thus we obtain equations of motion for $g(\mathbf{a},t)$ as well as for the  dynamics ``orthogonal to $\mathbf{A}$'', i.e.~for the evolution of $\mathcal{Q}^{\rm Z}\rho_N(\Gamma,t)$.
\begin{eqnarray}
  \mathcal{P}^{\rm Z}\frac{\partial}{\partial t} \rho_N(\Gamma,t) &=& -\mathcal{P}^{\rm Z}\ii\mathcal{L} \rho_N(\Gamma,t) = - \mathcal{P}^{\rm Z}\ii\mathcal{L} \left (\mathcal{P}^{\rm Z}\rho_N(\Gamma,t) + \mathcal{Q}^{\rm Z}\rho_N(\Gamma,t)\right)\label{eq:Zwanzig1} \\
  \mathcal{Q}^{\rm Z}\frac{\partial}{\partial t} \rho_N(\Gamma,t) &=& - \mathcal{Q}^{\rm Z}\ii\mathcal{L} \rho_N(\Gamma,t) = - \mathcal{Q}^{\rm Z}\ii\mathcal{L} \left (\mathcal{P}^{\rm Z}\rho_N(\Gamma,t) + \mathcal{Q}^{\rm Z}\rho_N(\Gamma,t)\right) \label{eq:Zwanzig2}\quad .
\end{eqnarray}
Recall that $\mathcal{P}^{\rm Z}\rho_N(\Gamma,t)$ is proportional to the density $g(\mathbf{a},t)$, i.e.~\eref{Zwanzig1} is the type of equation we are looking for. 
 To shorten the notation, we define a density corresponding to the ``orthogonal'' dynamics $\rho_{\perp}:=\mathcal{Q}^{\rm Z}\rho_N(\Gamma,t)$. Then we express \eref{Zwanzig2} in terms of the dynamics of $\mathcal{P}^{\rm Z}\rho_N(\Gamma,t)$ \cite{zwanzig1960}. An operator equation of the form 
\[
\frac{\partial}{\partial t} \mathcal{G}(t)\circ  = - \mathcal{Q}^{\rm Z}\ii\mathcal{L}\mathcal{G}(t)\circ
\]
is formally solved by
\[
\mathcal{G}(t)\circ = e^{-\mathcal{Q}^{\rm Z}\ii\mathcal{L}t}\mathcal{G}(0)\circ \quad .
\]
We apply this solution iteratively to \eref{Zwanzig2} (depending on context, the names Dyson, Duhamel or Picard are used when refering to this procedure) and obtain
\[
\rho_{\perp}(t) = e^{-\mathcal{Q}^{\rm Z}\ii\mathcal{L}t}\rho_{\perp}(0) - \int_0^t\dd \tau e^{-\tau\mathcal{Q}^{\rm Z}\ii\mathcal{L}}\mathcal{Q}^{\rm Z}\ii\mathcal{L}\mathcal{P}^{\rm Z}\rho_N(\Gamma, t-\tau) \quad .
  \]
Then we insert this expression into \eref{Zwanzig1}, note that the term $e^{-\mathcal{Q}^{\rm Z}\ii\mathcal{L}t}\rho_{\perp}(0)$ vanishes under the projection $\mathcal{P}^{\rm Z}$, and obtain
     \[
          \frac{\partial }{\partial t}\mathcal{P}^{\rm Z}\rho_N(\Gamma,t) = - \mathcal{P}^{\rm Z} \ii \mathcal{L} \mathcal{P}^{\rm Z}\rho_N(\Gamma,t) + \int_0^t d\tau \mathcal{P}^{\rm Z} \ii \mathcal{L} e^{-\mathcal{Q}^{\rm Z} \ii \mathcal{L}\tau}\mathcal{Q}^{\rm Z} \ii \mathcal{L} \mathcal{P}^{\rm Z} \rho_N(\Gamma,t-\tau)  \quad .  
        \]
        From this equation follows the equation of motion for the coarse-grained observables
        \begin{eqnarray}
            \frac{\partial g(\mathbf{a},t)}{\partial t} &=& - \sum_j \frac{\partial}{\partial a_j}
             \left(v_j(\mathbf{a})g(\mathbf{a},t)\right) \label{eq:ZwanzigEOM} \\
            & & + \int_0^t \; d\tau \; \int d\mathbf{a}^\prime \sum_{j,k} \frac{\partial}{\partial a_j}\left(W(\mathbf{a})K_{jk}(\mathbf{a},\mathbf{a}^\prime,\tau)\right)) \frac{\partial}{\partial a^\prime_k}\left(\frac{g(\mathbf{a}^\prime, t-\tau)}{W(\mathbf{a}^\prime)} \right) \nonumber
        \end{eqnarray}
        with the transport coefficients
        \[
          v_j(\mathbf{a}) := \int \dd \Gamma\; \delta(\mathbf{A(\Gamma)} - \mathbf{a})\ii\mathcal{L}\mathbf{A}_j(\Gamma) \coloneqq \left\langle \frac{dA_j}{dt} ; \mathbf{a} \right\rangle \]
        and
        \begin{multline}
          K_{jk}(\mathbf{a},\mathbf{a}^\prime,t) \coloneqq\\
          \int \dd \Gamma\;\left(\int \dd \Gamma'\; \ii\mathcal{L}\mathbf{A}_j(\Gamma)e^{-t\mathcal{Q}^{\rm Z}\ii\mathcal{L}}\mathcal{Q}^{\rm Z}\ii\mathcal{L}\mathbf{A}_k(\Gamma')\delta\left(\mathbf{A}(\Gamma') - \mathbf{a}^\prime \right)\right)\delta(\mathbf{A(\Gamma)} - \mathbf{a})\\
          = \left\langle \frac{dA_j}{dt} e^{-\mathcal{Q}^{\rm Z}\ii\mathcal{L}t}\mathcal{Q}^{\rm Z} \frac{dA_k}{dt} \delta\left(\mathbf{A} - \mathbf{a}^\prime \right); \mathbf{a} \right\rangle \nonumber \quad .
        \end{multline}
        The notation with angular brackets is Zwanzig's notation for the microcanonical average over all phase space points for which $\mathbf{A}(\Gamma) = \mathbf{a}$.
           In ref.~\cite{zwanzig:1961}, Zwanzig states that this equation is ``too complicated to be useful for any but the most formal applications''. Seen from today's perspective this assessment might be too negative, as equations similar to eq.~\ref{eq:ZwanzigEOM} can be tackled by numerical methods with modern computers (see sec.~\ref{sec:NumericalMemory}). In the 1960s, however, Zwanzig proceeded to discuss approximations to eq.~\ref{eq:ZwanzigEOM}.
The assumptions used in ref.~\cite{zwanzig:1961} to simplify eq.~\ref{eq:ZwanzigEOM}  are\begin{itemize}
         \item that $\mathbf{A}$ varies slowly in time compared to the microscopic degrees of freedom and therefore time derivatives $\frac{d^nA_k}{dt^n}$ can be neglected for orders $n>2$,
         \item that $g(\mathbf{a},t)$ remains sharp, i.e.~fluctuations around the mean remain small for all times.
         \end{itemize}
         We define ensemble averages 
          \[
            \alpha_j(t) \coloneqq \int d\mathbf{a}\; a_j\; g(\mathbf{a},t) \quad ,
            \]
          take their time derivatives, combine them with the equation of motion for $g(\mathbf{a},t)$, \eref{ZwanzigEOM}, and apply approximations according to the two assumptions given above to obtain an equation of motion for the observables
 
  \begin{multline}
\label{eq:TransportZwanzig}
    \frac{d \alpha_j(t)}{dt} = v_j(\alpha_j(t))\\
     + \int_0^t d\tau \sum_k \left[ K_{jk}(\mathbf{\alpha}(t-\tau),\tau) 
F_k(\mathbf{\alpha}(t-\tau)) + \frac{\partial}{\partial \alpha_k} K_{jk}(\mathbf{\alpha}(t-\tau),\tau)\right] 
\end{multline}
with the thermodynamic force $F_k(\mathbf{a}) \coloneqq \frac{\partial \ln{W(\mathbf{a})}}{\partial a_k}$.

If we compare \eref{TransportZwanzig} to the equations of motion for the Brownian particle, \eref{diffusion} and \eref{simpleLE}, which we discussed in the introduction, we notice a crucial difference: the non-locality in time. In general, the evolution of a coarse-grained observable depends on the entire history of the process, but the stochastic descriptions of Brownian motion introduced by Langevin, Smoluchowski and Einstein are Markovian (i.e.~time-local). Hence these descriptions can only be approximative.

The goal of Zwanzig's work was to generalize Onsager's reciprocal relations and to establish criteria for their validity. We recall the central statement of the reciprocal relations \cite{Onsager1931a, Onsager1931b}: for a system that is perturbed only slightly from equilibrium and that obeys a time-reversible microscopic equation of motion, the evolution of the observables is given by
\begin{equation}
\label{eq:TransportOnsager}
\frac{d \alpha_j(t)}{dt}=\sum_k L_{jk}F_k\left(\alpha_1,\alpha_2\ldots\alpha_n\right) \quad ,
\end{equation}
where the quantities $L_{jk}$ are {\it transport coefficients} and $F_k$ is defined as above. Direct comparison of \eref{TransportZwanzig} to \eref{TransportOnsager} shows that the reciprocal relations can be recovered as a limiting case of the equation of motion derived by Zwanzig, if one chooses the observables $\mathbf{A}$ such that all $v_j=0$ and $\frac{\partial}{\partial \alpha_k}K_{jk}=0$. Then the transport coefficients are given by
\[
L_{jk} = \int_0^\tau \dd t \; K_{jk}(t) \quad ,
\]
where $\tau$ is a time ``of macroscopic size''\cite{zwanzig:1961}, for which the effects of memory decay. In this sense, Onsager's result can be seen as the Markovian limit of Zwanzig's result.

\subsection{Projection Operator Formalism II: Mori's Projection Operator}
\label{sec:Mori}
The relation between the exact equation of motion, \eref{ZwanzigEOM}, and the Langevin Equation, \eref{simpleLE}, becomes clearer if we use a different projection operator. First, however, it is useful to introduce a notation that allows us to distinguish between the field on phase space $\mathbb{A}(\Gamma)$ defined by the observable and the value of the dynamical variable taken at a specific point in time on a specific trajectory of the system $A_t\coloneqq\mathbb{A}(\Gamma_t)$.\footnote{To simplify the equations, we consider the case of a single observable in this subsection. However, the entire derivation can be done analogously for a set of observables.} The time-dependence denoted by a subscript indicates the evolution along a specific trajectory (in contrast to the time-dependence on the level of an ensemble as e.g.~in the transport coefficients in \eref{TransportZwanzig}). This notation, which follows ref.~\cite{Meyer2019}, will become convenient later when we discuss explicitly time-dependent Liouvillians. In Mori's work \cite{mori:1965} the distinction between  $\mathbb{A}(\Gamma)$ and $A_t$ is not clearly made, because Mori intended to study steady-state dynamics.

In sec.~\ref{sec:Zwanzig} we applied the projection operator $\mathcal{P}^{\rm Z}$ to the equation of motion of the phase space density $\rho_N$, the Liouville equation. Here we instead apply a projection operator to the equation of motion of the observable 
\begin{equation}
  \label{eq:EOMHeisenberg}
\frac{\dd A_t}{\dd t} = \frac{\dd \mathbb{A}(\Gamma_t)}{\dd t} = \dot{\Gamma}_t\cdot\frac{\partial\mathbb{A}}{\partial \Gamma}(\Gamma_t)=\ii\mathcal{L}\mathbb{A}(\Gamma_t) \quad ,
\end{equation}
which is formally solved by $A_t=e^{\ii\mathcal{L}t}\mathbb{A}(\Gamma_0)$.
In analogy to quantum mechanics, we can say that we switch from the Schr\"odinger picture to the Heisenberg picture.
We take the time derivative of the solution $e^{\ii\mathcal{L}t}\mathbb{A}(\Gamma_0)$ and insert a projection operator (which we will specify later) to split the equation of motion into two parts
\begin{equation}
  \label{eq:Split}
  \frac{\partial}{\partial t}e^{\ii\mathcal{L}t}\mathbb{A}(\Gamma_0) = e^{\ii\mathcal{L}t}\mathcal{P}\ii\mathcal{L}\mathbb{A}(\Gamma_0) + e^{\ii\mathcal{L}t}\mathcal{Q}\ii\mathcal{L}\mathbb{A}(\Gamma_0) \quad ,
\end{equation}
with $\mathcal{Q}\circ = (1-\mathcal{P})\circ$. Then we make use of the fact that
\[
e^{\ii\mathcal{L}t}\circ = e^{ \mathcal{Q}\ii\mathcal{L}t}\circ + \int_0^t \dd \tau \; e^{\ii\mathcal{L}(t-\tau)}\mathcal{P}\ii\mathcal{L}e^{ \mathcal{Q}\ii\mathcal{L}\tau} \mathcal{Q}\ii\mathcal{L} \circ 
\]
and obtain
\begin{multline}
  \label{eq:Dyson}
\frac{\partial}{\partial t}e^{\ii\mathcal{L}t}\mathbb{A}(\Gamma_0) = e^{\ii\mathcal{L}t}\mathcal{P}\ii\mathcal{L}\mathbb{A}(\Gamma_0) + e^{\mathcal{Q}\ii\mathcal{L}t}\mathcal{Q}\ii\mathcal{L}\mathbb{A}(\Gamma_0)\\ + \int_0^t  e^{\ii\mathcal{L}(t-\tau)}\mathcal{P}\ii\mathcal{L}e^{\mathcal{Q}\ii\mathcal{L}\tau}\mathcal{Q}\ii\mathcal{L}\mathbb{A}(\Gamma_0) \dd \tau \quad .
  \end{multline}
  In this part of the derivation it becomes obvious that the convention of the physics community to name the time-evolution operator $\ii\mathcal{L}$ is rather inconvenient. We could now multiply all factors $\ii$ to simplify the equations, but then the structure of the equations would become unclear. Therefore we opt instead for leaving all factors $\ii$ in place and treating $\ii\mathcal{L}$ as one symbol.
  
  Depending on the type of problem one would like to study, one can now define a suitable projection operator and insert it into \eref{Dyson}. Mori defined
\begin{equation}
  \label{eq:MoriPO}
\mathcal{P}^{\rm M} \mathbb{X}(\Gamma) \coloneqq \frac{\left(\mathbb{X},\mathbb{A}\right)}{\left(\mathbb{A},\mathbb{A}\right)}\mathbb{A}(\Gamma) \quad ,
  \end{equation}
  where
  \begin{equation}
    \label{eq:innerProductEQ}
\left(\mathbb{X},\mathbb{Y}\right) \coloneqq \int\; \dd\Gamma\; \mathbb{X}(\Gamma)\mathbb{Y}(\Gamma) \rho_N^{\text{EQ}}(\Gamma) \quad , 
\end{equation}
i.e.~the correlation between two observables in the equilibrium ensemble is interpreted as an inner product on the space of phase space functions \cite{mori:1965,grabert:1982}. (This implies that we can only consider observables for which the integral in \eref{innerProductEQ} exists.)
Inserting $\mathcal{P}^{\rm M}$ into \eref{Dyson}, we obtain the {\it Generalized Langevin Equation} (GLE)\footnote{In the engineering literature, \eref{Dyson} is often referred to as the GLE rather than the more specific \eref{detGLE}.}
 \begin{equation}
   \label{eq:detGLE}
     \frac{\dd A_t}{\dd t} = \omega A_t + \int_0^t\; \dd\tau \; K(t-\tau)A_\tau + f_t \quad .
    \end{equation}
     The {\it drift} $\omega$ is defined as
    \[
      \omega \coloneqq \frac{\left(\ii\mathcal{L}\mathbb{A},\mathbb{A}\right)}{\left(\mathbb{A},\mathbb{A}\right)} \quad ,
    \]
    the {\it fluctuating force} $f_t$ is defined as
    \begin{equation}
      \label{eq:fluctforce}
      f_t \coloneqq e^{t\mathcal{Q}_M\ii\mathcal{L}}\mathcal{Q}_M\ii\mathcal{L}\mathbb{A}(\Gamma_0)
    \end{equation}
    and the {\it memory kernel} $K(t)$ is related to the fluctuating force by
    \begin{equation}
      \label{eq:EQmemoryKernel}
      K(t) \coloneqq \frac{\left(\ii \mathcal{L} e^{t\mathcal{Q}_M\ii\mathcal{L}}\mathcal{Q}_M\ii\mathcal{L}\mathbb{A},  \mathbb{A} \right)}{\left(\mathbb{A}, \mathbb{A} \right)} = -\frac{\left(f_0,f_t\right)}{\left(\mathbb{A},\mathbb{A}\right)} \quad .
    \end{equation}
    The time-dependence in parentheses indicates that $K(t)$, in contrast to $f_t$, is an ensemble-averaged quantity.
    \eref[Eq.]{EQmemoryKernel} is a generalized version of the fluctuation dissipation relation, \eref{simpleFDT}.
    If the evolution of the observable $A_t$ is slow compared to the evolution of the degrees of freedom we have integrated out, then the fluctuating force, $f_t$, which by construction is orthogonal to $\mathbb{A}$, evolves fast. Under these conditions, the kernel $K(t-\tau)$ approaches a delta distribution $\delta(t-\tau)$ and the similarity to the Langevin Equation, \eref{simpleLE}, becomes evident.
    
    We need to recall, however, that $f_t$ is deterministic and that the GLE as derived by Mori is not a stochastic differential equation. Often, in the literature,  $f_t$ is replaced by a continuous-time stochastic process $X_t$, which is chosen such that its correlations fulfill \eref{EQmemoryKernel} (see sec.~\ref{sec:NumericalMemory} for a discussion of methods to generate such processes). The stochastic equation of motion obtained in this way
    \begin{equation}
      \label{eq:GLE}
      \dd A_t  = \left(\omega A_{t}  + \int_{0}^{t} K(t-\tau)A_{\tau} \, \dd \tau\right) \dd t  + \dd X_t 
    \end{equation}
    is also called the Generalized Langevin Equation.
    We remark that the step from \eref{detGLE} to \eref{GLE} is non-trivial and needs to be taken with care.
    Langevin's mathematical negligence has been straightened out using It\=o calculus \cite{Oksendal2013,Kampen1992}. In principle, the same formalism can be applied to integro-differential equations of the Volterra form to obtain stochastic integro-differential equations such as \eref{GLE} \cite{padgett1972,berger1980}. However, there are conditions on the process $X_t$ and on the function $K(t-\tau)$, which need to be fulfilled for a unique solution to exist and to be stable against perturbation. In the physics literature, these conditions are usually not verified before \eref{GLE} is used.

    In summary, if we assume that the distribution of microstates is stationary, and that there is time-scale separation between the observable of interest and the other degrees of freedom, and if we interpret the fluctuating force as a stochastic process, the projection operator formalism allows us to derive the Langevin equation from Hamiltonian dynamics.

    \subsection{Nonlinear Versions of the Langevin Equation}
    \label{sec:nonlinear}
    In the introduction, sec.~\ref{sec:Brownian}, we recalled Langevin's arguments and wrote down a stochastic equation of motion for the velocity of a Brownian particle based on an analogy to Newtonian mechanics. We could, of course, also have used the position of the particle as the coarse-grained variable and could have suggested an equation of the form
    \begin{equation}
\label{eq:nlLE} m \ddot{x}_t = -\left.\frac{\dd U(x)}{\dd x} \right\vert_{x=x_t} - \gamma \dot{x}_t + \sqrt{2\gamma k_BT}\xi_t \quad , 
      \end{equation}
     where $\gamma$ is a friction coefficient, $U(x)$ is an effective potential, $k_BT$ is the thermal energy and $\xi_t$ is white Gaussian noise. Note that the effective potential $U(x)$ takes $x_t$ as input, hence in contrast to the equations discussed in sec.~\ref{sec:Mori}, this version of the GLE is {\it nonlinear}. We followed the convention of the physics literature and used $\ddot{x}$ and $\dot{x}$ to denote the time derivatives. This notation emphasizes the interpretation of \eref{nlLE} as the Newtonian equation of motion of a particle in a potential energy landscape -- an interpretation which seems to enjoy great popularity in introductory texts on biomolecular modelling. However, as said above, the terms $\ddot{x}$ and $\dot{x}$ need to be understood in the sense of stochastic processes, and the functional form of $U(x)$ depends on the properties of all other degrees of freedom of the system. Hence the analogy to Newtonian mechanics should not be overstretched.

     \subsubsection{One particle in a harmonic bath}
     \label{sec:harmonicBath}
     Like the Langevin Equation \eref{simpleLE}, also nonlinear differential equations of a form similar to \eref{nlLE} can be derived by means of projection operators \cite{chorin2002,Vrugt2020,Hijon2010,izvekov2017}.
Before we discuss a full derivation, we first consider a simple example as discussed in the book on non-equilibrium statistical mechanics by Zwanzig \cite{Zwanzig73,Zwanzig01}. We consider one particle  of mass $m$, position $x$ and momentum $p$, which is coupled to $N$ independent harmonic oscillators and an external potential $V_{\rm ext}(x)$. \footnote{There exist straightforward extensions of this type of analysis to nonlinear coupling \cite{Cortes85} and to systems under oscillatory external driving \cite{Bingyu18}.} The Hamiltonian of the particle is
\[
\mathcal{H}_{\rm particle}(x,p) \coloneqq \frac{p^2}{2m} + V_{\rm ext}(x) \quad ,
\]
and the Hamiltonian of the harmonic bath, including the coupling to the particle, is
\[
  \mathcal{H}_{\rm bath} \coloneqq \sum^N_{j=1} \left(\frac{p_j^2}{2} + \frac{1}{2}\omega_j^2
    \left(q_j - \frac{\gamma_j}{\omega_j^2}x \right)^2\right) \quad ,
\]
where $q_j$ and $p_j$ are the positions and momenta of the bath particles, $\omega_j$ are the frequencies of the oscillators and $\gamma_j$ are the coupling constants. The equations of motion of the bath can be solved exactly
\[
{q_{j}}_t - \frac{\gamma_j}{\omega_j^2}x_t = \left({q_{j}}_{0} - \frac{\gamma_j}{\omega_j^2}x_0\right)\cos \omega_jt + {p_{j}}_{0}\frac{\sin\omega_jt}{\omega_j} - \gamma_j \int_0^t\dd \tau \frac{p_\tau}{m} \frac{\cos\omega_j(t-\tau)}{\omega_j^2}\quad .
\]
If we insert the solutions ${q_{j}}_t$ into the equation of motion of the momentum of the particle, we obtain a GLE which looks similar to \eref{nlLE}, 
\begin{equation}
  \label{eq:harmonicBath}
\frac{\dd p_t}{\dd  t} = - \left. \frac{\dd }{\dd x}V_{\rm ext}(x)\right\vert_{x=x_t} - \int_0^t\dd \tau\; K^{\rm HB}(\tau) \frac{p_{t-\tau}}{m}  + f^{\rm HB}_t \quad .
\end{equation}
 The memory kernel $K^{\rm HB}$ is given explicitly by
\[
K^{\rm HB}(t) \coloneqq \sum_j\frac{\gamma_j^2}{\omega_j^2} \cos {\omega_jt} 
\]
and the fluctuating force is given by
\[
  f^{\rm HB}_t \coloneqq \sum_j\gamma_j{p_j}_0\frac{\sin {\omega_jt}}{\omega_j}
  + \sum_j\gamma_j\left({q_j}_0-\frac{\gamma_j}{\omega_j^2}x_0\right) \cos {\omega_jt}  \quad .
  \]
We have not yet specified the initial conditions on the bath degrees of freedom. If we use the canonical distribution and set
\[\rho(p_j,q_j,0)\propto e^{-\beta \mathcal{H}_{\rm bath}(x_0,{p_j}_0,{q_j}_0)} \quad ,\]
we obtain a fluctuation-dissipation relation similar to \eref{EQmemoryKernel}
\[\langle f^{\rm HB}(t) f^{\rm HB}(t') \rangle = k_BTK^{\rm HB}(t-t')\quad .\]
For suitably chosen coupling constants and frequencies, the support of the memory kernel becomes infinitesimally small and $K^{\rm HB}(t)\to\delta(t)$. Then \eref{harmonicBath} is reduced to the nonlinear Langevin Equation, \eref{nlLE}.

In summary, for the special case of a particle in a harmonic bath, an equilibrium initial distribution of the bath degrees of freedom and the correct choice of the system parameters, the form of \eref{nlLE} is justified.

\subsubsection{The Generalized Langevin Equation with a Potential of Mean Force}
\label{sec:nlGLE}
An equation of the same structure as \eref{nlLE} is widely used in materials science and biophysics to model the coarse-grained dynamics of complex systems \cite{snook2006}.
The nonlinear term $U(x)$ is then interpreted as a {\it potential of mean force}
  \[U^{\rm MF}(x) \coloneqq - k_BT \ln{\left(\rho_X^{\text{EQ}}(x)\right)} \quad ,\]
     where $\rho_X^{\text{EQ}}(x)$ is the probability of finding a value $x$ of the observable $\mathbb{X}(\Gamma)$ in the equilibrium ensemble
     \begin{equation}
       \label{eq:Defrhox}
       \rho_X^{\text{EQ}}(x) \coloneqq \int \dd  \Gamma\; \rho_N^{\text{EQ}}(\Gamma) \; \delta\left(\mathbb{X}\left(\Gamma\right)-x\right) \quad .
     \end{equation}
     Instead of the term {\it potential of mean force}, one also often encounters the terms {\it effective free energy} and {\it free energy landscape} denoted by $\Delta G(x)$. In equilibrium, these are identical, $\Delta G(x) \coloneqq U^{\rm MF}(x)$.

     Also integro-differential equations of a form similar to \eref{harmonicBath} are often used to model coarse-grained variables, such as 
\begin{equation}
  \label{eq:nlGLE}
\mu\frac{\dd ^2x_t}{\dd  t^2} = - \left. \frac{\dd U^{\rm MF}(x) }{\dd x}\right\vert_{x=x_t} - \int_0^t \dd  \tau \; K^{\rm NL}(t-\tau) \left. \frac{\dd  x_{t'}}{\dd  t'}\right\vert_{t'=\tau} + f^{\rm NL}_t \quad ,
\end{equation}
where $\mu$ is a generalized mass and we used the superscript ${\rm NL}$ to indicate that the kernel and the flucutating force enter a nonlinear GLE.
These are versions of the GLE in which the memory term is linear in the time derivative of the coarse-grained variable, but the drift term is replaced by a nonlinear expression. In most cases the systems which are modeled by these equations are much more complex than the example of the single particle in a hamornic bath given above (see ref.~\cite{hernandez1999_2,Bhadauria2015,lei:2016,daldrop:2017,wang2019,wang2020,Ozmaian2019,grogan2020} for a few examples in which this equation of motion is used.) Let us therefore check whether an equation of motion of the form of \eref{nlGLE} can be derived for the general case.

We will present this derivation in more detail than the discussions of the previous sections, because we are not aware of a similar analysis in the literature. To skip the details of the derivation, the reader may read the following brief summary and move to sec.~\ref{sec:Stochastic}.
\begin{itemize}
  \item
    In general, a nonlinear potential of mean force is accompanied by a nonlinear memory term. Equations of the form of \eref{nlGLE} are, at best, approximate.
    \item 
      There is no fluctuation-dissipation relation that would relate the memory kernel of the nonlinear GLE to the corresponding fluctuating force.
      \item Even if there is time-scale separation, the resulting equation of motion in general does not have the form of \eref{nlLE}. 
\item
  Probably, the frequent use of \eref{nlGLE} in models of complex systems is caused by a misunderstanding regarding the exchangeability of the derivations in ref.~\cite{zwanzig:1961,mori:1965} and \cite{Zwanzig01}.
  \end{itemize}

 We start from \eref{Dyson} for a set of observables $\mathbf{A}=\left\{\mathbb{A}_1(\Gamma),\ldots,\mathbb{A}_k(\Gamma)\right\}$, which we interpret as the components of a vector $(\mathbb{A}_1(\Gamma),\ldots,\mathbb{A}_k(\Gamma))^\top$
 \begin{multline}
  \label{eq:DysonTwoOP}
\frac{\partial}{\partial t}e^{\ii\mathcal{L}t}\mathbf{A}(\Gamma_0) = e^{\ii\mathcal{L}t}\mathcal{P}\ii\mathcal{L}\mathbf{A}(\Gamma_0) + e^{\mathcal{Q}\ii\mathcal{L}t}\mathcal{Q}\ii\mathcal{L}\mathbf{A}(\Gamma_0)\\ + \int_0^t  e^{\ii\mathcal{L}(t-\tau)}\mathcal{P}\ii\mathcal{L}e^{\mathcal{Q}\ii\mathcal{L}\tau}\mathcal{Q}\ii\mathcal{L}\mathbf{A}(\Gamma_0) \dd \tau \quad .
\end{multline}
The equation of motion we aim for, \eref{nlGLE}, is of second order in time and contains only one coarse-grained variable. To produce an equation of this form, we treat the observable $\mathbb{A}(\Gamma)$ and its time derivative as two observables $\mathbf{A}_t:=(\mathbb{A}(\Gamma_t), \dot{\mathbb{A}}(\Gamma_t))^\top$, and then use \eref{DysonTwoOP}. 

The next step is to choose a suitable projection operator. As \eref{Defrhox} defines a so-called {\it relevant density}, which relates a phase space function (here $\mathbb{X}(\Gamma)$) to the ``macroscopic value'' it takes in equilibrium (here $x$), the projection operator formalism by Zwanzig (sec.~\ref{sec:Zwanzig}) is a reasonable choice. However, it turns out that the self-evident definition
 \[
       \rho^{\rm EQ}_{\mathbf{A}}(\mathbf{a}) \coloneqq \int \dd  \vec{r}^N \dd  \vec{p}^N \rho_N^{\text{EQ}} \left(\vec{r}^N,\vec{p}^N\right) \; \delta\left( \mathbf{A}\left(\vec{r}^N,\vec{p}^N\right)-\mathbf{a}\right) 
     \]
     does not produce an equation of the desired form. Instead, we use a modified density as introduced by Lange and Grubm\"uller \cite{Lange2006}
     \[
	\rho_{\mathbf{A}}(\mathbf{a}) \coloneqq \int\dd \Gamma\,\rho_N^{\text{EQ}}(\Gamma)\delta\left(\mathbf{A}(\Gamma)-\mathbf{a}\right)\left\|\nabla_{r}\mathbb{A}(\Gamma)\mat{M}^{-1/2}\right\|^2\quad ,
      \]
       where $\mat{M}$ is the mass matrix with diagonal elements corresponding to the particle masses and $\nabla_r$ refers to the gradient with respect to all components of $\vec{r}^N$.
This definition is useful for the following reason:
   Assume the first observable, $\mathbb{A}(\vec{r}^N)$, depends only on the positions and not on the momenta $\vec{p}^N$. As the second observable is the time derivative of the first one, we obtain
   \[
     \mathbf{A}_2(\Gamma)=\ii\mathcal{L}\mathbf{A}_1(\Gamma)=\ii\mathcal{L}\mathbb{A}(\Gamma)= \left(\nabla_r\mathbb{A}(\Gamma)\right)\mat{M}^{-1}\vec{p}^N \quad .
   \]
   The factor $\left\|\nabla_{r}\mathbb{A}(\Gamma)\mat{M}^{-1/2}\right\|^2$ in the reduced density will absorb terms related to $\left(\nabla_r\mathbb{A}(\Gamma)\right)\mat{M}^{-1}$ in the next steps of the derivation.

   We define a projection operator
\begin{equation}
    \label{eq:ZwanzigProjector2}
	\mathcal{P}\mathbb{Y}(\Gamma) \coloneqq \frac{1}{\rho_{\mathbf{A}}\left(\mathbf{A}(\Gamma)\right)}\int\dd \Gamma'\,\rho_N^{\text{EQ}}(\Gamma')\delta\left(\mathbf{A}(\Gamma')-\mathbf{A}(\Gamma)\right)\left\|\nabla_{r}\mathbb{A}(\Gamma')\mat{M}^{-1/2}\right\|^2\mathbb{Y}(\Gamma')
   \quad .
\end{equation}
Note that our definition of the projection operator differs from the one given in ref.~\cite{Lange2006}, which is not normalized correctly.
To simplify the following steps, we assume that the microscopic dynamics is produced by a Hamiltonian of the form $\mathcal{H}(\Gamma)=\sum_i \vec{p}_i^2/(2m_i)+V\left(\vec{r}^N\right)$, i.e.~there is no coupling between positions and momenta.
The first term in \eref{DysonTwoOP} for the second observable (the so-called {\it conservative force term}) has the form
\begin{multline}
  \label{eq:ZwanzigFullDriftTerm}
	e^{\ii\mathcal{L}t}\mathcal{P}\ii\mathcal{L}\mathbf{A}_2(\Gamma) =\\ e^{\ii\mathcal{L}t}\frac{1}{\rho_{\mathbf{A}}\left(\mathbf{A}(\Gamma)\right)}\int\dd \Gamma'\,\rho_N^\text{EQ}(\Gamma')\delta\left(\mathbf{A}(\Gamma')-\mathbf{A}(\Gamma)\right)\left\|\nabla_{r}\mathbb{A}(\Gamma')\mat{M}^{-1/2}\right\|^2\\
	\times\left(\mat{M}^{-1}\vec{p}'^N\cdot\nabla_{r'}\left(\nabla_{r'}\mathbb{A}(\Gamma')\cdot\mat{M}^{-1}\vec{p}'^N\right)- \right.\\ \left. \nabla_{r'}V\left(\vec{r}'^N\right)\cdot\nabla_{r'}\mathbb{A}(\Gamma')\mat{M}^{-1}\right) \quad .
\end{multline}
If one intends to relate \eref{ZwanzigFullDriftTerm} to the gradient of a potential of mean force (as e.g.~the term $\frac{\dd }{\dd x}U^{\rm MF}(x)$ in \eref{nlGLE}), it is constructive to first compute $\partial \rho_\mathbf{A}/\partial a_1$. We start by introducing mass-weighted coordinates $\tilde{r}^N=\mat{M}^{1/2}\vec{r}^N$ and $\tilde{p}^N=\mat{M}^{-1/2}\vec{p}^N$. Hence, $\dd \Gamma=\dd \tilde{\Gamma}$ and we can write
\begin{equation}
	\frac{\partial \rho_A\left(\mathbf{a}\right)}{\partial a_1} = \int\dd \tilde{\Gamma}\,\tilde{\rho}_N^\text{eq}\left(\tilde{\Gamma}\right)\delta\left(\tilde{\mathbf{A}}_2\left(\tilde{\Gamma}\right)-a_2\right)\left\|\nabla_{\tilde{r}}\tilde{\mathbf{A}}_1\right\|^2\frac{\partial}{\partial a_1}\delta\left(\tilde{\mathbf{A}}_1\left(\tilde{\Gamma}\right)-a_1\right) \quad ,\label{eq:derivativeRelevantDensity}
\end{equation}
where all functions with a tilde correspond to the ones defined before, but for mass-weighted coordinates. In order to have a more compact notation we do not write out the dependencies on $\tilde{\Gamma}$ from now on. Next, we use
\[
	\left\|\nabla_{\tilde{r}}\tilde{\mathbf{A}}_1\right\|^2\frac{\partial}{\partial a_1}\delta\left(\tilde{\mathbf{A}}_1-a_1\right) = -\left(\nabla_{\tilde{r}}\tilde{\mathbf{A}}_1\right)\cdot\nabla_{\tilde{r}}\delta\left(\tilde{\mathbf{A}}_1-a_1\right)
\]
to rewrite the last term in \eref{derivativeRelevantDensity}, and we carry out a partial integration over $\tilde{r}^N$ to obtain
\begin{multline}
  \frac{\partial \rho_A\left(\mathbf{a}\right)}{\partial a_1} = \int\dd \tilde{\Gamma}\,\delta\left(\tilde{\mathbf{A}}_1-a_1\right) \nonumber
  \left\{\delta\left(\tilde{\mathbf{A}}_2-a_2\right)\left(\tilde{\rho}_N^\text{EQ}\,\nabla_{\tilde{r}}^2\tilde{\mathbf{A}}_1+ \nabla_{\tilde{r}}\tilde{\rho}^\text{EQ}\cdot\nabla_{\tilde{r}}\tilde{\mathbf{A}}_1 \right) \right.\\+\left.\tilde{\rho}_N^\text{EQ}\,\nabla_{\tilde{r}}\tilde{\mathbf{A}}_1\cdot\nabla_{\tilde{r}}\delta\left(\tilde{\mathbf{A}}_2-a_2\right)\right\} \nonumber \quad .
\end{multline}
Now we exploit the fact that $\mathbf{A}_2=\dot{\mathbb{A}}=\left(\nabla_r\mathbb{A}\left(\vec{r}^N\right)\right)\mat{M}^{-1}\vec{p}^N$, rewrite the last term using
\[
	\nabla_{\tilde{r}}\tilde{\mathbf{A}}_1\cdot\nabla_{\tilde{r}}\delta\left(\tilde{\mathbf{A}}_2-a_2\right) = \nabla_{\tilde{r}}\left(\nabla_{\tilde{r}}\tilde{\mathbf{A}}_1\cdot\tilde{p}^N\right)\cdot\nabla_{\tilde{p}}\delta\left(\tilde{\mathbf{A}}_2-a_2\right)
\]
and carry out a partial integration over $\tilde{p}^N$ to get
\begin{equation}
   \label{eq:drhoda}
	\frac{\partial \rho_A\left(\mathbf{a}\right)}{\partial a_1} =\int\dd \tilde{\Gamma}\,\delta\left(\tilde{\mathbf{A}}_1-a_1\right)\delta\left(\tilde{\mathbf{A}}_2-a_2\right)\left\{\nabla_{\tilde{r}}\tilde{\rho}_N^\text{EQ}\cdot\nabla_{\tilde{r}}\tilde{\mathbf{A}}_1- \nabla_{\tilde{p}}\tilde{\rho}_N^\text{EQ}\cdot\nabla_{\tilde{r}}\left(\nabla_{\tilde{r}}\tilde{\mathbf{A}}_1\right) \right\}.
\end{equation}
If we now use consider a canonical ensemble $\rho_N^\text{EQ}\propto e^{-\beta\mathcal{H}(\Gamma)}$, i.e.
\[
  \nabla_{\tilde{r}}\tilde{\rho}_N^\text{EQ} = (\nabla_{\tilde{r}}\tilde{V})\tilde{\rho}_N^\text{EQ} \quad , \quad \nabla_{\tilde{p}}\tilde{\rho}_N^\text{EQ} = \tilde{p}\;\tilde{\rho}_N^\text{EQ} 
\]
we see that the terms in the curly parentheses in \eref{drhoda} correspond to the second line of \eref{ZwanzigFullDriftTerm} multiplied by ${\rho}_N^\text{EQ}$. 

We define a potential of mean force
\[U(\mathbf{a}) \coloneqq  -k_BT\ln\rho_{\mathbf{A}}(\mathbf{a})\]
If $\mathbb{A}$ is linear in $\vec{r}^N$, then the term $\left\|\nabla_{r}\mathbb{A}(\Gamma')\mat{M}^{-1/2}\right\|^2$ contributes only a constant prefactor $\frac{1}{\mu}$ to \eref{ZwanzigFullDriftTerm} and $\nabla_{\tilde{p}}\tilde{\rho}_N^\text{eq}\cdot\nabla_{\tilde{r}}\left(\nabla_{\tilde{r}}\tilde{\mathbb{A}}\right) = 0$. Thus $U(\mathbf{a})$ only depends on $a_1$. This is the case, in particular, if the observable is the position of one particle or the center of mass position of a set of particles $\mathbb{X}(\Gamma)$.
Then we can simplify \eref{ZwanzigFullDriftTerm} to 
\[ 
  \exp(\ii\mathcal{L}t)\mathcal{P}\ii\mathcal{L}\mathbf{X}_2(\Gamma_0) =  \left. -\frac{1}{\mu}  \; \exp(\ii\mathcal{L}t)\frac{\partial U(x)}{\partial x}\right\vert_{x=\mathbb{X}(\Gamma_0)} =
  \left. - \frac{1}{\mu} \; \frac{\partial U(x)}{\partial x}\right\vert_{x=x_t} \quad . \]
We conclude that under these (rather specific) conditions the first term in the equation of motion for $\mathbf{A}_2$ is, indeed, proportional to a gradient of a potential of mean force and the prefactor $\frac{1}{\mu}$ is the inverse of a reduced mass.
 
  Next we consider the integrand in \eref{DysonTwoOP}
  \[
     e^{\ii\mathcal{L}(t-\tau)}\mathcal{P}\ii\mathcal{L}e^{\mathcal{Q}\ii\mathcal{L}\tau}\mathcal{Q}\ii\mathcal{L}\mathbf{A}_2(\Gamma_0) \quad .
  \]
  Following the work by Hijon and co-workers \cite{Hijon2010}, we can write this expression in a form in which the structure becomes slightly more evident if we consider that
  \begin{equation}
    \label{eq:Hijon}
    \mathcal{P}_j\ii\mathcal{L}e^{t\mathcal{Q}\ii\mathcal{L}}\mathcal{Q}\ii\mathcal{L}\mathbf{A}_k(\Gamma_0) = \sum_{l}\left( - B_{kl}(\mathbf{a},t)\frac{\partial U(\mathbf{a})}{\partial a_l} + k_BT \frac{\partial B_{kl}(\mathbf{a},t)}{\partial a_l}\right)
  \end{equation}
  with the memory matrix
  \[
B_{kl}(\mathbf{a},t) \coloneqq  \beta \mathcal{P}\left( \left(\ii\mathcal{L}\mathbb{A}_l \right) \left(e^{t\mathcal{Q}\ii\mathcal{L}}\mathcal{Q}\ii\mathcal{L}\mathbb{A}_k\right)\right) \quad .
\]
(Note that the $\mathbf{a}$-dependence in $B_{kl}(\mathbf{a},t)$ stems from the application of the projection operator.)

Again, we needed to demand that $\mathbf{A}$ is linear in $\vec{r}^N$ and to include the inverse mass in order to be able to use the potential of mean force.
(The work of Hijon and co-workers is more general than the derivation discussed here. They used a different projection operator, for which $\mathbf{A}$ does not need to be linear in $\vec{r}^N$, and \eref{Hijon} still holds. However, the conservative force cannot then be interpreted as a potential of mean force.)
Inserting this expression in \eref{DysonTwoOP}, we obtain the nonlinear GLE for
$\dot{A}_t$
\begin{multline}
  \label{eq:nlGLEcomplete}
  \mu \frac{\dd \dot{A}_t}{\dd t} =  -k_BT \left. \frac{\partial U(\mathbf{a})}{\partial a_1}\right\vert_{\mathbf{a}=\mathbf{A}_t}+ f^{\rm NL}_t\\ - \mu\int_0^t\dd  \tau \left([B](\mathbf{A}_{t-\tau},t)\cdot \left.\frac{\partial U(\mathbf{a})}{\partial \mathbf{a}}\right\vert_{\mathbf{a}=\mathbf{A}_{t-\tau}} -  k_BT \left. \frac{\partial [B](\mathbf{a},t)}{\partial \mathbf{a}}\right\vert_{\mathbf{a}=\mathbf{A}_{t-\tau}}\right)_2  \quad ,
\end{multline}
where $f^{\rm NL}_t$ is the second component of $\mathbf{f^{\rm NL}_t}=e^{\mathcal{Q}\ii\mathcal{L}t}\mathcal{Q}\ii\mathcal{L}\mathbf{A}(\Gamma_0)$, square brackets indicate matrices, and the notation $(\ldots )_2$ indicates the second component of the vector resulting from the operation in inside the parentheses.
  
  The memory term in this equation remains nonlinear in both components of $\mathbf{A}_t$. Hence we do not obtain an expression of the form $K^{\rm NL}(t-\tau)\dot{A}_\tau$ unless we make approximations. We conclude that \eref{nlGLE} is not exact. Note also that $\langle \mathbf{f}^{\rm NL}_0 \mathbf{f}^{\rm NL}_t \rangle$ is not proportional to the memory term. For this type of nonlinear GLE, there is no fluctuation-dissipation relation. 

  Under the approximation $[B](\mathbf{a},t) \approx [B'](\mathbf{a})\delta(t)$, we obtain a Markovian Langevin equation
  \[
  \mu \frac{\dd \dot{A}_t}{\dd t} =  -\left(\mu[B']_2(A_t,\dot{A}_t)+ k_BT \right) \left. \frac{\partial U(\mathbf{a})}{\partial a_1}\right\vert_{\mathbf{a}=\mathbf{A}_t} + \mu \;k_BT \left. \frac{\partial [B']_2(\mathbf{a})}{\partial \mathbf{a}}\right\vert_{\mathbf{a}=\mathbf{A}_{t}} + f'_t \quad .
\]
Note however, that this structure is still more complex than \eref{nlLE}. Thus even in the case of time-scale separation, care is needed when using a coarse-grained model which contains a potential of mean force.

Another common approximation of \eref{nlGLEcomplete} retains the non-locality in time, but removes the dependence of $[B](\mathbf{a},t)$ on the value of the observable $\mathbf{a}$.
  In particular, if $\mathbb{A}$ is the center of mass position $\mathbb{X}$ of a set of particles and $\dot{\mathbb{A}}$ is the corresponding velocity, it is common to assume that $[B](x,\dot{x},t)$  does not depend on the position, and to average the momentum-contribution to $[B](\dot{x},t)$ over the equilibrium distribution. This approximation is used frequently in the work of Voth, Karniadakis, Darve and co-workers \cite{izvekov2017,Li2017,Han2018,Lee2019}.

We have now seen that the memory term derived with a Zwanzig-type projection operator does not have the same structure as the one in \eref{harmonicBath}. To construct a linear memory term, we repeat the procedure with a Mori-type projection onto $\mathbf{A}=(\mathbb{A},\dot{\mathbb{A}})^\top$ (see sec.~\ref{sec:Mori}) as a cross-check. This provides us with an equation of the form
\begin{equation}
  \label{eq:GLEtwoOP}
\frac{\dd \mathbf{A}_t}{\dd t}=[\omega]\mathbf{A}_t + \int_0^t \dd \tau [K](t-\tau)\mathbf{A}_\tau + \mathbf{f}_t \quad ,
  \end{equation}
  where the definitions of $\omega$ and $K$ are the same as in sec.~\ref{sec:Mori}, but for vectorial quantities.
 The second component of the observable is the same as $\ii\mathcal{L}$ applied to the first component, hence the dynamics of the orthogonal degrees of freedom becomes easy to handle.
  We define \[
    k_{ij}(t) =  (\mathcal{Q}\ii\mathcal{L}\mathbf{A}_i, e^{t\mathcal{Q}\ii\mathcal{L}}Q\ii\mathcal{L}\mathbf{A}_j)= \left({(f_i)}_0, {(f_j)}_t\right)\]
  such that
  $[K] = -[k](t)(\mathbf{A},\mathbf{A})^{-1}$. 
  Then we use the fact that
  $(1-\mathcal{P})\ii\mathcal{L}\mathbb{A} = 0$ by construction
  and that $(\ii\mathcal{L}A,A)=0$, because $(\ii\mathcal{L}A,B)=-(A,\ii\mathcal{L}B)$,
  to show that
  \[\omega_{11}= \omega_{22}= 0 \; , \; \omega_{12}=1 \; , \; \omega_{21}=\frac{(\dot{\mathbb{A}},\dot{\mathbb{A}})}{(\mathbb{A},\mathbb{A})}\; , \; k_{11}(t)=k_{12}(t)=k_{21}(t)\equiv 0\; {\text{and}} \; f_1\equiv 0\; .\]
  From this follows that the first component of \eref{GLEtwoOP} is consistent with the trivial relation $\frac{\dd A_t}{\dd t} = \dot{A}_t$ and that the only non-trivial contribution to the memory kernel of the second component is
$K_{22}$
    For the second component of \eref{GLEtwoOP} we thus obtain
    \[
      \frac{\dd \dot{A}_t}{\dd t}=\omega_{21}A_t - \int_{0}^t K_{22}(t-\tau) \dot{A}_{\tau} \dd \tau + {(f_2)}_t\quad .
    \]
    The memory term of this equation is equivalent to the one of \eref{nlGLE}, but the drift term is linear by construction.

    In summary, we assume that a misunderstanding regarding the exchangeability of the derivations from ref.~\cite{zwanzig:1961,mori:1965} and \cite{Zwanzig01} underlies the frequent use of \eref{nlGLE} in models for complex coarse-grained systems. The conclusions drawn in ref.~\cite{Lange2006} regarding the structure of the equation of motion seem to be incorrect.
Another piece of work, which is often cited to justify the use of \eref{nlGLE}, is the derivation by Kinjo and Hyodo \cite{kinjo2007}. We do not discuss this derivation in detail here, but we note that it contains a slip similar to the one in ref.~\cite{Lange2006}. The operator defined in eq.~26 of ref.~\cite{kinjo2007} is not a projection operator (unfortunately, Kinjo and Hyodo did not write out the time-dependence explicitly in all terms of eq.~26 of ref.~\cite{kinjo2007}. However, to us it seems that the last term in eq.~26 should contain a time $t$, and not $t_0$, while the other terms contain a $t_0$.)

   For more details on projection operator formalism and in careful theoretical approaches to the coarse-graining problem, we recommend e.g.~ref.~\cite{Zwanzig01,chorin:2000,Vrugt2020,Hijon2010,izvekov2017}.

\subsubsection{Phase Field Models}
\label{sec:PhaseField}
The idea of using functions $f(\mathbb{A})$ as additional observables in order to produce different forms of nonlinear Langevin equations can be extended \cite{chorin:2000,nordholm1972,nordholm1975}. One might, for instance, aim for equations of motion in terms of powers of the observable and its spatial derivatives
\[
  \mathbf{A} = \{\mathbb{A}, \mathbb{A}^2, \ldots \mathbb{A}^n,\nabla\mathbb{A}, \ldots (\nabla\mathbb{A})^n, \nabla^2\mathbb{A}, \ldots ( \nabla^2\mathbb{A})^n,\ldots\} \quad .\]
Such equations of motion would be similar in structure to {\it phase-field models}.

We briefly recall the basic ideas of phase-field modelling \cite{boettinger2002,steinbach2009}: If we intend to model a phase transition process such as e.g.~the formation of liquid in a supersaturated vapor or the solidification of a melt after a temperature quench, it is useful to introduce an order parameter field $\phi_t(\vec{r})$ which quantifies the amount of stable phase formed at time $t$. As the driving force of the phase transition dynamics, phase field models use a Landau-type free energy density, i.e.~a polynomial in the order parameter and its spatial derivatives. The free energy of the system is then expressed as a functional $G^{\rm PF}[\phi(\vec{r})]$ of the form
\[
  G^{\rm PF}[\phi(\vec{r})] = \int \dd \vec{r}^\prime \left(a_1|\nabla \phi(\vec{r}^\prime)| + a_2| \nabla^2 \phi(\vec{r}^\prime)| +\ldots + b_1\phi(\vec{r}^\prime) + b_2 \phi^2(\vec{r}^\prime) + \ldots \right) \quad .
\]
In order to obtain a model for a specific material, the parameters $a_1,a_2,\ldots,b_1,b_2,\ldots$ are either fitted to material properties or extracted from computer simulation. 
The equation of motion for the order parameter field is assumed to be of the form
\begin{equation}
  \label{eq:Phasefield}
\frac{{\partial}\phi_t(\vec{r})}{\partial t} = \left. - \gamma\frac{\delta G^{\rm PF}[\phi(\vec{r})]}{\delta \phi}\right\vert_{\phi(\vec{r}) = \phi_t(\vec{r})} + \eta_t(\vec{r}) \quad ,
  \end{equation}
where $\gamma$ is a friction coefficient and $\eta_t(\vec{r})$ is a stochastic force-field. The literature on numerical methods to solve these equations for specific geometries and choices of $G^{\rm PF}$ is vast. However, to our knowledge,  \eref{Phasefield} has not yet been derived rigorously from first principles by means of a projection operator formalism. We assume that it should be possible to provide such a derivation following the lines of argument sketched in sec.~\ref{sec:nlGLE}, possibly in combination with the arguments given by Espa{\~n}ol and L\"owen in their derivation of dynamic DFT \cite{espanol2009}, sec.~\ref{sec:DDFT}. This would, in particular, allow to systematically assess the limits of applicability of phase-field models.

\subsection{Projection Operators and Stochastic Processes}
\label{sec:Stochastic}
Coarse-graining methods are not restricted to microscopic, deterministic dynamics. Often researchers begin with an existing coarse-grained model and intend to coarse-grain it further. A typical example is the simulation of biomolecules with an implicit solvent. Here the degrees of freedom of the biomolecule are treated as atomistic, while the solvent is coarse-grained, and the aim is to produce an effective model for the evolution of structural motifs. Another example is the simulation of colloidal particles in a suspension, in which again, the degrees of freedom of the solvent have already been integrated out to obtain the Brownian dynamics of the (``atomistic'') colloids, and the aim is to obtain an effective description of e.g.~a phase transition process.
Projection operator techniques can be used to derive the corresponding equations of motion for these additional steps of coarse-graining, too.

Espa{\~n}ol and V{\'a}zquez considered the dynamics of $m$ coarse-grained degrees of freedom $\mathbf{z} = \{z_1, \ldots z_m\}$, governed by the Liouville equation
    \[\partial_t \rho^{\rm CG}(\mathbf{z},t) = \ii \mathcal{L}^{\rm CG} \rho^{\rm CG}(\mathbf{z},t)\]
    with a Liouvillian of the form
    \[
\ii \mathcal{L}^{\rm CG}\circ \coloneqq -\sum_i\frac{\partial}{\partial z_i} v_i(\mathbf{z})\circ + \sum_{i,j}\frac{1}{2} k_B \frac{\partial}{\partial z_i}\frac{\partial}{\partial z_j}d_{ij}(\mathbf{z})\circ \quad ,
\]
where $d_{ij} = d_{ji}$, i.e.~the distribution of the coarse-grained variables $\rho^{\rm CG}$ evolves according to a Fokker-Planck equation with a drift vector $\mathbf{v}$ and a diffusion tensor $[d]$. 
Espa{\~n}ol and V{\'a}zquez applied a projection operator of the Zwanzig-type to the Fokker-Planck equation to project the dynamics onto a set  of ``more coarse-grained'' variables $\mathbf{a}(\mathbf{z}) = \{a_1(\mathbf{z}),\ldots , a_k(\mathbf{z})\}$. They obtained an equation of motion, which is similar in structure to \eref{ZwanzigEOM}. Under the assumption of time-scale separation, this equation of motion can again be approximated by a Fokker-Planck equation, i.e.~the structure of the equation of motion is invariant under consecutive steps of coarse-graining if there is time-scale separation at each step \cite{espanol2002}.

Kranz and co-workers started out from a stochastic description of a granular system in terms of a Langevin equation for the velocities of the granular particles. Via the BBGKY hierarchy, \eref{BBGKY}, they derived mode coupling equations to describe the glass transition \cite{kranz2013}. In this work, first the average over the stochastic contributions was taken and then the coarse-grained equations of motion for the averaged quantities $\langle A \rangle_{\rm \xi}(t)$ was obtained. 

Recently, Glatzel and co-workers showed that the projection operator formalism by Mori can be applied in a straightforward manner to a wide range of stochastic processes on the trajectory level, i.e.~without taking the average over the distribution of the stochastic process first \cite{glatzel2021}. The conditions on the process, for which the arguments given in ref.~\cite{glatzel2021} hold, are rather weak: The values of the stochastic variables need to be bounded and the stochastic process needs to be either time-discrete or, if it is time-continuous, it must be possible to approximate each realization by an analytic function. The resulting equation of motion for the coarse-grained variables is of the form \eref{detGLE}, as in the deterministic case. Therefore, in particular, the dynamics generated by molecular dynamics simulations using a thermostat or barostat can be described on a coarse-grained level by means of the GLE and all numerical methods developed to analyze \eref{detGLE} (see sec.~\ref{sec:NumericalMemory}) can be applied to data from molecular dynamics simulations.

\subsection{Numerical Methods}
\label{sec:Numerical}

\subsubsection{Blobs, United Atoms and Interaction Sites}
\label{sec:blobs}
 Arguably the most direct way to compute the static values of ensemble averaged observables is to sample the configurational part of the equilibrium distribution, $\rho^{\text{EQ}}_N(\vec{r}^N)$, by means of Monte Carlo (MC) techniques \cite{frenkel2001,allen2017}. These techniques are widely used to compute structure factors, order parameters, thermodynamic coefficients, interfacial tensions, etc.
 If, in addition, we are interested in the dynamics of observables we need to perform Molecular Dynamics (MD) simulations to numerically estimate transport coefficients, effective interactions between groups of particles, and memory functions. In the following we will briefly describe classes of coarse-graining methods that are used in MD simulation.
 
 In soft matter physics and, in particular, in the field of polymer simulation it is a common strategy to group atoms into larger units (which are called blobs, united atoms or interaction sites in the literature), to determine the effective forces between these units, and to propagate an approximative Langevin Equation on the coarse-grained level. In the notation introduced above, the set of observables $\mathbf{A}\left(\Gamma\right)$ then consists of the positions
 \[\mathbf{R}(\Gamma) \coloneqq \left(\vec{\mathbb{R}}_1\left(\Gamma\right), \ldots, \vec{\mathbb{R}}_M\left(\Gamma\right)\right) \]
 and momenta \[\mathbf{P}(\Gamma) \coloneqq\left(\vec{\mathbb{P}}_1\left(\Gamma\right), \ldots, \vec{\mathbb{P}}_M\left(\Gamma\right)\right) \] of $M$ coarse-grained units, preferably for $M\ll N$. 
 If the system is in thermal equilibrium it is clear, in principle, which equation of motion one would need to solve in order to predict the motion of the units; it is \eref{ZwanzigEOM} for the dynamics of the set of observables $\mathbf{A} = \left(\vec{\mathbb{R}}^M,\vec{\mathbb{P}}^M\right)$. However, this equation is very complex even for simple interaction potentials, because it contains memory and the effective interactions are not just pairwise.

 In practical applications it would be convenient if the equation of motion were similar in structure to the Newtonian equation of motion of the microscopic system, i.e.~if it were of the form
 \[
   M_i\frac{\dd ^2\vec{R}_i}{\dd t^2} = \frac{\dd \vec{P}_i}{\dd t} = \sum_i^M \vec{F}\left(\left|\vec{R}_i- \vec{R}_j\right|\right) 
   \]
   or at least of the form
   \begin{equation}
     \label{eq:blobdynamics}
    M_i\frac{\dd ^2\vec{R}_i}{\dd t^2} = \frac{\dd \vec{P}_i}{\dd t} = \vec{F}\left(\vec{R}_1,\ldots \vec{R}_M\right) + \vec{F}^{\rm ST}_i \quad ,
 \end{equation}
 where $M_i$ is the mass associated to unit $i$, $\vec{R}_i=\mathbb{R}_i(\Gamma_t)$, $\vec{P}_i=\mathbb{P}_i(\Gamma_t)$, $\vec{F}\left(\vec{R}_1,\ldots ,\vec{R}_M\right)$ is a many-body force and $\vec{F}^{\rm ST}_i$ is a stochastic force.
 The central questions in the field of coarse-grained soft matter simulation are therefore:
 \begin{itemize}
 \item Under which conditions can we use coarse-grained forces that contain only two-body (or at most three- or four-body) interactions?
 \item Under which conditions can the dynamics of the coarse-grained units be considered Markovian?
   \item If we numerically propagate the coarse-grained equation of motion, errors accumulate. How are these errors related to the microscopic evolution of the system?
 \item  If we wish to switch forward and backward dynamically between the microscopic and the coarse-grained scales, how should we proceed when breaking the coarse-grained units into their constituents?
 \end{itemize}
 These questions cannot be addressed in general. They need to be answered in the context of the class of systems and the materials properties that one would like to model. In particular, the practitioner needs to decide which of the thermodynamic properties, the symmetries and the transport coefficients of a system shall be reproduced to which level of accuracy. Good overviews of coarse-grained soft matter simulation methods can be found e.g.~in ref.~\cite{chen2011,potestio2014,ingolfsson2014,gartner2019} as well as in the introductions of ref.~\cite{Han2018,chaimovich2011,ruhle2009,praprotnik2008}. Here we briefly summarize the key concepts:

In an equilibrium ensemble, the interaction between coarse-grained units is given by the potential of mean force
  \begin{equation}
    U^{\rm MF}\left(\vec{R}_1,\ldots ,\vec{R}_M\right) = -k_BT\; \ln\left({\int \dd \Gamma\;\delta\left[\textbf{R}\left(\Gamma\right) - \left(\vec{R}_1,\ldots ,\vec{R}_M\right)\right] \rho_N^{\text{EQ}}(\Gamma)}\right)\quad . \nonumber
    \end{equation}
    The {\it Boltzmann inversion} or {\it inverse Monte Carlo} approach \cite{ercolessi1994,lyubartsev1995,tschop1998} is a method to  approximate the full potential of mean force $U^{\rm MF}$ by a two-body potential $U(|\vec{R}_i-\vec{R}_j|)$. The effective potential $U(|\vec{R}_i-\vec{R}_j|)$ is parameterized and then optimized by matching the two particle densities, i.e.~one carries out a numerical search for those values of the parameters which minimize the difference between the two-particle density of the coarse-grained model, $\rho^{\rm CG}_2\left(\vec{R}_i,\vec{R}_j\right)$, and the corresponding ``two-observable density'' of the microscopic model
    \[
      \rho^{\text{MIC}}_2\left(\vec{R}_i,\vec{R}_j\right) \coloneqq \frac{\int \dd \Gamma \rho_N^{\rm EQ}(\Gamma)\delta\left(\mathbb{R}_i\left(\Gamma\right) - \vec{R}_i\right) \delta\left(\mathbb{R}_j\left(\Gamma\right) -\vec{R}_j\right) }{\int \dd \Gamma \rho_N^{\rm EQ}(\Gamma)} \quad .
\]
 As the potential of mean force is a static quantity, both MD and MC can be used to sample it.
The method of {\it iterative Boltzmann inversion} is a refined version of Boltzmann inversion, in which the steps of matching the pair distribution functions and sampling the potential are iterated to improve the quality of the approximation \cite{reith2003,rosenberger2016}.   

    Instead of working with the potential of mean force, one can parameterize the effective forces $\vec{F}^{\rm CG}(\vec{R}_i- \vec{R}_j)$ between two coarse-grained units and then minimize
    \[
      \left|\vec{F}^{\rm MIC}\left(\vec{R}_i- \vec{R}_j\right) - \vec{F}^{\rm CG}\left(\vec{R}_i-\vec{R}_j\right)\right| \quad ,
    \]
    where
 \begin{multline}
\vec{F}^{\rm MIC}\left(\vec{R}_i-\vec{R}_j\right) = \\ - \int \dd \Gamma\;\sum_{\vec{r}_i \in u_i}\sum_{\vec{r}_j \in u_j}\frac{\partial V_{\rm int}(\vec{r}_i- \vec{r}_{j})}{\partial  \vec{r}_i} \delta\left(\mathbb{R}_i\left(\Gamma\right) - \vec{R}_i\right) \delta\left(\mathbb{R}_j\left(\Gamma\right) -\vec{R}_j\right)\rho_N^{\text{EQ}}(\Gamma) 
 \end{multline}
 is the sum of the forces that act between the constituents of the two units in the microscopic model averaged over a canonical ensemble \cite{ercolessi1994,izvekov2005}. When models for polymers are designed, additional three-body and four-body forces are usually included to take into account the bond angles and the torsion angles of the polymer backbone. Also other mesoscopic degrees of freedom as e.g.~the size of units in a responsive material can be incorporated in the effective model \cite{Lin2020}, but the methods to optimize the parameters, which we discuss here, remain the same.

 Methods in which the forces or the potential of mean force are approximated, such that the coarse-grained two-particle density is reproduced, are called {\it structure optimization methods}.
 Next to the idea of structure optimization, there is also an approach via the {\it optimization of thermodynamic properties}. In this case experimental thermodynamic reference data such as the bulk density or the bulk compressibility, the surface tension, the persistence length (in the case of polymers), the free energy of vaporizaton and hydration (in the case of lipids) and similar thermodynamic properties are used to determine the parameters of the force field \cite{nielsen2003,marrink2007,shinoda2007,baron2007,he2010,ouldridge2011}.
 
  In this review, we have opted to show derivations which start out from classical microscopic models. However, the derivations presented in sec.~\ref{sec:Zwanzig} and sec.~\ref{sec:Mori} can be applied equally well to quantum mechanical systems \cite{nakajima1958,zwanzig1960,mori:1965} and the resulting equations of motion have the same mathematical structure. Thus, if chemical details need to be taken into account in a coarse-grained model, forces can be fitted to results from ab-initio electronic structure calculations or quantum chemical simulations \cite{glaser1997,izvekov2004,lyubartsev2009,xu2018,heinz2016}.
 
An interesting variant of structure optimization is to minimize the relative entropy $S^{\rm REL}$ between the microscopic model and the coarse-grained model. Shell suggested to use the following definition for the relative entropy \cite{shell2008,chaimovich2011}
 \begin{eqnarray}
   S^{\rm REL} &\coloneqq& \nonumber \int \dd \Gamma \; \rho_N^{\text{EQ}}(\Gamma)\; \ln \left(\frac{\rho_N^{\text{EQ}}(\Gamma)}{\rho^{\rm CG}_M(\Gamma;U,\mathbf{R}^{-1})} \right) \\ && \nonumber + \frac{  \int \dd \vec{R}^M\; \rho_N^{\rm EQ}(\mathbf{R}^{-1}(\vec{R}^M))\; \left(\ln\left\{a\int \dd \Gamma \; \delta\left(\mathbf{R}(\Gamma) - \vec{R}^M\right)\right\} \right) }{\int \dd \Gamma \rho_N^{\rm EQ}(\Gamma)}\\
               &=& \left\langle \ln{\left(\frac{\rho_N^{\text{EQ}}(\Gamma)}{\rho^{\rm CG}_M(\Gamma;U,\mathbf{R}^{-1})}\right)} \right\rangle + \langle S^{\rm MAP} \rangle \quad .
                   \label{eq:Srel}
 \end{eqnarray}
 Here we used the short notation $\vec{R}^M \coloneqq\{\vec{R}_1,\ldots,\vec{R}_M\}$ for the positions of the coarse-grained units. $\rho^{\rm CG}_M(\Gamma;U,\mathbf{R}^{-1})$ is the probability of finding a microscopic state $\Gamma$ if one samples the coarse-grained model according to the equilibrium distribution for a given parameterization of the potential $U(\vec{R}_1\ldots \vec{R}_M)$, and one uses a map $\Gamma=\mathbf{R}^{-1}(\vec{R}_1\ldots\vec{R}_M)$ to break the coarse-grained units into their microscopic components. Further, we introduced a normalization factor $a$, to render the expression in the curly parentheses dimensionless. (This constant seems to be missing in ref.~\cite{chaimovich2011}.)
 
 The first term of \eref{Srel} is the Kullback–Leibler divergence between the equilibrium distribution of the microscopic model and $\rho^{\rm CG}_M(\Gamma;U,\mathbf{R}^{-1})$. The second integral in the first line runs over the volume accessible to the coarse-grained positions. This term is the ensemble average with respect to the equilibrium distribution of the micropscopic states, $\rho^{\rm EQ}_N(\Gamma)$,  over the entropy associated to the mapping between the models, $S^{\rm MAP}\coloneqq \ln\left\{a\int \dd \Gamma \; \delta\left(\mathbf{R}(\Gamma)- \vec{R}^M\right)\right\}$, which is caused by the degeneracy of $\mathbf{R}(\Gamma)$. As Rudzinzki and Noid pointed out, it is inconvenient that \eref{Srel} requires a map $\mathbf{R}^{-1}$ to break up the coarse-grained units \cite{Rudzinzki2011}. Given the degeneracy of $\mathbf{R}(\Gamma)$, the choice of such a map is inevitably ambiguous. Therefore a definition of the relative entropy on the level of the coarse-grained variables is preferable
 \[
S^{\rm REL}[U,\mathbf{R}] \coloneqq k_B \int \dd \vec{R}^M \;\rho_{\mathbf{R}}(\vec{R}^M)\; \ln{\left(\frac{\rho_{\mathbf{R}}(\vec{R}^M)}{\rho^{\rm CG}_M(\vec{R}^M)}\right)} \quad ,
\]
with the relevant density
\[
  \rho_{\mathbf{R}}(\vec{R}^M) \coloneqq \frac{\int \dd \Gamma  \rho_N^{\rm EQ}(\Gamma) \delta\left(\textbf{R}\left(\Gamma\right) - \vec{R}^M\right) }{\int \dd \Gamma \delta\left(\textbf{R}\left(\Gamma\right) - \vec{R}^M\right) } \quad .
  \]
Compared to the structure optimization approach on the two-particle level, this approach, which is based in probability theory rather than liquid state theory, has the advantage that it is quite general and applicable to a large range of different observables. 

The research field of designing suitable effective forces for coarse-grained MD simulations is large and we have obviously not given an exhaustive overview here. The intention of this brief summary was rather to show which assumptions and approximations enter the basic equations of motion which are solved by this type of coarse-grained simulation. Most importantly, these equations of motion are derived under the assumption of time-scale separation and equilibrium conditions (or non-equilibrium steady-state conditions). It is not straightforward to generalize any of the methods listed above to systems out of equilibrium.

\subsubsection{Dissipative Particle Dynamics}
\label{sec:DPD}
For simulations of polymer mixtures, polyelectrolytes, oil-water-surfactant mixtures as well as lipid-membranes, {\it Dissipative Particle Dynamics} (DPD) is a popular method \cite{groot1997,pivkin2011,espanol2017}.  
 DPD works on the level of coarse-grained units (\eref{blobdynamics}) and uses a specific functional form of the forces $\vec{F}\left(\vec{R}_i,\ldots \vec{R}_M\right) + \vec{F}^{\rm ST}_i$ between the units such that the resulting equation of motion has the form
 \begin{multline}
   \label{eq:DPD}
   M_i\frac{\dd ^2\vec{R}_i}{\dd t^2} = \left. -\frac{\partial U(\vec{R}^M)}{\partial \vec{R}}\right\vert_{\vec{R}=\vec{R}_i} \\-
   \sum_{j\ne i, j=1}^{M}\gamma\omega^D(R_{ij})(\vec{v}_{ij}\cdot\vec{e}_{ij})\vec{e}_{ij} + \sum_{j\ne i, j=1}^{M}\sigma\nu^R(R_{ij})\vec{e}_{ij}\xi_{ij} \quad ,
   \end{multline}
   where the sums run over all other other coarse-grained units in the system, $U(\vec{R}^M)$ is the potential energy of the coarse-grained model, $R_{ij} \coloneqq |\vec{R}_i-\vec{R}_j|$ is the distance between two coarse-grained units, $\vec{v}_{ij}\coloneqq \frac{\dd \vec{R}_i}{\dd t} -  \frac{\dd \vec{R}_j}{\dd t}$ is their relative velocity, $\vec{e}_{ij} \coloneqq \frac{\vec{R}_i-\vec{R}_j}{|\vec{R}_i-\vec{R}_j|}$, and $\xi_{ij}$ is white Gaussian noise, which is independent for different pairs of units. This choice implies that the dynamics of the positions $\vec{R}_i$ is derived from a potential energy, while the dynamics of the corresponding velocities is dissipative.  
   The functions $\omega^D(R_{ij})$ and $\nu^R(R_{ij})$ have a finite support in order to restrict the model to local dissipative interactions. To ensure thermodynamic consistency, the friction coefficient $\gamma$ is related to $\sigma$ by $\sigma^2=2\gamma k_BT$ and the weight functions are related by $\omega^D(R_{ij})=\left(\nu^R(R_{ij})\right)^2$. 
   
   Interestingly \eref{DPD} has been used for two rather distinct purposes: the coarse-grained simulation of macromolecules and colloidal systems by means of DPD on the one hand \cite{holm2008} and the numerical solution of the Navier-Stokes equation on the other hand \cite{brady1988,malevanets1999,gompper2009}. In the latter context, the idea is to exploit the fact that \eref{DPD} conserves momentum and therefore correctly models hydrodynamic transport through the coarse-grained fluid. In particular, if we choose the weight function $\omega^D(R_{ij})$ such that it restricts the dissipative interaction to particles within a cuboid cell, and if we replace the dissipative term by a rotation of the relative velocity, we obtain the method called {\it Multi-Particle Collision Dynamics} (MPCD) or {\it Stokesian Rotation Dynamics} (SRD). The numerical integration of \eref{DPD} is then split into two parts, a ``streaming step'' and a ``collision step''. In the streaming step the positions of all units are propagated according to
   \[
\vec{R}_i(t+\Delta t) = \vec{R}_i(t) + \vec{V}_i(t)\Delta t \quad ,
\]
where $\vec{V}_i$ is the velocity of unit $i$.
In the  collision step the velocities are updated according to
\[
\vec{V}_i(t+\Delta t)= \vec{u}(t) + [R]\cdot\left(\vec{V}_i- \vec{u}\right) \quad ,
\]
where $\vec{u}$ is the average velocity of the units in the cell that contains unit $i$, and $[R]$ is a rotation matrix.
The rotation angle, the duration $\Delta$, the number density of units in the collision cell and the mass of the units are free parameters which can be used to fix the transport coefficients of the fluid such as the shear viscosity and the thermal conductivity \cite{malevanets1999,pooley2005}. If other particles are embedded in the fluid, e.g.~colloidal particles or polymers, the interactions between the coarse-grained units of the solvent and these particles can be used to tune their diffusivity.

The {\it Lattice Boltzmann} (LB) method is a closely related approach to model the mesoscopic properties of a fluid. Just as MPCD and SRD, it consists in a ballistic streaming step and a diffusive collision step. However, as the method is based on lattice automata, the continuum fluid is discretized into effective particles which move on a grid rather than in continuous space. The body of literature on the manifold variants of the LB method is quite large. We will therefore not discuss it here, but refer the reader to two very useful review articles, ref.~\cite{succi2001,dunweg2009}, and note that the LB method has been combined with an integral equation solver to take memory effects in the coarse-grained dynamics into account, thus allowing to simulate glass-forming fluids \cite{papenkort2014,papenkort2015}.
   
The specific form of the coarse-grained interaction in the DPD method, \eref{DPD}, has been suggested by means of an educated guess rather than a derivation from first principles \cite{hoogerbrugge1992}. MPCD and LB are based on a more controlled approximation  as they have been constructed to solve the Boltzmann equation, \eref{Boltzmann}, but generally, all methods listed in this section are subject to the same limitations as the ones discussed in sec.~\ref{sec:nlGLE}. Therefore the users of DPD, MPCD, SRD, LB and related simulation methods need to make sure that the Markovian approximation as well as the assumption of pairwise contributions $\omega^D(R_{ij})$ to the dissipative force (resp.~of contributions of the form $[R]\cdot(\vec{V}_i(t)-\vec{u})$) are reasonable for each system that they simulate.

\subsubsection{Backmapping}
Often coarse-grained models are used to speed up simulations of certain processes, such as self-assembly, structure formation or conformational changes of large molecules, but once the process has occurred, the microscopic degrees of freedom are of interest again. Then the coarse-grained units need to be broken up in such a way that a realistic microscopic configuration is produced. This procedure is called {\it backmapping}. It can be carried out with different intentions: to compare information from the atomistic model with experimental data such as e.g.~a structure factor from neutron scattering, to seed a new set of atomistic simulation trajectories  or to use atomistic positions as part of an adaptive multi-resolution scheme.  Accordingly the requirements for a ``good'' backmapping scheme differ between applications. 

In 1998, Tsch\"op and co-workers introduced a backmapping procedure to generate atomistic simulation configurations of a polymer melt for a comparison with neutron scattering data after the melt had been equilibrated in a coarse-grained simulation \cite{tschop1998_2}. The procedure consisted in constructing templates for the groups of atoms, which had been coarse-grained into single units, placing them such that the contour of the atomistic chain matched the one of the coarse-grained chain and then letting the restraints on the atoms go such that the atoms could relax into the nearest local energy minima. A similar strategy had also been employed for a lattice model by Kotelyanskii and co-workers in 1996 \cite{kotelyanskii1996}, who generated self-avoiding random walks on a cubic lattice as templates for polymer backbones and then decorated them with atoms, which they let relax into local minima by MD simulation. With improved methods of placing the atoms, this type of backmapping strategy is still widely used. Atoms can either be placed according to tabulated fragments of molecules \cite{hess2006,peter2008,peter2009}, using geometrical interpolation \cite{gopal2010,wassenaar2014} or simply randomly \cite{rzepiela2010}. These methods are not restricted to equilibrium configurations. They can also be applied to systems under shear \cite{chen2009}.

Recently, Bayesian Inference \cite{peng2019}  as well as machine learning methods \cite{li2020,an2020} have been used to determine optimized atomistic positions. In the machine learning approach a network is trained on a set of atomistic simulation snapshots together with the corresponding coarse-grained configurations. When the trained network is then given unknown coarse-grained configurations it produces atomistic snapshots which contain correlations caused by the coarse-grained structure, such as e.g.~orientational correlations between different groups of atoms in a polymer which are imposed by the global structure of the backbone. On the one hand, this is an advantage over the methods listed above, in which such correlations either had to be included by hand or they had to form as a result of the relaxation procedure. On the other hand, so far the quality of machine learned backmapping procedures still seems to depend strongly on the system under study \cite{li2020,an2020}.   

Bayesian Inference makes use of Bayes' theorem \cite{bayes1763}, i.e.~the probability of a microscopic configuration $\Gamma$ stemming from a coarse-grained configuration $\mathbf{R}$ can be expressed as 
\[
P(\Gamma|\vec{R}^M) = \frac{P(\vec{R}^M|\Gamma)}{P(\vec{R}^M)}P(\Gamma) =\frac{P(\vec{R}^M|\Gamma)}{\rho_{\mathbf{R}}(\vec{R}^M)}\rho_N^{\rm EQ}(\Gamma) \quad ,
\]
 where $P(\vec{R}^M|\Gamma)$ is determined by the map $\vec{R}^M = \mathbf{R}(\Gamma)$. Obviously, if we could compute $\rho_{\mathbf{R}}(\mathbf{R})$ and $\rho_N^{\rm EQ}(\Gamma)$ analytically, we would not need the simulation in the first place. Thus the method relies on using suitable approximate expressions for these probabilities.

Adaptive resolution models couple microscopic and coarse-graind degrees of freedom within one simulation, and they coarse-grain and map back on the fly. Such models have been proposed for soft materials \cite{praprotnik2005, praprotnik2007,praprotnik2008,potestio2013} as well as solids \cite{rudd2000,csanyi2004,Lu2006}. They are used, in particular to study interfaces between phases or crack propagation, as the bulk of a material can often be modelled on a rather coarse scale, while interfaces need a more detailed model. The conditions which the protocol of changing scales should fulfill depend on the system under study. In solids, mostly forces or energies are matched, while schemes for soft materials match potentials of mean force or thermodynamic transport coeffcients. In general, however, care needs to be taken when applying coarse-graining and backmapping methods to phase transformation processes, crack propagation and other types of non-equilibrium processes. After all, the derivations of most commonly used coarse-grained models require either the assumption of equilibrium or of Markovianity.

\subsubsection{Markov State Models}
\label{sec:MSM}
 In the field of biomolecular modelling the systems of interest usually contain components that relax on a large range of different time-scales. If one is interested e.g.~in the dynamics of a protein in aequous solution, the time-scale needed for conformational changes is many orders of magnitude larger than the time-scale needed for the relaxation of the water molecules or the oscillations of the bonds between individual atoms in the protein. For many biomolecules it make sense to assume that most degrees of freedom are equilibrated almost instantenously compared to the motion of the structural motifs, and that this motion can therefore be understood as a stochastic process in a free energy landscape.
In addition, in many cases this landscape consists of local minima which are separated by high free energy barriers.
 Then Kramers theory can be used to simplify the dynamics even further, i.e.~the Langevin equation for the motion of the coarse-grained variables in the free energy landscape is replaced by a set of rate equations which describe the transitions between the minima \cite{Kramers1940}.

 To construct a Markov State Model (MSM), we partition the configuration space of the system into cells which contain one minimum and its basin of attraction each (as in the case of liquid structure theory, sec.~\ref{sec:DFT}, we restrict the partitioning to configurations, because the distribution of the momenta can be factored out). If $P_i(t)$ is the probability of finding the system in a cell $i$ at time $t$ and $k_{ji}$ is the rate of going from cell $j$ to cell $i$, the probability evolves according to
 \[
P_{i}(t+\dd  t) = P_i(t)\left(1-\sum_{j\ne i}k_{ij}\dd  t\right)+ \sum_{j\ne i}P_j(t)k_{ji}\dd  t \quad .
   \]
   Here we assumed that the time-scale on which we model the system is sufficiently coarse-grained for the process of barrier crossing to be Markovian. For the entire system we then have a master equation of the form
   \[
     \frac{\dd \mathbf{P}(t)}{{\dd }t} = \mathbf{P}(t)\cdot[K] \quad ,
   \]
   where $K_{ii}=-\sum_{j\ne i}k_{ij}$ and $K_{i\ne j}= k_{ji}$. 
   This type of equation can be analyzed by well-established means of the theory of continuous-time Markov processes \cite{Feller1957,Feller1971}.
   
Thus in Markov State Modelling the task of coarse-graining is split into the following steps: discretization of phase space in terms of suitable basins in the free energy landscape, computation of the transition rates between these basins, and solving the resulting master equations \cite{Rohrdanz13,Bowman13a,prinz2011,dill1997,onuchic1997,wales2003,Hummer05,sittel2018}. For a biomolecule in water, the microscopic configurational space has a very high dimension. Hence the identification of a set of coarse-grained coordinates, which describe the system well and in terms of which the free energy landscape can be partitioned, is the most difficult step.
To start out with, it is usually (but not always) reasonable to exclude the degrees of freedom of the water from the analysis. However, the remaining state-space of a peptide then still contains dozens to hundreds of degrees of freedom, that of a protein a factor ten to a thousand times more. To a certain extent, researchers can rely on empirical knowledge to make educated guesses on structural motifs that might guide the partitioning. More systematically, linear reduction techniques such as principal component analysis can be used to find structure in the simulation data, as well as nonlinear reduction techniques and machine learning (for a review see e.g.~\cite{sittel2018}.) The quality of the partitioning can be improved additionally by including information on the dynamics of the molecule \cite{Molgedey1994,perez2013,paul2019}.

Once clusters in configuration space have been identified, the transition rates can be determined. In some cases this is possible by means of ``brute-force simulation'', i.e.~by running long molecular dynamics trajectories and counting the transitions between pairs of basins. However, often enhanced sampling methods are needed such as e.g.~Umbrella Sampling \cite{torrie1977}, Wang Landau Sampling \cite{landau2004} or transition path sampling \cite{bolhuis2002}.

The approach we sketched here, which is practical and guided by empirical knowledge, is used by many researchers in the biomolecular modelling community. However, Markov State Modelling can also be put on a mathematical basis. We briefly summarize the very instructive and readable overview given by Koltai and co-workers in ref.~\cite{koltai2018}.
We assume that certain degrees of freedom of the system have already been integrated out, e.g.~the  degrees of freedom of the water molecules, and that we start out from a stochastic model for the remaining degrees of freedom, e.g.~the positions of the atoms in the biomolecule. We consider a continuous-time homogeneous stochastic process $\left\{\vec{x}_t\right\}_{t\ge 0}$ defined on the space of the configurations of $N$ atoms, $X \subset \mathbb{R}^{3N}$. The process is described by a probability density $p:X\times X \rightarrow\mathbb{R}_{\ge 0}$, where $p(\vec{x},\vec{y};t)$ is the conditional probability density of the particles being at positions $\vec{y}$ at time $t$ given that they were initialized at $\vec{x}$ at time $0$, and we normalized  $p(\vec{x},\vec{y};t)$ such that
\[
P(\vec{x}_t\in X|\vec{x}_0=\vec{x}) = \int_X p(\vec{x},\vec{y};t) \; \dd \vec{y} \quad .
\]
In the following, we only consider ergodic processes for which there is a unique stationary density $\mu(\vec{x})$ and for which $p(\vec{x},\vec{y};t)$ is continuous in both arguments. If the process is reversible, $\mu$ fulfills the detailed balance condition
\[
\mu(\vec{x})\;p(\vec{x},\vec{y};t) = \mu(\vec{y})\;p(\vec{y},\vec{x};t) \quad \forall \; \vec{x},\vec{y} \in X, t \ge 0 \quad .
\]
Now we would like to study the dynamics of an observable $\mathbb{A}(\vec{x})$.
The observable is propagated by the Koopman operator $\mathcal{K}(t)$ for a lag time $t$
\[
\mathcal{K}(t) \mathbb{A}(\vec{x})\coloneqq \int_X p(\vec{x},\vec{y};t)\;\mathbb{A}(\vec{y})\;\dd \vec{y} \quad .
\]
This equation is the equivalent of \eref{EOMHeisenberg} for a system with stochastic dynamics.
As we imposed detailed balance, the Koopman operator is identical to the Perron-Frobenius transfer operator $\mathcal{T}(t)$, which propagates a probability density $u(\vec{x})$ with respect to the stationary density as
\[
\mathcal{T}(t) u(\vec{x}) \coloneqq \frac{1}{\mu(\vec{x})}\int_X u(\vec{y})\;\mu(\vec{y})\; p(\vec{y},\vec{x};t)\;\dd \vec{y} \quad .
\]
Hence we can move easily between the ``Heisenberg picture'' and the ``Schr\"odinger picture'' and use properties of $\mathcal{T}_t$ when we analyze the dynamics of observables. If we consider a system under time-dependent external driving, there is no such simple relation between the Koopman operator and the Perron-Frobenius operator anymore (see sec.~\ref{sec:HuguesWork} for the deterministic case and \ref{sec:neMSM} for the stochastic case). In essence, this is why non-equilibrium Markov State Modelling is difficult. 

$\mathcal{K}(t)$  is self-adjoint, i.e.
\[
\left( \mathcal{K}(t) f,g\right)_\mu = \left( f,\mathcal{K}(t) g\right)_\mu \quad ,
  \]
  where $f$ and $g$ are elements of  $L^2_\mu$, the $L^2$-Hilbert space of functions which are integrable with respect to $\mu$, and similar to \eref{innerProductEQ} we defined an inner product
  \[
    \left( f,g \right)_\mu \coloneqq \int_X \; f(\vec{x}) g(\vec{x}) \; \mu(\vec{x})\; \dd \vec{x} \quad .
    \]
    Thus the eigenvalues $\lambda_i^t$ of $\mathcal{K}(t)$ are real-valued, and the eigenfunctions form an orthogonal basis of $L^2_{\mu}$. We call the normalized eigenfunctions of $\mathcal{K}(t)$ $\phi_i$ and expand functions $f \in L^2_\mu$ as
    \[
f=\sum_{i=0}^\infty \left( f,\phi_i\right)_\mu\phi_i \quad .
      \]
      With this we obtain an equation for the dynamics of the observable $\mathbb{A}(\vec{x})$ in terms of the modes $\left( \mathbb{A}(\vec{x}),\phi_i\right)$
      \begin{equation}
        \label{eq:Koopman}
\mathcal{K}(t) \mathbb{A}(\vec{x})= \sum_{i=0}^\infty \lambda_i^t\left( \mathbb{A}(\vec{x}),\phi_i\right)_\mu\phi_i\ \quad .
\end{equation}
    As we imposed the condition of ergodicity, the largest eigenvalue of $\mathcal{K}(t)$ is 1.  We sort the eigenvalues in descending order
$1=\lambda_1^t> \lambda_2^t \ge \lambda_3^t \ge \ldots $.
      If there is a {\it spectral gap}, i.e.~if there is a $k$ for which
      \[
1-\lambda_k^t \ll \lambda_k^t-\lambda_{k+1}^t \quad ,
\]
we consider $\{\lambda_1^t\ldots \lambda_k^t\}$ the {\it dominant spectrum} of $\mathcal{K}(t)$ and truncate the series \eref{Koopman} at order $k$.

The eigenvalues $\lambda_i^t$ are associated with decay times $\tau_i\coloneqq -t/\ln(\lambda_i^t)$. As $\mathcal{K}(t)$ and $\mathcal{T}(t)$ are identical, the existence of a spectral gap in $\mathcal{T}(t)$ implies that the dynamics of any observable, \eref{Koopman}, is dominated by the $k$ longest decay times $\tau_1,\ldots,\tau_k$. The eigenfunctions of $\mathcal{T}(t)$, $\phi_1\ldots\phi_k$, define a set of reaction coordinates which allow us to partition the state space and to build a model.
Expressed in these terms, the task of Markov State Modelling is to estimate the spectrum of  the Perron-Frobenius operator, to identify the spectral gap and to suggest a $k\times k$-matrix $[K]$, the spectrum of which approximates the true spectrum as closely as possible.

Numerically, the optimal matrix $[K]$ can be found by means of a variational principle \cite{noe2013,mcgibbon2015,wu2020}, as e.g.~in the VAMP algorithm introduced by No\'e and co-workers \cite{nuske2014}. Even data obtained from systems out of equilibrium (e.g.~from non-equilibrium molecular dynamics simulation trajectories) can be used to estimate the transition rates of an equilibrium MSM \cite{nuske2017,wan2020}.

Most of the work on practical applications of Markov State Modelling that we found in the literature used thermostats when generating the molecular dynamics trajectories, based on which the configuration space was partitioned and the transition rates were determined. Thus an equilibrium assumption already entered the choice of simulation method. In addition, the metastable states were usually interpreted as being located in the minima of a free energy landscape, i.e.~the implicit assumption was made that the dynamics is governed by an equation form of \eref{nlLE}. Or, at least interchangeability of the Koopman operator and the propagator of the density of states was implicitly assumed when the relevant time-scales  were analyzed. However, Markov State Modelling is not neccessarily an equilibrium method, as the coarse-grained states can, in principle, be defined on a purely dynamical basis. We will discuss non-equilibrium extensions to Markov State Modelling in sec.~\ref{sec:neMSM}.

\subsubsection{Steady States with Memory}
\label{sec:NumericalMemory}
The methods discussed in the previous subsections require time-scale separation between the microscopic and the coarse-grained model for the Markovian resp.~the time-local approximation to hold.\footnote{In the literature one frequently encounters the term {\it Markovian} used in the context of deterministic, time-local equations of motion on the coarse-grained scale such as \eref{detGLE} with $K(t-\tau) \propto \delta (t-\tau)$. Here we follow this convention and imply that both types of dynamics, stochastic and deterministc, are considered, unless specified otherwise.} If this separation is not given, first the memory kernel $K(t-\tau)$ of the effective equation of motion needs to be determined in the simulation of the microscopic system. Then an integro-differential equation rather than a differential equation needs to be solved. 

Before we discuss numerical methods to solve these two tasks, we consider the autocorrelation function of a coarse-grained observable
\begin{equation}
  \label{eq:DefCorrelation} 
C(t) \coloneqq \langle A(t)A(0)\rangle = \int \dd \Gamma\;\rho^{\text{EQ}}_N(\Gamma) \left( e^{i \mathcal{L} t}\mathbb{A}(\Gamma)\right) \;\mathbb{A}(\Gamma)  = \left(e^{i \mathcal{L} t}\mathbb{A},\mathbb{A}\right)\quad .
\end{equation}
Inserting this definition into \eref{detGLE} we obtain an equation of motion for $C(t)$
\begin{equation}
  \label{eq:GLECorrelation}
  \frac{\dd  C(t)}{\dd t} = \omega C(t) + \int_0^t K(t-\tau) \; C(\tau) \rm d \tau \quad .
\end{equation}
Conveniently, the fluctuating force term is averaged to zero. Thus the form of \eref{GLECorrelation} is independent of whether we use the deterministic version of the GLE, \eref{detGLE}, or the stochastic version, \eref{GLE}, to derive it. In \eref{GLECorrelation} the drift $\omega$ is the same constant as in \eref{detGLE} and the memory kernel $K(t)$ is the same function. Hence we have the choice of determining $K(t)$ either from correlation data or from individual trajectories. 

 Note that the relationship between the correlation function  $C(t)$ and the kernel $K(t)$ is a simple convolution.
In principle, it is therefore clear that the memory kernel can be obtained by means of a Laplace transform or a Fourier transform of the observed autocorrelation function of the variable of interest. In practice, however, the number of sampled trajectories, the time-resolution with which $C(t)$ is sampled and the physics of the specific problem (which determines the functional form of the memory kernel) pose conditions on the type of numerical method that can be used. As this is a typical linear inverse problem, the body of mathematical literature on suitable numerical methods is large and cannot be reviewed here.

Instead we give an overview over the physics literature. There is a wide range of work on specific systems in which memory kernels are extracted either from simulation data or from experimental data. Some authors determine the kernel directly in the time domain from discretized measurements of the autocorrelation function \cite{berkowitz:1981,straub1987,Lange2006,shin:2010,carof:2014,torres2015,brennan2018,kowalik2019}, or of its Fourier transform \cite{Yamaguchi2002,townsend2018} or of its Laplace/z-transform \cite{kneller:2001,satija2019}. Other authors use parameterizations and then perform fits; again this is done in the time domain \cite{daldrop:2017}, on the Fourier transform \cite{gottwald:2015} or on the Laplace transform \cite{lei:2016}. Recently also machine learning has been used to carry out these fits \cite{wang2020}. Other authors expand the kernel  \cite{fricks2009,meyer2020,meyer2020num,zhu2020,amati2019} or its Laplace transform \cite{Ma2016,grogan2020} and numerically determine the coefficients of the expansion. Often, these procedures are guided by additional information on the functional form of $K(t)$, e.g.~the limiting behavior at $t=0$ and $t \to \infty$, which is known from theoretical considerations for specific systems and observables \cite{Grebenkov2011,lesnicki2016,townsend2018,vinales2020}. Similar to the iterative Boltzmann inversion method for fitting effective forces, the fit of the memory kernel can also be improved by an iterative procedure \cite{jung2017,wang2020}. A recent review on the application of these approaches in the context of polymer modelling can be found in ref.~\cite{klippenstein2021}.

Once the kernel is known, the corresponding equation of motion needs to be solved. Compared to the fit of the kernel, this is a more difficult problem \cite{chorin2002}. In ref.~\cite{parish2017}, Parish and Duraisamy give a very instructive overview of numerical methods used to approximate memory kernels and to propagate the GLE in the context of hydrodynamic simulations.  
One widely used method, the $t$-model method \cite{chorin:2000,hald2007}, consists in approximating the dynamics orthogonal to the variable of interest, $e^{\mathcal{Q} \ii\mathcal{L}t}\circ$, such that a time-local equation of motion results.\footnote{Ref.~\cite{basu2018} contains a similar idea, but seems to have been developed independently.} To motivate the $t$-model, we first note that the last term in \eref{DefCorrelation} is an inner product of the form that enters the definition of the projection operator $\mathcal{P}^{\rm M}$. We recall \eref{Dyson}
\[
  \frac{\partial}{\partial t}e^{\ii\mathcal{L}t}\mathbb{A}(\Gamma_0) = e^{\ii\mathcal{L}t}\mathcal{P}\ii\mathcal{L}\mathbb{A}(\Gamma_0) + e^{\mathcal{Q}\ii\mathcal{L}t}\mathcal{Q}\ii\mathcal{L}\mathbb{A}(\Gamma_0) + \int_0^t  e^{\ii\mathcal{L}(t-\tau)}P\ii\mathcal{L}e^{\mathcal{Q}\ii\mathcal{L}\tau}\mathcal{Q}\ii\mathcal{L}\mathbb{A}(\Gamma_0) \dd \tau 
\]
 and see that \eref{GLECorrelation} can be obtained from \eref{Dyson} by application of $\mathcal{P}^{\rm M}$.  
Under the approximation of the $t$-model
\begin{equation}
  \label{eq:tmodel}
\mathcal{P}\int_0^te^{\ii\mathcal{L}(t-\tau)}\mathcal{P}\ii\mathcal{L}e^{\mathcal{Q}\ii\mathcal{L}\tau}\mathcal{Q}\ii\mathcal{L}\mathbb{A}_j(\Gamma_0)\dd \tau \approx t e^{t \ii \mathcal{L}}\mathcal{P} \ii \mathcal{L} Q \ii \mathcal{L}\mathbb{A}_j(\Gamma_0)
\end{equation}
the memory term in \eref{GLECorrelation} is simplified to a term linear in time $t$, i.e.~the problem is made Markovian and standard solvers for partial differential equations can be applied. Improvements of the $t$-model consist e.g.~in estimating the temporal support and magnitude of the memory kernel in order to produce time-dependent correction factors for the LHS of \eref{tmodel} \cite{stinis2013,parish2017}. 

Another common approach is not to propagate \eref{GLECorrelation} directly, but to use a GLE of the form of \eref{GLE} on the level of individual trajectories instead and to enforce that in the ensemble average \eref{GLECorrelation} is fullfilled. To this end, the memory kernel is expanded in a suitable series. By means of an additional set of stochastic processes a fluctuating force is then generated which fulfills the fluctuation dissipation relation for each term in the series \cite{hauge1973}.  If, for instance, the kernel can be approximated by a series of the form
\begin{equation}
  \label{eq:Expfit}
K(t) = \sum_{k=1}^N\frac{c_k}{\tau_{k}} \exp{\left(-\frac{t}{\tau_k}\right)} \quad ,
\end{equation}
the corresponding noise $X_t$ can simply be generated by a sum over processes $\xi^k_t$ governed by an autoregressive model (AR) of order 1 
  \[
X_t = \sum_{k=1}^N\xi^k_t {\rm dt} \quad ,
    \]
which are correlated such that $K(t) \propto \langle X_0X_t \rangle$ is fulfilled.
The time-discretized version of the GLE for the observable, \eref{GLE}, can then be integrated in a straightforward manner. We define memory modes
\[
\kappa^{i,k}_t \coloneqq \int_0^t\frac{c_k}{\tau_{k}}\exp{\left(-\frac{t-t^\prime}{\tau_k}\right)}A_{t^\prime}\dd t^\prime
\]
evaluate these modes at each time-step and integrate the equation of motion
\begin{equation}
  \label{eq:GLEexpdecay}
\dd A_t= \omega A_t \dd t + \sum_{k=1}^N\kappa^{i,k}_t \dd t + X_t \quad .
\end{equation}
\eref[Eq.]{GLEexpdecay} can easily be integrated numerically, e.g.~by a Runge-Kutta method \cite{cicotti1980,berkowitz1983,Xiang1991,tuckerman1991,guardia1985,smith1990,Wan1998,gordon2008,baczewski2013}. Bockius and co-workers recently pointed out that the same kind of approach can also be applied directly to a series expansion of the autocorrelation function rather than the kernel \cite{Bockius2021}. For a detailed discussion on how to determine the parameters of the AR process such that the memory is optimally reproduced, see e.g.~ref.~\cite{kneller:2001,fricks2009}.   

Similar in spirit is an approach that has been introduced in the context of solid state physics \cite{stella2014}. Kantorovich and co-workers considered a system that is coupled to a harmonic bath, where in addition to the case discussed by Zwanzig (see sec.~\ref{sec:nonlinear}) the bath degrees of freedom are also coupled to each other by means of harmonic potentials.
The kernel then contains cross-terms between all bath degrees of freedom, but the general structure of the kernel, a series of cosine-functions, remains the same.
In order to solve the GLE for the system numerically, Kantorovich and co-workers suggested extending the state space by a small number of auxiliary degrees of freedom (small compared to the number of bath degrees of freedom), which are coupled linearly to the degrees of freedom of the system. The auxiliary degrees of freedom evolve according to Langevin equations. The idea of the method is to choose the parameters that enter these Langevin equations such that the effect of the auxiliary degrees of freedom on the dynamics of the system approximates the effect of the original bath as closely as possible. The total system (system and auxiliary degrees of freedom) by construction then obeys a set of Markovian Langevin equations, which can be propagated by standard methods. In principle, the justification via the physics of the harmonic bath is not required. If one replaces \eref{Expfit} by a Fourier series (as done e.g.~in ref.~\cite{gottwald:2015}), the two methods are identical.   

In the context of modelling soft matter, the method of extending the state space by auxiliary variables in order to obtain Markovian dynamics is also frequently used \cite{Lange2006,Schaudinnus15,Yoshimoto2017,Li2017,jung2017,lei:2016,Han2018,Lickert20}. In particular, for coarse-grained polymer models, as described in sec.~\ref{sec:blobs}, non-Markovian contributions to the equations of motion for the positions of the units are often taken into account by a memory kernel of the form
\[
  K(t) = \sum_{j}\kappa_j\exp{\left({-\frac{1}{\tau_j}}\right)}\cos\left(\omega_jt - \phi_j\right)
  \]
  i.e.~a sum of exponentially damped oscillators. Again, the equations of motion for the auxiliary variables are Langevin equations and the parameters are fit to approximate the memory kernel.
  
  Many of these approaches seem to have been developed independently in different parts of the physics community. In essence, however, they can all be deduced from the work of Berkowitz and co-workers \cite{berkowitz1983}, which we briefly summarize here.
  We aim at constructing a stochastic process $X_t$, which fulfills the condition $K(t) \propto \langle X_0 X_t \rangle$, to replace the fluctuating force $f_t$ in the GLE, \eref{detGLE}. The spectral density $S(\omega)$ of $X_t$ can be expressed in terms of the Fourier transform of the memory kernel as 
  \[
 S(\omega) \coloneqq  \int_{-\infty}^{\infty}\dd t\; \langle X_0 X_t \rangle e^{-i\omega t} = 2 k_BT  \int_{-\infty}^{\infty}\dd t\; K(t) e^{-i\omega t} \quad .
    \]
 The Wiener-Khinchin theorem states that $S(\omega)$ is related to the power spectrum of $X_t$, $\tilde{X}_\omega \coloneqq   \int_{-\infty}^{\infty}\dd t\;  X_t  e^{-i\omega t}$, by
  \[
\int_{-\infty}^{\infty}\dd t\; \langle X_0 X_t \rangle e^{-i\omega t} = \left\vert \int_{-\infty}^\infty \dd t \; X_t e^{-i\omega t}\right\vert^2 = |\tilde{X}_\omega|^2 \quad ,
\]
 
Hence $K(t)$ determines the amplitude of the Fourier transform of  
$X_t$. If we generate stochastic processes $z_\omega$ according to a zero-mean normal distribution with unit variance and superpose them as
\[
  X_t = \frac{2\pi}{\sqrt{2k_BT}}\int_{-\infty}^\infty \dd \omega \; \sqrt{S(\omega)} z_\omega e^{i\omega t} 
\]
we obtain the ``auxiliary degrees of freedom'' that generate the the correct memory kernel.

If we had to take into account infinitely many processes $z_\omega$ in practice, this approach would not be useful. This class of methods generally relies on the assumption that the expansion of the kernel can be truncated after a few terms. In other words, they rely on the existence of a Markovian embedding for the coarse-grained dynamics, which has a dimension much smaller than the dimension of the original microscopic system. In the case of a system coupled linearly to a harmonic bath, the effect of the relaxation times of the bath modes on the memory kernel can be assessed rather easily. This allows us to approximate the ``orthogonal dynamics''  $e^{\ii\mathcal{L}(t-\tau)}\mathcal{P}\ii\mathcal{L}e^{i\mathcal{Q}\mathcal{L}\tau}Q\ii\mathcal{L}\mathbb{A}(\Gamma_0)$ in a well-controlled way by a small number of additional equations of motion.
However, in the case where the degrees of freedom that we integrate out are those of a large set of interacting particles rather than a bath with simple interactions, we cannot, of course, expect to find such an embedding in general. Then the treatment of the additional degrees of freedom will become too complex for this class of method to be useful in practice.

\section{Non-Equilibrium}
The methods we have discussed in sec.~\ref{sec:Equilibrium} require the microscopic density $\rho_N(\Gamma,t)$ either to be stationary  or to relax into equilibrium. Stationarity is a precondition for the arguments presented in sec.~\ref{sec:DFT}, sec.~\ref{sec:Mori} and sec.~\ref{sec:Numerical}, while the arguments presented in the first part of sec.~\ref{sec:Zwanzig} as well as the derivation of \eref{nlGLEcomplete} in sec.~\ref{sec:nonlinear} still hold in the case of relaxation to equilibrium. This restriction is not always mentioned explicitly in the references we cited. Often the numerical methods presented in sec.~\ref{sec:Numerical} are applied to systems out of equilibrium and even to systems under time-dependent external driving or to systems containing active particles. The author has the impression that many users of simulation software are not aware of the restricted applicability of the coarse-grained models they employ. This is unfortunate, because there exist exact methods to coarse-grain systems out of equilibrium as well as methods to construct systematic and well-controlled approximations. In the following we will first discuss methods to coarse-grain systems which relax into equilibrium, then methods to coarse-grain systems with explicitly time-dependent Liouvillians, and then numerical methods to construct coarse-grained non-equilibrium dynamics based on simulation data.

\label{sec:Noneq}
\subsection{Relaxation to Equilibrium}
\label{sec:Relaxation}
\subsubsection{The Kinetics of Gases}
\label{sec:kinetic}
We return to the BBGKY-hierarchy introduced in sec.~\ref{sec:DFT}, which provides us with a strategy to systematically coarse-grain from the microscopic $N$-particle density $\rho_N(\vec{r}^N,\vec{p}^N,t)$ to the $n$-particle density
\[
\rho_n(\vec{p}^n,\vec{r}^n,t) = \frac{N!}{(N-n)!} \int  \dd\vec{r}^{N-n}\;\dd\vec{p}^{N-n}\; \rho_N(\vec{p}^N,\vec{r}^N,t) \quad .
\]
DFT constructs approximate solutions for the stationary, configurational densities $\rho_n(\vec{r}^n)$. Let us now look for approximations, which might allow us to tackle the time-dependent problem. We recall \eref{BBGKY}
\begin{multline}
   \label{eq:BBGKY2}
    \left( \frac{\partial}{\partial t} + \sum_{i=1}^{n} \frac{ \vec{p}_i}{m} \cdot \frac{\partial}{\partial \vec{r}_i} - \sum_{i=1}^{n}\left(\frac{\partial V_{\rm ext}(\vec{r}_i)}{\partial \vec{r}_i} + \sum_{j=1}^{n} \frac{ \partial V_{\rm int}(\vec{r}_i, \vec{r}_j)}{\partial  \vec{r}_i} \right) \cdot \frac{\partial}{\partial \vec{p}_i} \right) \rho_n \\ = \sum_{i=1}^{n} \int  \frac{\partial V_{\rm int}(\vec{r}_i, \vec{r}_{n+1})}{\partial  \vec{r}_i} \cdot \frac{\partial \rho_{n+1}}{\partial \vec{p}_i} \; d\vec{r}_{n+1} \; d\vec{p}_{n+1} \quad .
    \end{multline} 
    In the 1870s, when Boltzmann studied the kinetics of gases, he considered an equation similar to \eref{BBGKY2} for $n=1$ and approximated the two-body interaction term  by interactions due to uncorrelated collisions between pairs of particles
    \[\int \frac{\partial V_{\rm int}(\vec{r}_1, \vec{r}_{2})}{\partial  \vec{r}_1} \cdot \frac{\partial \rho_{2}(\vec{r}_1, \vec{p}_1,\vec{r}_{2},\vec{p}_{2},t)}{\partial \vec{p}_1} \; d\vec{r}_{2} \; d\vec{p}_{2} \approx \left(\frac{\partial \rho_1(\vec{r}_1,\vec{p}_1,t)}{\partial t}\right)_{\rm coll} \quad .\]
    In this way he obtained an equation of motion for the single particle density \cite{boltzmann1872}
      \begin{multline}
    \label{eq:Boltzmann}
 \left( \frac{\partial}{\partial t} + \frac{ \vec{p}_1}{m} \cdot \frac{\partial}{\partial \vec{r}} - \frac{\partial V_{\rm ext}(\vec{r})}{\partial \vec{r}} \cdot \frac{\partial}{\partial \vec{p}_1} \right) \rho_1(\vec{r},\vec{p}_1,t)\\ = \frac{1}{m}\int \dd \vec{p}_2 \dd \Omega\; \sigma(\Omega, |\vec{p}_2-\vec{p}_1|) \left(\rho_1(\vec{r},\vec{p}^{\;\prime}_1,t)\rho_1(\vec{r},\vec{p}^{\;\prime}_2,t)- \rho_1(\vec{r},\vec{p}_1,t)\rho_1(\vec{r},\vec{p}_2,t) \right) \quad ,
\end{multline}
where $\vec{p}_1$ and $\vec{p}_2$ are the momenta of two particles before they collide at position $\vec{r}$, $\vec{p}_1^{\;\prime}$ and $\vec{p}^{\;\prime}_2$ are the momenta after the collision, and $\sigma(\Omega, |\vec{p}_2-\vec{p}_1|)$ is the cross section for scattering into the solid angle $\Omega$. Despite this strong simplification of the interactions, \eref{Boltzmann} is still a nonlinear integro-differential equation and hence difficult to solve analytically. For a very readable review and discussion of the existence of solutions, although not up to date, see ref.~\cite{diperna1989}. Numerically, the Boltzmann equation can be solved by the Lattice Boltzmann technique \cite{succi2001,dunweg2009}, finite element methods \cite{zienkiewicz2006} or semi-Lagrangian schemes \cite{dimarco2018}.  
    
    \eref[Eq.]{Boltzmann} is not particularly accurate for systems which are dominated by long-ranged electrostatic forces such as plasmas. Vlasov therefore suggested modelling plasmas by a collision-free interaction that takes into account the electromagnetic field generated by the single particle density instead \cite{Vlasov_1968}
    \begin{multline}
      \label{eq:Vlasov}
 \left( \frac{\partial}{\partial t} + \frac{ \vec{p}}{m} \cdot \frac{\partial}{\partial \vec{r}} - \frac{\partial V_{\rm ext}(\vec{r})}{\partial \vec{r}}  \cdot \frac{\partial}{\partial \vec{p}} \right) \rho_1(\vec{r},\vec{p},t) =\\ -\left(\int\dd \vec{r}' \dd \vec{p}' \vec{F}_{\rm em}(\vec{r}^{\;\prime}, \vec{r},t)\rho_1(\vec{r}^{\;\prime},\vec{p}^{\;\prime},t)\right) \cdot \frac{\partial}{\partial \vec{p}}  \rho_1(\vec{r},\vec{p},t) \quad .
\end{multline}
Here we replaced the derivative of the potential $V_{\rm int}$ by the effective electromagnetic force $\vec{F}_{\rm em}$. Using Lorentz's force law for a particle of charge $q$ in an electric field $\vec{E}$ and a magnetic field $\vec{B}$, $\vec{F} = q(\vec{E} + \vec{v}\times\vec{B})$, the electric part of $\vec{F}_{\rm em}$ can be expressed in terms of the single particle density $\rho_1(\vec{r},\vec{p},t)$ via Gauss's law \cite{jackson91}
\[
  \nabla\cdot\vec{E}(\vec{r},t)\approx
  \frac{1}{\epsilon_0} q \int\dd\vec{p}\; \rho_1(\vec{r},\vec{p},t) \quad ,
  \]
where $\epsilon_0$ is the vacuum permittivity and we approximated the $N$-particle structure of the gas by $\rho_1(\vec{r},\vec{p},t)$.
The magnetic part of $\vec{F}_{\rm em}$ is related to $\rho_1(\vec{r},\vec{p},t)$ via the current density and
\[
\nabla\times\vec{B}(\vec{r},t)\approx\mu_0\int\dd\vec{p}\; \frac{\vec{p}}{m} \rho_1(\vec{r},\vec{p},t) + \mu_0\epsilon_0\frac{\partial \vec{E}(\vec{r},t)}{\partial t} \quad ,
\]
where $\mu_0$ is the vacuum permeability.

\eref[Eq.]{Vlasov} follows from \eref{BBGKY2} if we approximate $\rho_2(\vec{p}_1,\vec{p}_2,\vec{r}_1,\vec{r}_2,t)$ by\\ $\rho_1(\vec{p}_1,\vec{r}_1,t) \rho_1(\vec{p}_2,\vec{r}_2,t)$. Despite these approximations the Vlasov equation, \eref{Vlasov}, is not easy to solve. The search for exact solutions as well as the design of numerical methods are topics of active research. Analytical solutions are known for certain classes of problems in plasma physics (see ref.~\cite{chavanis2006b,colonna2016} for an overview) as well as for relativistic and quantized versions in the context of nuclear physics and particle physics \cite{perepelkin2020,mach2021}. If the reader's interest is drawn to the kinetics of gases, we recommend the instructive series of articles by Chavanis on the kinetics of systems with long-ranged interactions \cite{chavanis2006a,chavanis2006b,chavanis2008a,chavanis2008b,chavanis2008c}.

In the 1960s and 70s, the BBGKY hierarchy has been used to derive transport coefficients and time-dependent correlation functions for systems which are close to equilibrium \cite{ortoleva1969,kadanoff1968,lebowitz1969,gross1972}. 
In the context of coarse-graining, the question is whether these approaches can be taken further to construct more complex non-equilibrium models. \eref[Eq.]{Boltzmann} and \eref{Vlasov} constitute only the first order in the hierarchy and they contain severe approximations. Given that they are nevertheless challenging nonlinear integro-differential equations, the author doubts that much progress can be made on the direct route through \eref{BBGKY2}. Instead, the derivation of variational principles such as the one introduced by Gross \cite{gross1972}, the one recently proposed by Sereda and Ortoleva \cite{sereda2013}, or the extensions of DFT, which we will discuss in sec.~\ref{sec:DDFT} and sec.~\ref{sec:PowerFunctionals}, are promising routes to non-equilibrium. However, before we discuss these variational approaches, we consider a different route via a projection operator formalism.

\subsubsection{Time-dependent projection operators I: Grabert's extension of Zwanzig's method}
\label{sec:TimeDependentProjector}
Projection operator formalisms can be extended to systems out of thermal equilibrium if one allows for an explicit time-dependence of the projector \cite{nordholm1975,grabert1978,grabert:1982}. In the 1970s Grabert generalized Zwanzig's projection operator formalism (sec.~\ref{sec:Zwanzig}) by using a time-dependent density of microstates to define the relevant density. We recall the basic steps of his derivation.
In sec.~\ref{sec:nonlinear} we introduced the relevant density corresponding to a set of observables $\mathbf{A}(\Gamma)$
as \[
  \rho_{\mathbf{A}}(\mathbf{a}) = \int \dd \Gamma \; \rho^{\rm eq}_N(\Gamma)\;\delta(\mathbf{A}(\Gamma)-\mathbf{a})
  \quad ,
\]
 i.e.~the relevant density is the average of the equilibrium distribution $\rho_N^{\rm eq}(\Gamma)$ over all points $\Gamma$ in phase space at which the observables $\mathbf{A}(\Gamma) = \{\mathbb{A}_1(\Gamma), \ldots, \mathbb{A}_m(\Gamma)\}$ have the values $\mathbf{a} = \{a_1,\ldots,a_m\}$. Thus $\rho_{\mathbf{A}}(\mathbf{a})$ defines the relation between the observables and their ``macroscopic equilibrium values'' $\mathbf{a}$. This concept can be generalized to the case in which $\rho_N(\Gamma,t)$ evolves in time as it relaxes to equilibrium. (The case of a time-dependent Liouvillian is more difficult and will be discussed in sec.~\ref{sec:HuguesWork}.)
We define a relevant density
\begin{equation}
  \label{eq:rhoExp}
\rho_{\mathbf{A}}(t) \coloneqq  \frac{1}{Z(t)}e^{-\sum_i\lambda_i(t)\mathbb{A}_i(\Gamma)} \quad ,
  \end{equation}
  where $\lambda_i$ is the thermodynamic conjugate to $\mathbf{A}_i$ and $Z(t)$ is a normalization factor. Note that $\rho$ is a function of a phase space point $\Gamma$ via $\mathbf{A}(\Gamma)$. We follow the notation of Grabert here and do not write this dependence explicitly to avoid cluttering up the equations. $Z(t)$ and $\lambda_i$ are set such that
  \[
    \int\dd \Gamma \; \rho_{\mathbf{A}}(t) = 1  \quad , \quad \int\dd \Gamma \;\mathbb{A}_i(\Gamma)\; \rho_N(\Gamma,t) = \int\dd \Gamma\; \mathbb{A}_i(\Gamma)\; \rho_{\mathbf{A}}(t) = a_i(t) \quad \forall t \quad ,
    \]i.e.~the parameters $\lambda_i$ fix the ``macroscopic values'' $\mathbf{a}$, which we will observe when we measure $\mathbf{A}$ at time $t$.
  Then we define a projection operator
  \begin{equation}
    \label{eq:DefPOGrabert}
    \mathcal{P}(t) \mathbb{X}(\Gamma) \coloneqq \int \dd \Gamma'\; \rho_{\mathbf{A}}(t)\; \mathbb{X}(\Gamma') +
    \sum_i\left(\mathbb{A}_i(\Gamma)-a_i(t)\right) \int \dd \Gamma'\; \frac{\partial \rho_{\mathbf A}(t)}{\partial a_i(t)} \mathbb{X}(\Gamma') \quad .
  \end{equation}
  (We use the definition from ref.~\cite{grabert1978}, which is similar, but not identical to Robertson's projection operator \cite{robertson1966}. The interesting aspect of \eref{DefPOGrabert} is that this projection operator allows us to derive separate equations of motion for the averages of the observables and for the fluctuations around the averages, $\delta \mathbf{A}_t\coloneqq\mathbf{A}_t-\mathbf{a}(t)$.)
  
  The Dyson-Duhamel identity, which we used to obtain \eref{Dyson}, can be generalized to time-dependent projection operators provided we apply $\ii \mathcal{L}$ and $\mathcal{P}(t)$ in the correct order
  \begin{equation}
    \label{eq:DysonTimeDep}
    e^{\ii \mathcal{L}t}\circ = e^{\ii \mathcal{L}t}\mathcal{P}(t)\circ + \int_0^t\dd \tau \; e^{\ii \mathcal{L}\tau}\left(\mathcal{P}(\tau)\ii \mathcal{L}\mathcal{Q}(\tau) - \dot{\mathcal{P}} \right) G^+(\tau,t)\circ + \mathcal{Q}(0)G^+(0,t)\circ \quad ,
    \end{equation}
    where $\mathcal{Q}(t)\circ=(1-\mathcal{P}(t))\circ$ and $G^+(\tau,t)$ is the positively time order exponential operator
    \[
      G^+(\tau,t)\circ \coloneqq  \exp_+\left(\int_\tau^t\dd t'\; \ii \mathcal{L}\;\mathcal{Q}(t')\circ \right) \quad .
    \]
    We recall that the equation of motion for the components of $\mathbf{A}$,
        \[ 
\frac{\dd \left(A_i\right)_t}{\dd t} = \frac{\dd \mathbb{A}_i(\Gamma_t)}{\dd t} = \ii\mathcal{L}\mathbb{A}_i(\Gamma_t) \quad ,
\]
is formally solved by $\left(A_i\right)_t = e^{\ii\mathcal{L}t} \mathbb{A}_i(\Gamma,0)$. 
    Applying \eref{DysonTimeDep} to $\mathbb{A}_i(\Gamma,0)$
and inserting the definition of the projection operator, \eref{DefPOGrabert}, we  obtain a Generalized Langevin Equation of the form
      \begin{equation}\label{eq:EOMZwanzigTimeDep}
\frac{\dd (A_i)_t}{\dd t}=v_i(t) + \Omega_{ij}(t)(\delta A_j)_t + \int_0^t\dd \tau \left(K_i(t,\tau) + \phi_{ij}(t,\tau)(\delta A_j)_{\tau} \right) + (f_i)_{t,0} \quad .
\end{equation}
In analogy to the transport coefficients of \eref{ZwanzigEOM}, we defined the {\it organized drift}
\[
v_i(t)\coloneqq \int\dd \Gamma \; \rho_{\mathbf{A}}(t)\;\ii \mathcal{L} \mathbb{A}_i(\Gamma) \quad ,
\]
 the {\it effective frequencies}
  \[
    \Omega_{ij}(t)\coloneqq  \int\dd \Gamma \; \frac{\partial \rho_{\mathbf{A}}(t)}{\partial a_j(t)}\ii\mathcal{L}\mathbb{A}_i(\Gamma) \quad ,
  \]
  the {\it after-effect functions}
  \[
    K(t,\tau) \coloneqq  \int \dd \Gamma \; \rho_\mathbf{A}(\tau) \ii \mathcal{L} \mathcal{Q}(\tau) G^+(\tau,t) \ii \mathcal{L} \mathbb{A}_i(\Gamma) \quad ,
  \]
  the memory functions
  \[
    \phi_{ij}(t,\tau) \coloneqq  \int \dd \Gamma \; \frac{\partial \rho_{\mathbf{A}}(\tau)}{\partial a_j(\tau)} \ii \mathcal{L}\mathcal{Q}(\tau)G^+(\tau,t) \ii \mathcal{L} \mathbb{A}_i(\Gamma) \quad ,
  \]
  and the fluctuating force
  \[
(f_i)_{t,0} \coloneqq  \mathcal{Q}(0)G^+(0,t)\ii \mathcal{L}\mathbb{A}_i(\Gamma_0) \quad .
\]
  Interestingly, the specific structure of Grabert's projection operator leads to a nonlinear equation of motion for the averages $\mathbf{a}(t)$ and a linear equation of motion for the fluctuations $\delta \mathbf{A}_t=\mathbf{A}_t-\mathbf{a}(t)$.
  
In a series of articles in the 1990s, Shea and Oppenheimer applied this type of projection operator formalism to a system consisting of a small number of ``heavy'' particles and a large number of ``light'' particles \cite{shea1996,shea1997,shea1998}. They expressed the resulting equations of motion in terms of the ratio of particle masses, which served as a small parameter for a series expansion. This allowed them to analyze deviations from a system of ideal Brownian particles systematically and to derive a nonlinear theory of hydrodynamic interaction between Brownian particles immersed in a fluid. We presume that this approach might be a useful starting point, if one intends to construct a generalized, non-equilibrium version of the DPD method.

\subsubsection{Dynamic Density Functional Theory (DDFT)}
\label{sec:DDFT}
Another interesting field of application for the type of time-dependent projection operator formalism we discussed in sec.~\ref{sec:TimeDependentProjector} is density functional theory.
The classical version of DFT, which we introduced briefly in sec.~\ref{sec:DFT}, is a well-established tool to predict the structure and thermodynamic equilibrium properties of liquids. (We use the term ``classical'' here to distinguish this type of DFT from the one which is used in electronic structure calculation.) Clearly, it would be desirable to use density functionals, which have been developed for the equilibrium case and thoroughly tested, to predict also dynamical properties. There are several routes to a time-dependent version of DFT \cite{marconi2000,archer2004,espanol2009}. Here we will summarize an elegant route via a time-dependent projection operator formalism, which Espa{\~n}ol and L{\"o}wen introduced about ten years ago \cite{espanol2009}.

We start out by noting that the definition of the relevant density given in sec.~\ref{sec:TimeDependentProjector} can be used to construct a route to DFT (the static version) which is slightly different from the route via the BBGKY hierarchy sketched in sec.~\ref{sec:DFT}.
We take an equilibrium, grand canonical version of \eref{rhoExp}
\begin{equation}
\rho_{\mathbf{A}}(\Gamma) \coloneqq  \frac{1}{Z(\lambda)}\rho^{\rm EQ}(\Gamma)e^{-\beta\sum_i\lambda_i\mathbb{A}_i(\Gamma)} \quad ,
\end{equation}
where  $\rho^{\rm EQ}$ is the grand canonical distribution of microstates. Next,
we replace the set of observables ${\mathbb{A}_i}$ by the microscopic density operator
\[
\mathbf{n}_{\vec{r}}(\Gamma) \coloneqq  \sum_{i=1}^N\delta(\vec{r}_i-\vec{r}) \quad .
  \]
   In contrast to the observables of the previous chapters, $\mathbf{n}_{\vec{r}}$ is not a countable set, as we replaced the discrete index $i$ by $\vec{r} \in \mathbb{R}^3$ and turned the coarse-grained model into a field theory. However, most steps in the projection operator formalism can be applied analogously. The corresponding relevant density is
  \[
\rho_{\mathbf{n}}(\Gamma) = \frac{1}{\Xi[\lambda]}\rho^{\rm EQ}(\Gamma)\exp\left\{ -\beta \int \dd \vec{r} \lambda(\vec{r}) \mathbf{n}_{\vec{r}}(\Gamma)\right\}
\]
with 
\[
  \Xi[\lambda]\coloneqq \sum_{N=0}^\infty \frac{\exp{\left(\beta \mu N\right)}}{N! h^{3N}}
  \int \dd \vec{r}_1\dd \vec{p}_1\ldots \dd \vec{r}_N \dd \vec{p}_N \exp{\left(-\beta \mathcal{H}_N -\beta \sum_{i=1}^N\lambda(\vec{r}_i) \right)} 
\]
and the $N$-particle Hamiltonian $\mathcal{H}_N$. For $\lambda=0$, $\Xi[\lambda]$ is the grand partition function.

In analogy to the effective free energy (or potential of mean force), we can define an effective grand potential as a functional of $\lambda(\vec{r})$
\[
\Omega[\lambda] \coloneqq  -k_BT\ln\Xi[\lambda] \quad .
  \]
  The one-particle density from sec.~\ref{sec:DFT} is given by the grand canonical average over the microscopic density operator, $\rho_1(\vec{r})= \langle \mathbf{n}_{\vec{r}}\rangle_{\lambda}$. Hence we can obtain $\rho_1(\vec{r})$ by means of variation of the effective grand potential with respect to $\lambda(\vec{r})$
  \[
\frac{\delta\Omega[\lambda]}{\delta \lambda(\vec{r})} = \langle \mathbf{n}_{\vec{r}}\rangle_{\lambda} = \rho_1(\vec{r}) \quad ,
\]
where the subscript $\lambda$ indicates the average over $\rho_{\mathbf{n}}$ for a specific choice of $\lambda(\vec{r})$.
Finally, to relate these expressions to an equation of the form of \eref{DFT}, we take the Legendre transform of $\Omega[\lambda]$ to define the {\it density functional}
\begin{equation}
  \label{eq:DefOmegaBar}
\bar{\Omega}[\rho_1] \coloneqq \Omega[\lambda[\rho_1]] - \int \dd \vec{r} \; \rho_1(\vec{r}) \lambda(\vec{r})[\rho_1] 
\end{equation}
such that
 \[
\frac{\delta\bar{\Omega}[\rho_1]}{\delta \rho_1(\vec{r})} = - \lambda(r) \quad .
\]
Recall that $\lambda=0$ at equilibrium, thus we have recovered the condition for the equilibrium one-particle density, which DFT imposes.

These notions can now be generalized to describe the dynamics of relaxation of the one-particle density towards its equilibrium value. We allow for time-dependent fields $\lambda(\vec{r},t)$ and define a relevant density
 \[
\rho_{\mathbf{n}}(\Gamma,t) \coloneqq \frac{1}{\Xi[\lambda(t)]}\rho^{\rm EQ}(\Gamma)\exp\left\{ -\beta \int \dd \vec{r} \lambda(\vec{r},t) \mathbf{n}_{\vec{r}}(\Gamma)\right\} \quad ,
\]
such that
\[
  \rho_1(\vec{r},t) = \int \dd \Gamma \rho_{\mathbf{n}}(\Gamma,t)\mathbf{n}_{\vec{r}}(\Gamma)
  \]
With this choice of $\rho_{\mathbf{n}}(\Gamma,t)$ we could now, in principle, define a time-dependent projection operator as in sec.~\ref{sec:TimeDependentProjector}, and insert it in  
\eref{DysonTimeDep} to derive an equation of motion of the form of \eref{EOMZwanzigTimeDep} for the density field. However, this turns out to be rather involved. Therefore Espa{\~n}ol and L{\"o}wen assumed that there was time-scale separation between the observable and all other variables and neglected terms of order $\mathcal{O}((\ii \mathcal{L} \mathbf{n})^3)$ and higher. This approximation removes the memory term from \eref{EOMZwanzigTimeDep}. If the observables were a countable set $\mathbf{A}=\{\mathbb{A}_1(\Gamma),\ldots,\mathbb{A}_m(\Gamma)\}$, we would obtain an approximate equation of motion of the form
\begin{equation}
  \label{eq:DDFTMarkov}
\frac{\dd (A_i)_t}{\dd t}=v_i(t) +  \sum_j D_{ij}(t)\lambda_j(t) \quad ,
\end{equation}
with
\[
D_{ij}(t)\coloneqq \int_0^\tau \dd t' \int \dd \Gamma\; \rho_{\mathbf{A}}(t)\mathcal{Q}\ii \mathcal{L} \mathbb{A}_j(\Gamma)e^{\ii \mathcal{L}t'}\mathcal{Q}\ii \mathcal{L} \mathbb{A}_i(\Gamma) \quad .
  \]
  The time $\tau$ needs to be intermediate between the time scale on which the integrand decays to zero and the time scale on which the observables evolve.
  
  If we now use the density field $\mathbf{n}_{\vec{r}}$ as the observable instead, \eref{DDFTMarkov} is replaced by a partial differential equation
  \[
\frac{\partial \rho_1(\vec{r},t)}{\partial t} =  \int \dd \vec{r}'\; \nabla_r \nabla_{r'} [D](\vec{r},\vec{r}',t) \lambda(\vec{r}',t) \quad ,
    \]
    where the organized drift $v_i(\vec{r},t)$ has vanished because of time-reversal symmetry and, taking into account the continuity equation for the density field
    \[\ii \mathcal{L}\mathbf{n}_{\vec{r}}(\Gamma) = -\nabla_r\sum_i \vec{v}_i \delta(\mathbb{R}_i(\Gamma)- \vec{r}) =: -\nabla_r \vec{J}_{\vec{r}}(\Gamma) \quad , \]
we have replaced the dissipative matrix $[D]$ by the diffusion tensor 
    \[
[D](\vec{r},\vec{r}',t) = \int_0^\tau \dd t' \int \dd \Gamma \rho_{\mathbf{n}}(t)\vec{J}_{\vec{r}}(\Gamma)\otimes e^{\ii \mathcal{L}t'}\vec{J}_{\vec{r}}(\Gamma) \quad .
\]
Using the definition of $\bar{\Omega}$, \eref{DefOmegaBar}, and integrating by parts, we obtain an evolution equation for the density field. The driving force for the dynamics is the effective grand potential,
\begin{equation}
  \label{eq:DDFT}
\frac{\partial \rho_1(\vec{r},t)}{\partial t} = \int \dd \vec{r}' [D](\vec{r},\vec{r}',t)\nabla_{r'}\frac{\delta\bar{\Omega}[\rho_1]}{\delta \rho_1(\vec{r}',t)} \quad .
\end{equation}
This equation is the starting point for DDFT. From here, approximations of the diffusion tensor $[D]$ or the density functional $\bar{\Omega}[\rho_1]$ can be introduced to obtain models for specific systems.

\subsubsection{Time-dependent projection operators II: Mori}
\label{sec:TimeDependentProjectorMori}
 Coarse-graining over a time-dependent density of microstates $\rho_N(\Gamma,t)$ can also be achieved by means of a linear projection operator \cite{berne1977,snook2006,Meyer2017}.  
 Again, we start out from the equation of motion of the observable \eref{EOMHeisenberg}, but we combine it with a version of the Dyson-Duhamel identity, which differs from \eref{DysonTimeDep} in the time-ordering
  \[
    e^{\ii \mathcal{L}t}\circ = e^{\ii \mathcal{L}t}\mathcal{P}(t)\circ + \int_0^t\dd \tau\; e^{\ii \mathcal{L}\tau}\mathcal{P}(\tau)\left(\ii \mathcal{L}- \dot{\mathcal{P}}(\tau)\right)\mathcal{Q}_\tau  G^-(\tau,t)\circ + \mathcal{Q}(0)G^-(0,t)\circ \quad .
  \]
  Here $G^-(\tau,t)$ is the negatively time-ordered exponential operator
   \[
      G^-(\tau,t)\circ \coloneqq  \exp_-\left(\int_\tau^t\dd t'\; \ii \mathcal{L}\;\mathcal{Q}(t')\circ \right) \quad .
    \]
  We obtain the equation of motion
  \begin{multline}
    \label{eq:EOMMoriRelax}
 \frac{\dd A_t}{\dd t}=e^{\ii\mathcal{L}t}\mathcal{P}(t)\ii \mathcal{L}\mathbb{A}(\Gamma,0) + \int_0^t\dd \tau\;
 e^{\ii\mathcal{L}\tau}\mathcal{P}(\tau)(\ii \mathcal{L} - \dot{\mathcal{P}}(\tau))\mathcal{Q}(\tau) G^-(\tau,t) \ii \mathcal{L}\mathbb{A}(\Gamma,0)\\ + \mathcal{Q}(0)G^-(0,t)\ii\mathcal{L}\mathbb{A}(\Gamma,0)\; .
 \end{multline}
 Here we have again restricted the derivation to a single observable to avoid cluttering the equations, but all statements hold for sets of observables as well.
 We define a time-dependent version of Mori's projection operator
 \begin{equation}
   \label{eq:TimeDependentMoriPO}
\mathcal{P}(t)\mathbb{X}(\Gamma) \coloneqq \frac{\left(\mathbb{X},\mathbb{A}\right)_t}{\left(\mathbb{A},\mathbb{A}\right)_t}\mathbb{A}(\Gamma) \end{equation}
with the inner product
\begin{equation}
  \label{eq:TimeDependentInnerProduct}
\left(\mathbb{X},\mathbb{Y}\right)_t \coloneqq \int\; d\Gamma\; \left(e^{\ii \mathcal{L}t}\mathbb{X}(\Gamma)\right)\;\left(e^{\ii \mathcal{L}t}\mathbb{Y}(\Gamma)\right)\; \rho_N(\Gamma,0) \quad . 
\end{equation}
We have used parentheses to indicate which objects the operators act on. In the following we will drop the parentheses and assume that operators act only on the phase space fields right next to them unless stated otherwise.

    The definition of the projection operator, \eref{TimeDependentMoriPO}, is rather natural in the sense that it reflects the way simulation data or experimental data is usually analyzed -- we initialize a set of systems with a certain distribution of microstates $\rho_N(\Gamma,0)$, let them evolve to a time $t$ and measure either ensemble averages of observables
    \[
      \langle X_t \rangle_{\rm NEQ} \coloneqq \int \dd \Gamma \;  e^{\ii \mathcal{L}t} \mathbb{X}(\Gamma) \rho_N(\Gamma,0)
    \]
    or of correlation functions
    \[\langle X_t Y_t \rangle_{\rm NEQ} \coloneqq \int\; d\Gamma\; e^{\ii \mathcal{L}t}\mathbb{X}(\Gamma)\;e^{\ii \mathcal{L}t}\mathbb{Y}(\Gamma)\; \rho_N(\Gamma,0) = \left(\mathbb{X},\mathbb{Y}\right)_t \quad .\]
    The angle brackets indicate the average taken over the so-called ``bundle'' or ``swarm'' of non-equilibrium trajectories which have evolved from the initial distribution $\rho_N(\Gamma,0)$.
    In analogy to Mori's work, sec.~\ref{sec:Mori}, we define the generalized drift
    \[
      \omega(t) \coloneqq \frac{\left(\mathbb{A}, \ii \mathcal{L}\mathbb{A} \right)_t}{\left(\mathbb{A}, \mathbb{A} \right)_t} \quad ,
    \]
    the memory kernel
    \begin{equation}
      \label{eq:timeDependentKernel}
K(\tau,t) \coloneqq \frac{\left(\mathbb{A}, (\ii \mathcal{L} - \dot{\mathcal{P}}(\tau))\mathcal{Q}(\tau) G^-(\tau,t)\ii \mathcal{L} \mathbb{A} \right)_\tau}{\left(\mathbb{A}, \mathbb{A} \right)_\tau}
      \end{equation}
      and the fluctuating force
      \begin{equation}
        \label{eq:timeDependentFF}
f_{t,t'}\coloneqq e^{\ii \mathcal{L}t'}\mathcal{Q}(t')G^-(t',t)\ii\mathcal{L}\mathbb{A}(\Gamma,0) \quad .
\end{equation}
Inserting these definitions into \eref{EOMMoriRelax}, we obtain the {\it non-stationary, linear, Generalized Langevin Equation} (nsGLE)
\begin{equation}
  \frac{\dd A_{t}}{\dd t} = \omega(t)A_{t} + \int_{0}^{t} K(\tau,t)A_{\tau} \, \dd \tau + f_{0,t} \label{eq:nsGLE} \quad .
\end{equation}
This equation resembles the GLE, \eref{detGLE}, but similarly to the organized drift $\vec{v}(t)$ of \eref{EOMZwanzigTimeDep}, the generalized drift $\omega(t)$ depends explicity on time. Also similar to the after effect function of \eref{EOMZwanzigTimeDep}, the memory kernel $K(\tau,t)$ depends on two times (and not just on the difference $t-\tau$), and the fluctuating force depends on the initial conditions as well as on time. These time-dependences reflect the fact that the distribution of microstates is not stationary.

If the dynamics of $A_t$ is much slower than the dynamics of the other degrees of freedom, $K(\tau,t) \to \gamma(t)\delta(t-\tau)$, i.e.~we obtain the time-local (``Markovian'') limit of \eref{nsGLE} with a time-dependent friction coefficient $\gamma(t)$ (in contrast to e.g.~the time-local  limit of \eref{Hijon}, where the friction coefficient depends on the value of the observable rather than on time due to the choice of a nonlinear projection operator).

In sec.~\ref{sec:NumericalMemory} we remarked that it can be useful to consider the autocorrelation function of the observable if one intends to reconstruct the memory kernel from measured data. We define the non-stationary equivalent to \eref{DefCorrelation}
\begin{equation}
  \label{eq:DefCorrelationTimeDep} 
C(t',t) \coloneqq \langle A_{t'}A_t\rangle_{\rm NEQ} = \int\; \dd \Gamma \;   e^{i \mathcal{L} t'}\mathbb{A}(\Gamma) \; e^{i \mathcal{L} t}\mathbb{A}(\Gamma) \; \rho_N(\Gamma,0)\quad ,
\end{equation}
and insert $C(t',t)$ in \eref{nsGLE}. Due to the linearity of the projection operator, the fluctuating force term is averaged to zero at all times. We obtain an equation of motion for the autocorrelation function
\begin{equation}
  \label{eq:nsGLECorrelation}
  \frac{\dd  C(t',t)}{\dd t} = \omega(t) C(t',t) + \int_{t'}^t C(t',\tau)\;K(\tau,t) \; \dd \tau \quad .
\end{equation} 
In contrast to \eref{GLECorrelation}, the integral in \eref{nsGLECorrelation} is not a convolution. Hence we cannot determine $K(\tau,t)$ from measured data by means of a numerical inverse Laplace or Fourier transform (as done for the stationary case in the references listed in sec.~\ref{sec:NumericalMemory}). In sec.~\ref{sec:neNumericalMemory} we will discuss methods to analyze \eref{nsGLECorrelation} numerically. 

Next we check whether $K(\tau,t)$ and $f_{0,t}$ are related by a fluctuation-dissipation relation. If we rearrange the memory kernel, \eref{timeDependentKernel}, as
\begin{multline}
  K(\tau,t) =
  \frac{\left(\mathbb{A},\ii \mathcal{L}\mathcal{Q}(\tau) G^-(\tau,t)\ii\mathcal{L}\mathbb{A}\right)_\tau}{\left(\mathbb{A}, \mathbb{A} \right)_\tau} -  \frac{\left(\mathbb{A}, \dot{\mathcal{P}}(\tau)\mathcal{Q}(\tau) G^-(\tau,t)\ii\mathcal{L}\mathbb{A}\right)_\tau}{\left(\mathbb{A}, \mathbb{A} \right)_\tau}\nonumber\\ =
 - \frac{\left(\ii\mathcal{L}\mathbb{A}, \mathcal{Q}(\tau) G^-(\tau,t)\ii \mathcal{L} \mathbb{A} \right)_\tau}{\left(\mathbb{A}, \mathbb{A} \right)_\tau} \quad ,\nonumber
\end{multline}
insert a $\mathcal{P}\circ+\mathcal{Q}\circ$ in the last term and use the fact that the spaces $\mathcal{P}$ and $\mathcal{Q}$ project onto are orthogonal to each other
\begin{multline}
K(\tau,t) = - \frac{\left(\mathcal{P}(\tau)\ii\mathcal{L}\mathbb{A}, \mathcal{Q}(\tau) G^-(\tau,t)\ii \mathcal{L} \mathbb{A} \right)_\tau}{\left(\mathbb{A}, \mathbb{A} \right)_\tau}- \frac{\left(\mathcal{Q}(\tau)\ii\mathcal{L}\mathbb{A}, \mathcal{Q}(\tau) G^-(\tau,t)\ii \mathcal{L} \mathbb{A} \right)_\tau}{\left(\mathbb{A}, \mathbb{A} \right)_\tau}\nonumber\\ = -\frac{\left(\mathcal{Q}(\tau)\ii\mathcal{L}\mathbb{A}, \mathcal{Q}(\tau) G^-(\tau,t)\ii \mathcal{L} \mathbb{A} \right)_\tau}{\left(\mathbb{A}, \mathbb{A} \right)_\tau}\quad ,\nonumber
 \end{multline} 
we see that $K$ can be expressed in terms of the fluctuating force, \eref{timeDependentFF}, as  
\[
  K(\tau,t) = -\frac{\langle f_{\tau,\tau}f_{\tau,t}\rangle}{ \langle \left\vert A_{\tau}\right\vert^2\rangle} \quad .
\]
This expression is already similar to a fluctuation-dissipation relation. With a few additional transformations, which are listed in detail in ref.~\cite{Meyer2019}, it can be brought into the more familar form
\begin{equation}
  \label{eq:generalizedFDT}
 K(\tau,t)= -\frac{\langle f_{0,\tau}f_{0,t}\rangle}{ \langle \left\vert A_{\tau}\right\vert^2\rangle} \quad .
  \end{equation}
  Thus, as a consequence of the linear projection operator, the memory kernel and the fluctuating force of the linear nsGLE, \eref{nsGLE}, are related by a generalized version of the second fluctuation-dissipation theorem. However, this statement only refers to the mathematical structure of \eref{generalizedFDT}. There is no direct interpretation of $K(\tau,t)$ as a friction, i.e.~$\int \dd \tau K(\tau,t)$ is not proportional to the work transferred from $\mathbb{A}$ to the other degrees of freedom (the ``bath'') during the relaxation process.

  \subsubsection{The Generalized Fokker-Planck Equation}
  \label{sec:neFP}
In the previous sections we mostly discussed equations of motion which have structures similar to the Langevin Equation. \eref[Eq.]{ZwanzigEOM} is different in that it is an equation of motion for the probability distribution $g(a,t)$ rather than the value of the observable $A_t$, i.e.~it is an equation for the probability that the observable $\mathbb{A}(\Gamma_t)$ has the value $a$ at time $t$.\footnote{Again, in this subsection we consider one observable instead of a set $\mathbf{A}$ to simplify the discussion, but all statements hold equally for sets of observables.} As the derivation of \eref{ZwanzigEOM} does not require equilibrium assumptions and also holds for relaxation into equilibrium, let us briefly consider it more closely than in the introductory sec.~\ref{sec:Zwanzig}.
If the dynamics of the coarse-grained variable is Markovian, $g(a,t)$ is governed by the Fokker-Planck equation 
\begin{equation}
\frac{\partial g(a,t)}{\partial t} = \frac{\partial}{\partial a} \left[ a_{1}(a)g(a,t)  \right] + \frac{\partial^{2}}{\partial a^{2}} \left[ a_{2}(a) g(a,t)  \right] \quad ,
\end{equation}
where $a_{1}$ and $a_{2}$ are called drift and diffusion coefficient \cite{risken:1996}. Similar to the Langevin equation, this equation is also often interpreted in terms of a free energy landscape or potential of mean force
\begin{equation}\frac{\partial g(a, t)}{\partial t} =  \frac{\partial}{\partial a}  \left[D(a)  \left(  \frac{\partial }{\partial a}   + \beta  \frac{\partial \Delta G(a)}{\partial a} \right) g(a,t)  \right] \label{eq:FP} \quad ,
\end{equation}
where the diffusion coefficient $D$ can in general depend on $a$ and $\Delta G(a)=U^{\rm MF}(a)$ is defined as in sec.~\ref{sec:nonlinear} via \eref{Defrhox} and the equilibrium distribution $\rho_N^{\text{EQ}}(\Gamma)$.
In the case of ergodic dynamics without external driving, $g(a,t)$ reaches the equilibrium distribution $g^{\text{EQ}}(a) = \int\dd \Gamma \; \delta (\mathbb{A}(\Gamma)-a)\rho_N^{\text{EQ}}(\Gamma)$ in the long-time limit.

As for the case of the Langevin equation, also here the full equation of motion, \eref{ZwanzigEOM}, is considerably more complex than \eref{FP}. 
 If we intend to compare the two equations, the restriction to the microcanoncial ensemble which was used in the derivation of \eref{ZwanzigEOM} is unfavourable. In order to allow for a more general setting, we again follow a derivation by Grabert \cite{grabert:1982} and use a projection operator slightly different from \eref{DefPZ}
\[
  \mathcal{P}\mathbb{X}(\Gamma) \coloneqq \int \dd a \; \delta(\mathbb{A}(\Gamma)-a) \frac{1}{\rho_A(a)}\int \dd \Gamma'\;\rho_N^{\rm EQ}(\Gamma')\delta(\mathbb{A}(\Gamma')-a)\mathbb{X}(\Gamma') 
\]
with the reduced density \[\rho_A(a)\coloneqq\int \dd \Gamma \;\rho_N^{\rm EQ}(\Gamma)\;\delta(\mathbb{A}(\Gamma)-a)\quad .\]
We insert this projection operator into \eref{Dyson} and obtain
\begin{multline}
  \label{eq:FokkerPlanckFull}
  \frac{\partial g(a,t)}{\partial t} = -\frac{\partial}{\partial a}\left(\omega(a)g(a,t) \right) \\ + \int_0^t\dd \tau \; \left\{\frac{\partial}{\partial a}
    \int \dd a'\; \left[ D(a,a',t-\tau)\rho_A(a')\frac{\partial}{\partial a'}
    \left(\frac{g(a',\tau)}{\rho_A(a')}\right)\right] \right\} \quad ,
\end{multline}
where
\[\omega(a)\coloneqq \frac{1}{\rho_A(a)} \int \dd \Gamma\; \rho_N^{\text{EQ}}(\Gamma) \;\delta(\mathbb{A}(\Gamma)-a)\;\ii\mathcal{L}\mathbb{A}(\Gamma) \quad ,\]
\begin{equation}
  \label{eq:defD}
D(a,a',t) \coloneqq \frac{1}{\rho_A(a)} \int \dd \Gamma \; \rho_N^{\text{EQ}}(\Gamma)\; R_a(\Gamma,t)R_{a'}(\Gamma,0)
  \end{equation}
  and
  \[
 R_a(\Gamma,t)\coloneqq\mathcal{Q}e^{\ii\mathcal{L}\mathcal{Q}t}\delta(\mathbb{A}(\Gamma)-a)\ii\mathcal{L}\mathbb{A}(\Gamma) \quad .
\]
Thus $\omega(a)$ and $D(a,a',t)$ are equivalent to the transport coefficients $v(a)$ and $W(a)K(a,a',t)$ in \eref{ZwanzigEOM}, but we have allowed for a more general choice of stationary distribution $\rho_N^{\text{EQ}}(\Gamma)$.

Now we focus our discussion on the case in which the observable depends only on the positions $\mathbb{A}(\vec{r}^N)$ and not on the momenta, such that $\omega(a)$ vanishes due to symmetry. Further, we restrict the analysis to the canonical ensemble, $\rho_N^{\rm EQ}(\Gamma) \propto \exp(-\beta\mathcal{H}(\Gamma))$ and $\Delta G(a)\coloneqq -k_BT \ln(\rho_A(a))$, noting that other ensembles can be analyzed using a similar line of arguments.
The dependence on $a$ and $a'$ in the transport coefficients $D(a,a',t)$ is inconvenient. It can be removed by means of a Kramers-Moyal expansion, which is carried out in detail in ref.~\cite{Kuhnhold2019}. Here we just note that the functions $R_a$ can be combined into functions $d_k(a,a',t)$ such that
\[
D(a,a',t) = \sum_{k=0}^{\infty}\frac{\partial^{k} }{\partial a^{k}} \left(  d_{k}(a,a',t) \delta(a-a') \right)\quad .
\]
We define functions $\tilde{d}_{k}(a,t) := d_{k}(a, a,t)$ and note that 
\[
  \rho_A(a)\frac{\partial}{\partial a}\left(\frac{g(a,t)}{\rho_A(a)} \right)
  = \left(\frac{\partial}{\partial a} + \beta\frac{\partial \Delta G(a)}{\partial a} \right)
  g(a,t) \quad .
  \]
Thus we can write the equation of motion of the probability distribution, \eref{FokkerPlanckFull}, as
\begin{equation}
      \frac{\partial g(a, t)}{\partial t} =  \int_{0}^{t} d\tau  \sum_{k=0}^{\infty}  \frac{\partial^{k+1}}{\partial a^{k+1}}  \left[\tilde{d}_{k}(a,t-\tau)  \left(  \frac{\partial }{\partial a}   + \beta \frac{\partial \Delta G(a)}{\partial a} \right) p(a,\tau)  \right] \label{eq:fullFP} \quad .
    \end{equation}
    There are two major differences between \eref{fullFP}, which is exact, and the Fokker-Planck equation, \eref{FP}:
    \begin{itemize}
      \item
        The functions $\tilde{d}_{k}(a,t)$ act as memory kernels, which implies that the evolution of $g(a, t)$ is in general non-Markovian.
      \item Terms with $k \ge 1$ are in general not equal to zero, which implies that the evolution of $g(a, t)$ is not diffusive.
      \end{itemize}
As in the case of the nonlinear Langevin equation, one needs to keep these differences in mind when using a Fokker-Planck equation with a potential of mean force resp.~a free energy landscape to model coarse-grained dynamics. In particular, theories for phase transition processes which are based on \eref{FP}, such as e.g.~classical nucleation theory, are by construction approximative.      
  
\subsection{Time-dependent Liouvillians}
\label{sec:timeDependence}
\subsubsection{Power Functionals}
\label{sec:PowerFunctionals}
\eref[Eq.]{DDFT} provides us with an approximate method to compute the evolution of the one-particle density by means of a variation of an equilibrium thermodynamic potential, the density functional $\bar{\Omega}[\rho_1(\vec{r},t)]$. In this sense DDFT extends the principle of DFT to systems out of equilibrium. However, the strategy of exploiting a minimization principle in order to obtain an evolution equation for $\rho_1(\vec{r},t)$ can also be pursued in a different way, which is perhaps even closer to the original spirit of DFT and, most importantly, which is in principle exact. Schmidt and Brader suggested replacing the thermodynamic potential by its non-equilibrium analogue, the power functional \cite{schmidt2013,schmidt2018,heras2018}. They first introduced the concept for Brownian dynamics \cite{schmidt2013}, i.e.~for a type of dynamics, which is already coarse-grained. Recently they formulated a version which is based on microscopic Hamiltonian dynamics \cite{heras2018}. As we are interested in coarse-graining from deterministic mircoscopic dynamics, we will review the latter version here.

The idea of power functional theory is to construct a functional $G_t[\rho_1,\vec{J},\dot{\vec{J}}]$ such that the time derivative of the non-equilibrium current $\dot{\vec{J}}_{0}(\vec{r},t)$ is determined by the condition\footnote{Note that the subscript $0$ in $\dot{\vec{J}}_{0}$ does not refer to an initial condition here but to the solution of the stationarity condition \eref{minimizationPFT}. $\dot{\vec{J}}_{0}$ is the non-equilibrium analogue to the equilibrium density distribution $\rho^{\rm EQ}$.} 
\begin{equation}
  \label{eq:minimizationPFT}
\left.\frac{\delta G_t[\rho_1,\vec{J},\dot{\vec{J}}]}{\delta \dot{\vec{J}}(\vec{r},t)}\right\vert_{\dot{\vec{J}}=\dot{\vec{J}}_{0}}=0 \quad .
\end{equation}
The functional is non-local in space and it depends on the value of its arguments up to the time at which the variation is performed.

We consider a system of $N$ identical particles of mass $m$ subject to the Hamiltonian
\[
\mathcal{H} = \sum_i\frac{\vec{p}_i\cdot \vec{p}_i}{2m} + V_{\rm int}(\vec{r}^N) + \sum_iV_{\rm ext}(\vec{r}_i,t) \quad .
\]
We will use the same definitions for the one-particle operators as in the previous section,
\[
  n_{\vec{r}}(\Gamma) \coloneqq \sum_i\delta(\vec{r}-\vec{r}_i) \quad {\text{and}} \quad \vec{J}_{\vec{r}}(\Gamma) \coloneqq \sum_i\delta(\vec{r}-\vec{r}_i)\frac{\vec{p}_i}{m} \quad ,\]
as well as for the averages
\[\rho_1(\vec{r},t) \coloneqq \langle n_{\vec{r}}(\Gamma,t)\rangle  \quad {\text{and}} \quad \vec{J}(\vec{r},t) \coloneqq \langle \vec{J}_{\vec{r}}(\Gamma,t) \rangle  \quad .\]
The angle brackets indicate the ensemble average with respect to $\rho_N(\Gamma,t)$, i.e.~$\langle \mathbb{X} \rangle = \int \dd \Gamma \rho_N(\Gamma,t) \mathbb{X}(\Gamma)$ is a time-dependent quantity.

 Similar to the grand potential, $G_t[\rho_1,\vec{J},\dot{\vec{J}}]$ can be split into an intrinsic part and a term that accounts for interactions with an external field  
\[G_t[\rho_1,\vec{J},\dot{\vec{J}}]  =: G_t^{\rm int}[\rho_1,\vec{J},\dot{\vec{J}}] - \int \; \dd \vec{r} \; \dot{\vec{J}}\cdot \nabla V_{\rm ext} \quad ,
  \]
  such that we obtain from \eref{minimizationPFT} an Euler-Lagrange equation of the form
  \[
-\frac{\delta G_t^{\rm int}}{\delta{\dot{\vec{J}}}} - \nabla V_{\rm ext}=0 \quad .
\]
 The internal contribution can in turn be split into an ideal gas part and an excess part, $G_t^{\rm int} = G_t^{\rm id} +G_t^{\rm exc}$. The ideal gas part is known
  \[
    G_t^{\rm id}[\rho_1,\vec{J},\dot{\vec{J}}] = \int \dd \vec{r} \frac{\dot{\vec{J}}}{\rho_1}\cdot\left(\frac{m\dot{\vec{J}}}{2} - \nabla \cdot [\tau]^{\rm id} \right) \quad ,
  \]
  where $[\tau]^{\rm id}\coloneqq -m\frac{\vec{J}\otimes\vec{J}}{\rho_1}$ is the factorized contribution to the kinematic stress tensor of the ideal gas. The difficulty  is to account for transport effects beyond the factorization and, as in the case of DFT, to construct the excess functional $G_t^{\rm exc}$.

  In ref.~\cite{schmidt2018}, Schmidt shows how to obtain the correct functional by means of a constrained minimization. We define the microscopic accelerations $\vec{a}_i$, which will act as trial variational fields, and the {\it microscopic power rate functional}
  \begin{equation}
    \label{powerRate}\mathcal{G}_t[\rho_1,\vec{J},\dot{\vec{J}}]  \coloneqq \int \dd \Gamma \sum_i\frac{\left(\vec{f}_i - m\vec{a}_i\right)^2}{2m}\rho_N(\Gamma,t) - \int \dd \vec{r} \frac{m}{2\rho_1(\vec{r},t)} \left\langle \frac{\dd \vec{J}_{\vec{\mathbf{r}}}}{\dd t}\right\rangle^2 \quad ,
   \end{equation}
   where the force that acts on particle $i$ is $\vec{f}_i=-\nabla_{\vec{r}_i} V_{\rm int}(\vec{r}^N) - \nabla_{\vec{r}_i}V_{\rm ext}(\vec{r}_i,t)$.
   Note that memory is introduced here as the one-body fields need to be known at all times prior to t in order to reconstruct the many-body distribution $\rho_N(\Gamma,t)$.
   
   The true dynamics of the system is the one for which $\vec{f}_i = m\vec{a}_i$, and hence
   \[
\frac{\delta \mathcal{G}_t}{\delta \vec{a}_i(t)}=0 \quad .
     \]
     We introduce the microscopic kinematic stress operator
     \[[\tau]_{\vec{\mathbf{r}}}(\Gamma,t) \coloneqq -\frac{\vec{p}_i\otimes\vec{p}_i}{m} \delta(\vec{r}-\vec{r}_i)\]
     and note that due to Newton's second law
     \[
m\frac{\dd \vec{J}_{\vec{\mathbf{r}}}}{\dd t}=\sum_i\delta(\vec{r}-\vec{r}_i)\vec{f}_i + \nabla \cdot \sum_i [\tau]_{\vec{\mathbf{r}}} \quad .
\]
To make the transition from the microscopic description on the level of the accelerations  to the coarse-grained description on the level of the one-particle acceleration density $\dot{\vec{J}}$, we now impose the condition
\[
\dot{\vec{J}}(\vec{r},t) = \left\langle \sum_i\vec{a}_i\delta(\vec{r}-\vec{r}_i)+\frac{\nabla\cdot[\tau]_{\vec{\mathbf{r}}}(\vec{r},t)}{m}\right\rangle \quad .
\]
The functional $G_t[\rho_1,\vec{J},\dot{\vec{J}}]$ then follows from a constrained search for the minimum of $\mathcal{G}_t$
\[
G_t[\rho_1,\vec{J},\dot{\vec{J}}] = \min_{\vec{a}_i \to \rho_1,\vec{J},\dot{\vec{J}}}\mathcal{G}_t \quad .
  \]
Finally, to identify $G_t^{\rm exc}$, we note that   
on the level of the one-body current $\dot{\vec{J}}(\vec{r},t)$, Newton's second law can be written in terms of the internal force density $\vec{f}_{\rm int}$ and the kinematic stress tensor $[\tau]$ as 
\[
m\dot{\vec{J}}(\vec{r},t) = \vec{f}_{\rm int}(\vec{r},t)+\nabla\cdot[\tau](\vec{r},t) -\nabla V_{\rm ext}\rho(\vec{r},t) \quad ,
\]
where 
\[
\vec{f}_{\rm int}(\vec{r},t)=-\left\langle \sum_i \delta(\vec{r}-\vec{r}_i)\nabla_{\vec{r}_i}V_{\rm int}(\vec{r}^N)\right\rangle \quad  {\rm and} 
\quad 
[\tau](\vec{r},t) \coloneqq \left\langle \sum_i [\tau]_{\vec{\mathbf{r}}}(\Gamma,t) \right\rangle \quad .
\]
Hence the non-equilibrium current is determined by the condition
\[
\left. \frac{\delta G_t^{\rm exc}[\rho,\vec{J},\dot{\vec{J}}]}{\delta \dot{\vec{J}}(\vec{r},t)}\right\vert_{\dot{\vec{J}}=\dot{\vec{J}}_{0}} = \frac{\vec{f}_{\rm int}(\vec{r},t) + \nabla \cdot ([\tau](\vec{r},t)-[\tau]^{\rm id}(\vec{r},t))}{\rho_1(\vec{r},t)}
  \]

  \subsubsection{Mori's projection operator for full non-equilibrium}
  \label{sec:HuguesWork}
  It is rather difficult to extend the approach via Zwanzig's (resp.~Robertson's) projection operator sketched in sec.~\ref{sec:TimeDependentProjector} to explicitly time-dependent microscopic dynamics. In contrast, the linear projection operator introduced by Mori (sec.~\ref{sec:Mori}) turns out to be well-suited to this task. In the next paragraphs we summarize our own work on this extension \cite{Meyer2017,Meyer2019}, noting that some of the ideas have already been sketched by Nordholm in his PhD thesis~\cite{nordholm1972}, that a similar approach has been used by McPhie and co-workers for the case of relaxation into a non-equilibrium steady state \cite{McPhie01}, that te Vrugt and Wittkowski have developped a similar approach to driven quantum mechanical systems independently and at the same time as we published our work \cite{vrugt2019}, and most notably, that Kawai and Komatsuzaki had already developed a nearly identical approach eight years before us  \cite{Kawai11}. Regrettably, we only discovered their work after ref.~\cite{Meyer2019} had been published.

  In sec.~\ref{sec:TimeDependentProjectorMori} we generalized Mori's projection operator to the case of relaxation into equilibrium starting out from the equation of motion of the observable in the ``Heisenberg picture''. The ideas presented there can be applied to systems under time-dependent external driving as well as to active systems and to explicitly time-dependent observables, if we consider that the equation of motion for the observable needs to be modified accordingly.

  The following line of reasoning might be unfamiliar if the reader is used to handling Hamiltonian systems. If so, it might help to visualize the dynamics in terms of trajectories in phase space, $\Gamma_t$. The dynamics of an individual trajectory is given by an evolution equation, which assigns a time derivative $\dot{\Gamma}$ to each state $\Gamma$. Thus we can visualize the dynamics by ``stream lines'', $\dot{\Gamma}(\Gamma)$, in analogy to the stream lines of a flowing liquid. If the system contains active particles or is subject to time-dependent driving, the time at which the system reaches a certain state $\Gamma$ matters explicitly, i.e.~the stream lines themselves depend on time, $\dot{\Gamma}(\Gamma,t)$.
  
  We begin with the equation of motion for the microscopic density, a generalization of the Liouville equation, \eref{Liouville}, and take the time-dependence $\dot{\Gamma}(\Gamma,t)$ into account
  \begin{equation}
  \label{eq:Liouville2}
\frac{\partial}{\partial t} \rho_N(\Gamma,t) = - \frac{\partial}{\partial \Gamma}\cdot[\dot{\Gamma}(\Gamma,t)\rho_N(\Gamma,t)] =: -\ii \revL(t)  \rho_N(\Gamma,t) \quad .
\end{equation}
We introduced the symbol $\ii \revL(t)$ to indicate the fact that $ \frac{\partial}{\partial \Gamma}\cdot[\dot{\Gamma}(\Gamma)\circ]$ is not the same as the operator
  \[
 \dot{\Gamma}(\Gamma,t)\cdot\frac{\partial}{\partial{\Gamma}} \circ \coloneqq \ii \mathcal{L}(t) \circ \quad ,
\]
which will enter the equation of motion of the observable below. 
\eref[Eq.]{Liouville2} is formally solved by
    \[
      \rho_N(\Gamma,t) =  \exp_+\left(-\int_0^t\dd t'\; \ii \revL(t')\right) \rho_N(\Gamma,0)\quad .
    \]
    With this expression we obtain the equation of motion for an ensemble averaged observable $\mathbb{A}(\Gamma)$ in the Schr\"odinger picture
    \begin{equation}
      \label{eq:timeDepSchroedinger}
\langle A(t) \rangle = \int \dd \Gamma\; \mathbb{A}(\Gamma) \exp_+\left(-\int_0^t\dd t'\; \ii \revL(t')\right) \rho_N(\Gamma,0) \quad .
      \end{equation}
    As $\ii\revL(t) \neq \ii\mathcal{L}(t)$, the equation of motion in the Heisenberg picture does not follow from \eref{Liouville2} as directly as in the Hamiltonian (conservative) case. We denote by $\gamma(\Gamma, 0; t)$ the point in phase-space reached at time $t$ given that system was in state $\Gamma$ at time $0$, i.e.
    \[
\frac{\dd \gamma(\Gamma, 0; t)}{\dd t} = \left. \dot{\Gamma}(\gamma(\Gamma, 0; t),\tau)\right\vert_{\tau=t} \quad ,
      \]
      and we define the corresponding time-evolution operator
      \[
\gamma(\Gamma, 0; t)  =: \mathcal{U}(t,t')\gamma(\Gamma, 0; t')\quad .
        \]    
         $\mathcal{U}$ acts on an observable $\mathbb{A}(\Gamma)$ through $\Gamma$, i.e.
        \begin{equation}
          \label{eq:AandU}
\mathbb{A}(\gamma(\Gamma, 0; t)) = \mathcal{U}_{0,t}\mathbb{A}(\Gamma) \quad .
\end{equation}
We take the time derivative
\[
\frac{\dd}{\dd t}\mathcal{U}(0,t)\mathbb{A}(\Gamma) = \mathcal{U}(0,t)\left[\dot{\Gamma(\Gamma,t)}\cdot \frac{\partial}{\partial \Gamma}\mathbb{A}(\Gamma)\right]
\]
and see that 
\[
\frac{\dd}{\dd t}\mathcal{U}(0,t)\circ= \mathcal{U}(0,t)\ii \mathcal{L}(t)\circ \quad .
  \]
  Note that here the operator we are interested in, $\mathcal{U}(0,t)$, stands left of $\ii\mathcal{L}(t)$ -- in contrast to the density $\rho_N(\Gamma,t)$ which stands right of $\ii\revL(t)$ in \eref{Liouville2}.
  This equation is formally solved by
  \[
U(0,t)\circ = \exp_-\left(\int_0^t\dd t'\;\ii\mathcal{L}(t')\circ\right) \quad ,
\]
where $\exp_-(\circ)$ is the negatively time-ordered  exponential operator (also called {\it left-time ordered exponential operator} \cite{Vrugt2020} or {\it left-hand side time exponential} \cite{holian1985}).
In the Heisenberg picture, we would like to express the evolution of the ensemble averaged observable $\langle A(t) \rangle$ as
\begin{equation}
  \label{eq:TimeDependentHeisenberg}
  \langle A(t) \rangle = \int \dd \Gamma \; \rho_N(\Gamma,0) \exp_-\left(\int_s^t\ii\mathcal{L}(t')\dd t'\right) \mathbb{A}(\Gamma).
\end{equation}
Using \eref{AandU}, we obtain
\begin{eqnarray}
  \langle A(t) \rangle &=& \nonumber \sum_{n=0}^\infty\int_0^t\dd t_1 \ldots \int_0^{t_{n-1}}\dd t_n\int\dd\Gamma\; \rho_N(\Gamma,0)\;\ii\mathcal{L}(t_n)\ldots\ii\mathcal{L}(t_1)\mathbb{A}(\Gamma)\\
  &=&  \sum_{n=0}^\infty-\int_0^t\dd t_1 \ldots \int_0^{t_{n-1}}\dd t_{n}\int\dd\Gamma\;\ii\revL(t_n)\;\rho_N(\Gamma,0)\ii\mathcal{L}(t_{n-1})\ldots\ii\mathcal{L}(t_1)\mathbb{A}(\Gamma)\nonumber\\
  &=& \ldots =  \int \dd \Gamma\;  \mathbb{A}(\Gamma) \exp_+\left(-\int_0^t\dd t'\; \ii \revL(t')\right) \rho_N(\Gamma,0) \quad ,
\end{eqnarray}
where the ellipsis indicates repeated integration by parts, such that the operators $\ii \mathcal{L}(t_i)$ are shuffled to the left of $\rho_N(\Gamma,0)$ and replaced by $\ii \revL(t_i)$. 
        This is one of various pathways found in the literature  to show that
        \[
          \left[\exp_-\left(\int_s^t\dd t'\;\ii\mathcal{L}(t')\circ\right)\right]^\dagger = \exp_+\left(-\int_s^t\dd t'\;\ii\revL(t')\circ\right) \quad ,
          \]
          i.e.~\eref{TimeDependentHeisenberg} is valid also in the case of explicitly time-dependent microscopic dynamics \cite{holian1985,Meyer2019,vrugt2019,Vrugt2020}.

          We use \eref{AandU} and \eref{TimeDependentHeisenberg} as a starting point for the projection operator formalism
          \begin{equation}
            \label{eq:FullyTimeDepObservable}
            \frac{\dd A_t}{\dd t} = \mathcal{U}(0,t)\ii\mathcal{L}(t)\mathbb{A}(\Gamma,0) = \mathcal{U}(0,t)\left[\mathcal{P}(t)\ii \mathcal{L}(t)\mathbb{A}(\Gamma,0) +\mathcal{Q}(t)\ii \mathcal{L}(t)\mathbb{A}(\Gamma,0)  \right]\quad .
            \end{equation}
            Again, we need a generalized version of the Dyson-Duhamel identity to handle the dynamics in the space orthogonal to the projected dynamics. We define an operator $\mathcal{Z}(t)\circ\coloneqq\mathcal{U}(0,t)\mathcal{Q}(t)\circ$ and express its time-evolution as
            \[
\dot{\mathcal{Z}}(t)\circ=\mathcal{Z}(t)\ii\mathcal{L}(t)\mathcal{Q}(t)\circ+\mathcal{U}(0,t)\mathcal{P}(t)\left(\ii\mathcal{L}(t)-\dot{\mathcal{P}}(t)\right)\mathcal{Q}(t)\circ \quad ,
              \]
              i.e.
              \[
\mathcal{Z}(t)\circ=\mathcal{Z}(t')G^-(t',t)\circ+\int_{t'}^t\dd \tau \; \mathcal{U}(0,\tau)\mathcal{P}(\tau)\left(\ii\mathcal{L}(\tau)-\dot{\mathcal{P}}(\tau)\right)\mathcal{Q}(\tau)G^-(\tau,t)\circ
\]
with
\[
G^-(t',t)\circ \coloneqq \exp_-\left( \int_{t'}^t\dd \tau \; \ii \mathcal{L}(\tau)\mathcal{Q}(\tau)\circ\right) \quad .
\]
Then, in analogy to \eref{TimeDependentInnerProduct} we define
\begin{equation}
  \label{eq:DefInnerProduct2}
\left(\mathbb{X},\mathbb{Y}\right)_t\coloneqq\int\dd \Gamma\;  \left(\mathcal{U}(0,t)\mathbb{X}\right)\left(\mathcal{U}(0,t)\mathbb{Y}\right)\rho_N(\Gamma,0)\
  \end{equation}
  and with this definition apply \eref{TimeDependentMoriPO}
  \[ \mathcal{P}(t)\mathbb{X}(\Gamma) \coloneqq \frac{\left(\mathbb{X},\mathbb{A}\right)_t}{\left(\mathbb{A},\mathbb{A}\right)_t}\mathbb{A}(\Gamma)\]
  to \eref{FullyTimeDepObservable}. The resulting equation of motion has exactly the same structure as \eref{nsGLE}
  \[
  \frac{\dd A_{t}}{\dd t} = \omega(t)A_{t} + \int_{0}^{t} K(\tau,t)A_{\tau} \, \dd \tau + f_{0,t}  \quad ,
\]
but $A_t$ equals $\mathcal{U}(0,t)\mathbb{A}(\Gamma,0)$ rather than $e^{\ii\mathcal{L}t}\mathbb{A}(\Gamma,0)$, and the definitions of the drift, the memory kernel and the fluctuating force need to be adapted
 \[
      \omega(t) \coloneqq \frac{\left(\mathbb{A}, \ii \mathcal{L}(t)\mathbb{A} \right)_t}{\left(\mathbb{A}, \mathbb{A} \right)_t} \quad ,
    \]
    \begin{equation}
      \label{eq:FullyTimeDependentKernel}
K(\tau,t) \coloneqq \frac{\left(\mathbb{A}, (\ii \mathcal{L}(\tau) - \dot{\mathcal{P}}(\tau))\mathcal{Q}(\tau) G^-(\tau,t)\ii \mathcal{L} \mathbb{A} \right)_\tau}{\left(\mathbb{A}, \mathbb{A} \right)_\tau} = -\frac{\left( \ii \mathcal{L}\mathbb{A},\mathcal{Q}(\tau)G^-(\tau,t)\ii\mathcal{L}(t)\mathbb{A}\right)_\tau}{\left(\mathbb{A}, \mathbb{A} \right)_\tau}
      \end{equation}
      \begin{equation}
        \label{eq:FullyTimeDependentFF}
f_{t',t}\coloneqq \mathcal{U}(0,t')\mathcal{Q}(t')G^-(t',t)\ii\mathcal{L}(t)\mathbb{A}(\Gamma,0) \quad .
\end{equation}
Even the generalized fluctuation-dissipation relation, \eref{generalizedFDT}, is still valid.

The fact that the structure of the nsGLE remains intact even for explicitly time-dependent dynamics implies that the numerical methods designed to extract the memory kernel and to propagate the coarse-grained dynamics of \eref{nsGLE} can be applied to active matter and systems under time-dependent external driving \cite{meyer2020,meyer2020num}.

\subsubsection{Zwanzig's projection operator for full non-equilibrium}
\label{sec:neZwanzig}
As remarked above, the projection operator of the Zwanzig type is not well suited to be applied to a time-dependent Liouvillian. However, we can take a detour via a Mori-type projector. In two resourceful pieces of work, Kawai and Komatsuzaki \cite{Kawai11} as well as Izvekov \cite{izvekov2013} delineate how one can formally relate \eref{nsGLE} to a nonlinear equation based on a relevant density by means of a suitable expansion.
To illustrate the construction of the expansion, we first consider the equilibrium case. We recall the projection operator formalisms introduced in sec.~\ref{sec:Zwanzig} and  sec.~\ref{sec:Mori}. Using the inner product, 
\begin{equation}
  \label{eq:innerProductEq2}
\left(\mathbb{X},\mathbb{Y}\right)=\int \dd \Gamma \; \rho^{\rm EQ}(\Gamma)\;\mathbb{X}(\Gamma)\mathbb{Y}(\Gamma) \quad ,
  \end{equation}
we write Zwanzig's projection operator as
\[
\mathcal{P}^Z\mathbb{X}(\Gamma)= \frac{\int\;  d \Gamma^\prime \; \rho^{\rm EQ}(\Gamma')\delta (\mathbb{A}(\Gamma^\prime)- \mathbb{A}(\Gamma)) \mathbb{X}(\Gamma^\prime)}{\int\; d \Gamma^\prime\; \rho^{\rm EQ}(\Gamma')\delta (\mathbb{A}(\Gamma^\prime)- \mathbb{A}(\Gamma))}=\left.\frac{\left(\delta(\mathbb{A}(\Gamma)-a),\mathbb{X}(\Gamma)\right)}{\left(\delta(\mathbb{A}(\Gamma)-a),1\right)}\right\vert_{a=\mathbb{A}(\Gamma)}\quad .
\]
Zwanzig's intention was to split the dynamics into a part that depends on $\Gamma$ solely through the observable and a remaining part. Therefore we focus now on functions of the form $g(\mathbb{A}(\Gamma))$. Note that \eref{innerProductEq2} defines an inner product on the space of these functions, too. 
 We take an orthonormal basis $\{\phi_1(\mathbb{A}(\Gamma)), \phi_2(\mathbb{A}(\Gamma)),\ldots\}$ of this space, which we could construct e.g.~by the Gram-Schmidt procedure. Due to completeness, we have
\[
\sum_i\phi_i(\mathbb{A}(\Gamma))\phi_i(\mathbb{A}(\Gamma')) = \left.\frac{\delta(\mathbb{A}(\Gamma)-\mathbb{A}(\Gamma'))}{\left(\delta(\mathbb{A}(\Gamma)-a),1\right)}\right\vert_{a=\mathbb{A}(\Gamma)}\quad ,
\]
thus we can express Zwanzig's projection operator as
\[
\mathcal{P}^Z\mathbb{X}(\Gamma) = \sum_i\left(\mathbb{X}(\Gamma),\phi_i(\mathbb{A}(\Gamma))\right)\;\phi_i(\mathbb{A}(\Gamma)) \quad .
\]
However, this is nothing but a Mori projection, \eref{MoriPO}, on the set of observables $\mathbf{F}\coloneqq\left\{\phi_1(\mathbb{A}(\Gamma)),\phi_2(\mathbb{A}(\Gamma)),\ldots \right\}\;$. (Note that the normalization $\left(\mathbf{F},\mathbf{F}\right)^{-1}$ which appears in the Mori projector is unity, because $\left(\phi_i(\mathbb{A}(\Gamma)),\phi_j(\mathbb{A}(\Gamma))\right)=\delta_{i,j}$.) 
We carry out the linear projection in each component and obtain a set of equations of the form of \eref{detGLE}
\begin{equation}
  \label{eq:infDimMori}
\frac{\dd \mathbf{F}_t}{\dd t} = [\omega]\mathbf{F}_t + \int _0^t\dd \tau \; [K](t-\tau) \mathbf{F}_{\tau} + \mathbf{f}_t \quad ,
\end{equation}
where the drift and the memory kernel are square matrices of infinite dimension.
So far we have derived an equation for $\mathbf{F}$ rather than $\mathbb{A}$. To obtain a nonlinear GLE for $\mathbb{A}$, we use a set of polynomials as basis functions $\phi_i$ and set $\phi_1\propto \mathbb{A}$. Then the first component of \eref{infDimMori} is the equation of motion we have been looking for -- we have used a ``Mori-type'' linear projection operator formalism to derive a ``Zwanzig-type'' nonlinear equation of motion!

As the projection is linear, there is still a fluctuation-dissipation relation between the kernel and the fluctuating force
\[
\left\langle \mathbf{f}_{t}\otimes \mathbf{f}_{0}\right\rangle= -[K](t)\left(\mathbf{F},\mathbf{F}\right) \quad .
  \]
However, this relation does not hold separately for each individual term in the series and thus, in particular, not for the part that is linear in the observable $\mathbb{A}$. (This is, in essence, why we could not construct a fluctuation-dissipation relation for the nonlinear GLE discussed in sec.~\ref{sec:nlGLE}.)

 Now we extend this approach to non-equilibrium. First we get rid of the explicit time dependence in the Liouvillian and ``augment'' phase space by one additional dimension, time, i.e.~$\Gamma^{\rm a} = (\Gamma,t)$, where the superscript a stands for ``augmented phase space''.
We denote observable fields on the augmented phase space by $\mathbb{A}^{\rm a}(\Gamma^{\rm a})$. The equivalent to the Liouville operator is
\[
\ii\mathcal{L}^{\rm a}\circ \coloneqq \dot{\Gamma}^{\rm a}(\Gamma^{\rm a})\cdot\frac{\partial}{\partial \Gamma^{\rm a}}\circ
  \] and observables  evolve according to the equation
\begin{equation}
  \label{eq:EOMAugmented}
A_t(\Gamma^{\rm a}) = e^{\ii\mathcal{L}^{\rm a}t}\mathbb{A}^{\rm a}(\Gamma^a)\quad .
\end{equation}

Next we introduce an inner product on the augmented phase space
  \begin{multline}
\left(\mathbb{X}^{\rm a},\mathbb{Y}^{\rm a}\right)^{\rm a}_t \coloneqq \int\dd \Gamma^{\rm a} \; \mathbb{X^{\rm a}}(\Gamma^{\rm a})\mathbb{Y^{\rm a}}(\Gamma^{\rm a})\;\rho^{\rm a}_N(\Gamma^{\rm a},t)\nonumber\\ = \nonumber \int \dd \Gamma^{\rm a} \;\rho^{\rm a}_N(\Gamma^{\rm a},0) \left(e^{\ii\mathcal{L}^{\rm a}t}\mathbb{X}^{\rm a}(\Gamma^{\rm a})\right)\;e^{\ii\mathcal{L}^{\rm a}t}\mathbb{Y}^{\rm a}(\Gamma^{\rm a})\;\quad ,
\end{multline}
where the notation  $\rho^{\rm a}_N(\Gamma^{\rm a},t)$ indicates that we have synchronized the phase space distribution such that $\rho^{\rm a}_N(\Gamma^{\rm a},t)= \rho_N(\Gamma,t)\delta(\tau-t)$. As above, we define an orthonormal basis $\{\phi^{\rm a}_i (\mathbb{A}^{\rm a},\tau)\}$ such that
\[
\left(\phi^{\rm a}_i (\mathbb{A}^{\rm a},\tau),\phi^{\rm a}_j (\mathbb{A}^{\rm a},\tau)\right)_t=\delta_{i,j} \quad \forall t\quad .
  \]
  Note that we will, in general, need a different set of basis functions for each time $t$.
  As a generalized version of $\mathcal{P}^{\rm Z}$ we define 
  \[
\mathcal{P}^{\rm Z}(t)\mathbb{X}(\Gamma) \coloneqq \frac{\int \dd \Gamma' \; \rho_N(\Gamma',t) \delta(\mathbb{A}(\Gamma')-\mathbb{A}(\Gamma))\mathbb{X}(\Gamma')}{\int \dd \Gamma' \; \rho_N(\Gamma',t) \delta(\mathbb{A}(\Gamma')-\mathbb{A}(\Gamma))} \quad .
    \]
    Using the basis set and the augmented phase space, this projector can be brought into the form
    \begin{multline}
\mathcal{P}^{\rm Z}(t)\mathbb{X}^{\rm a}(\Gamma^{\rm a}) = \sum_i\left(\int \dd {\Gamma^{\rm a}}'\;\rho_N^{\rm a}({\Gamma^{\rm a}}',t)\phi^{\rm a}_i (\mathbb{A}^{\rm a}({\Gamma^{\rm a}}'),\tau)\mathbb{X}^{\rm a}({\Gamma^{\rm a}}') \right)\phi^{\rm a}_i (\mathbb{A}^{\rm a}(\Gamma^{\rm a}),t)\\ = \sum_i\left(\int \dd {\Gamma^{\rm a}}'\;\rho_N^{\rm a}({\Gamma^{\rm a}}',0)e^{\ii\mathcal{L}^{\rm a}t}\phi^{\rm a}_i (\mathbb{A}^{\rm a}({\Gamma^{\rm a}}'),t)\mathbb{X}^{\rm a}({\Gamma^{\rm a}}')\right)\phi^{\rm a}_i (\mathbb{A}^{\rm a}(\Gamma^{\rm a}),t) \quad .
      \end{multline}
      Note that this expression is not of the form
      \[\frac{\left(\mathbb{X}^{\rm a},\phi^{\rm a}(\mathbb{A}^{\rm a})\right)_t}{\left(\phi^{\rm a}(\mathbb{A}^{\rm a}),\phi^{\rm a}(\mathbb{A}^{\rm a})\right)_t}\phi^{\rm a}(\mathbb{A})\quad ,\] i.e.~it is not a Mori projection operator on the augmented space. However, as it is very similar in structure, it can still be inserted straightforwardly into the Dyson-Duhamel identity. We obtain the equation of motion for $\mathbb{A}$
      \begin{equation}
        \label{eq:nsnlGLE}
\frac{\dd A_t}{\dd t} = \sum_i\left(\omega_i(t)\phi^{\rm a}_i (A_t,t) + \int_{s}^t\dd \tau \;K_i(t,\tau)\phi^{\rm a}_i (A_\tau,\tau)\right) + f_{s,t} \quad ,
        \end{equation}
        with
        \[
          \omega_i(t) = \int \dd {\Gamma^{\rm a}}'\; \rho^{\rm a}_N( {\Gamma^{\rm a}}',0) e^{\ii\mathcal{L}^{\rm a}t}\left( \phi^{\rm a}_i(\mathbb{A}^{\rm a}({\Gamma^{\rm a}}'),t)\ii\mathcal{L}^{\rm a}\mathbb{A}^{\rm a}({\Gamma^{\rm a}}')\right) \quad ,
      \]
      \[
K_i(t,\tau) = \int \dd {\Gamma^{\rm a}}'\; \rho^{\rm a}_N({\Gamma^{\rm a}}',0) e^{\ii\mathcal{L}^{\rm a}\tau}\left( \phi^{\rm a}_i(\mathbb{A}^{\rm a}({\Gamma^{\rm a}}'),t)(\ii\mathcal{L}^{\rm a}-\dot{\mathcal{P}^{\rm Z}}(\tau))\mathcal{Q}^{\rm Z}(\tau) \mathcal{G}^-(\tau,t)\ii\mathcal{L}^{\rm a}\mathbb{A}^{\rm a}({\Gamma^{\rm a}}')\right)
        \]
        and
        \[
f_{s,t}=e^{\ii\mathcal{L}^{\rm a}s}\mathcal{Q}^{\rm Z}(s)\mathcal{G}^-(s,t)\ii\mathcal{L}^{\rm a}\mathbb{A}^{\rm a}({\Gamma^{\rm a}}') \quad .
\]
We used the same definitions for $\mathcal{Q}$ and $\mathcal{G}^-$ as in sec.~\ref{sec:TimeDependentProjectorMori} but such that the operators act on the augmented phase space.

If we again impose the condition, that $\phi_1^{\rm a}(\mathbb{A}^{\rm a})\propto \mathbb{A}^{\rm a}$, we could, in principle, truncate the sum in \eref{nsnlGLE} at $i=1$ in order to obtain a memory term linear in $\mathbb{A}$. However, this sum is not an expansion in a small parameter. We do, in general, not know how large the contributions of the infinitely many other terms are. Thus this kind of truncation does not produce a controlled approximation to the true dynamics.

In ref.~\cite{Kawai11}, Kawai and Komatsuzaki show a derivation which is similar in spirit to the one discussed here. Surprisingly, in the summary of their results, they write that the ``extended fluctuation dissipation theorem holds generally for any irreversible system'', although this is not the case for eq.~(64) of ref.~\cite{Kawai11}, resp.~for \eref{nsnlGLE}. We presume that this statement refers to the approximation which they present in the last part of ref.~\cite{Kawai11}. There, Kawai and Komatsuzaki prove that the irreversible GLE, which had been introduced  by Hernandez and Somer based on phenomenological arguments \cite{hernandez1999_1,hernandez1999_2,hernandez1999_3,hernandez1999_4,Hernandez99}, is obtained if one truncates the expansion in $\phi_i(\mathbb{A}^{\rm a},t)$ at the first order. For this approximation, there is indeed a version of the fluctuation-dissipation relation, but as in the case of the truncation discussed above, the resulting equation of motion does not produce a controlled approximation to the exact dynamics.

In summary, nonlinear versions of the nsGLE can be derived by means of a suitable series expansion. However the relation of the ``drift'' term $\sum_i\omega_i(t)\phi^{\rm a}_i (A_t,t)$ to a time-dependent potential of mean force is not obvious, and the memory term $\sum_i \int_{s}^t\dd \tau \;K_i(t,\tau)\phi^{\rm a}_i (A_\tau,\tau)$ contains polynomials of all orders in the observable.

\subsection{Numerical Methods}

\subsubsection{Non-Equilibrium Memory Kernels}
\label{sec:neNumericalMemory}
The formalism of sec.~\ref{sec:HuguesWork} can be used to show that the structure of the equation of motion for the autocorrelation function of an observable, $C(t',t) = \langle A_{t'}A_t\rangle_{\rm NEQ}$, remains invariant even in the case of explicitly time-dependent dynamics
\begin{equation}
  \label{eq:neCorr2}
  \frac{\dd  C(t',t)}{\dd t} = \omega(t) C(t',t) + \int_{t'}^t C(t',\tau)\;K(\tau,t) \; \dd \tau \quad .
\end{equation}
Unfortunately, most of the numerical methods presented in sec.~\ref{sec:NumericalMemory} rely on numerical deconvolution schemes thus they require the integral term to be a convolution product of the form $\int_0^t\dd \tau K(t-\tau)C(\tau)$.

However, it is still possible to extract $K(\tau,t)$ numerically from experimental data or simulation data of $C(t',t)$ efficiently.
We begin by constructing the memory kernel through a Picard iteration \cite{tricomi1985}. To this end we define functions $j_{0}(t',t)$, $S_{n}(t',t)$ and $J(t',t)$ as
		\begin{align}
		j_{0}(t',t) &\coloneqq \frac{1}{C(t',t')} \frac{\partial }{\partial t'} \left[ C(t',t') - C(t',t)	 \right] \nonumber \\
		S_{0}(t',t) &\coloneqq \frac{1}{C(t',t')} \frac{\partial C(t',t) }{\partial t'} \nonumber \\ 
		 S_{n+1}(t',t) &\coloneqq \int_{t'}^{t} \dd\tau S_{n}(t',\tau)S_{0}(\tau,t) \label{eq:recursive_S}\\
		J(t',t) &\coloneqq j_{0}(t',t) + \sum_{n=0}^{\infty}\int_{t'}^{t} \dd\tau S_{n}(t',\tau)j_{0}(\tau,t) \label{eq:J_exp}
		\end{align}
		The memory kernel $K(t',t)$ in \eref{neCorr2} is the time derivative of $J(t',t)$ 
		\[
		K(t',t) = \frac{\partial J(t',t)}{\partial t} \quad .
              \]
              Now we split the recursion relation for $S_{n}$ in \eref{recursive_S} into a ``forward part'' $(t'>t)$ and a ``backward part'' $(t'<t)$
              \begin{equation}
                \label{eq:Snsplit}
		S_{n+1}(t',t) = \left\{
		\begin{tabular}{ll}
		$\int_{0}^{T}\dd \tau\; \Theta(\tau-t')S_{n}(t',\tau) \Theta(t-\tau) S_{0}(\tau,t)$ & if $t'>t $ \\
		$-\int_{0}^{T}\dd\tau \; \Theta(t'-\tau)S_{n}(t',\tau) \Theta(\tau-t) S_{0}(\tau,t)$ & if $t'<t $ 
		\end{tabular} 
		\right.
		\end{equation}
		where $\Theta$ is the Heaviside function, and $T$ is the total duration for which we have measured the correlation function. We define the functions $S_{n}^{\triu}$ and $S_{n}^{\tril}$ (the symbols indicate functions that will later be replaced by upper triangular and lower triangluar matrices) 
		\begin{align}
		S_{n}^{\triu}(t',t) \coloneqq& S_{n}(t',t)\Theta(t-t') \nonumber \\
		S_{n}^{\tril}(t',t) \coloneqq& S_{n}(t',t)\Theta(t'-t) \nonumber \quad .
		\end{align}
		With these definitions, we write \eref{Snsplit} as 
		\begin{equation}
		\label{eq:recursion_pm}
		S_{n+1}(t',t) = \left\{
		\begin{tabular}{ll}
		$\int_{0}^{T}\dd \tau \;S^{\triu}_{n}(t',\tau) S^{\triu}_{0}(\tau,t)$ & if $t'>t $ \\
		$-\int_{0}^{T}\dd\tau \; S^{\tril}_{n}(t',\tau) S^{\tril}_{0}(\tau,t)$ & if $t'<t $ 
		\end{tabular} 
		\right.
              \end{equation}
When we use data from computer simulations or experiments, it has usually been measured at discrete times. For simplicity, we assume that the measurements were timed evenly with an increment $\Delta t$. Hence all two-time functions introduced above can be interpreted as matrices, e.g.~$S_{n}(t',t) = S_{n}(i\Delta t, j\Delta t) = [S]_{n_{i,j}}$.  Then \eref{recursion_pm} can be written as
		\begin{align}
		[S]_{n+1}^{\triu} =& \nonumber [S]_{n}^{\triu} [S]_{0}^{\triu} \Delta t \quad \longrightarrow \quad [S]_{n}^{\triu} = {[S]_{0}^{\triu}}^{n+1} (\Delta t)^{n} \\
		[S]_{n+1}^{\tril} =& \nonumber - [S]_{n}^{\tril} [S]_{0}^{\tril} \Delta t \quad \longrightarrow \quad  [S]_{n}^{\tril} = \left( -1 \right)^{n}  {[S]_{0}^{\tril}}^{n+1} (\Delta t)^{n}
		\end{align}
		If we exclude the diagonal elements from either the lower or the upper triagonal matrices, then $[S]_{n} = [S]_{n}^{\triu} + [S]_{n}^{\tril}$.
		In summary, we can write \eref{J_exp} as
		\begin{multline}
		[J]^{\triu} = \sum_{n=0}^{\infty}  \left([S]_{0}^{\triu}\Delta t\right)^{n} [j]_{0}^{\triu}\nonumber \\ = \nonumber \left( \mathbb{I} -  [S]_{0}^{\triu} \Delta t \right)^{-1} [j]_{0}^{\triu}	
	\quad {\rm resp.} \quad  		
		[J]^{\tril} = \left( \mathbb{I} +  [S]_{0}^{\tril}\Delta t \right)^{-1} [j]_{0}^{\tril} \quad ,
		\end{multline}
                where $\mathbb{I}$ is the identity matrix. This provides us with a simple recipe to compute the memory kernel:
                \begin{itemize}
                \item Take a numerical time derivative of $C(t',t)$.
                \item Compute $[S]_{0}^{\triu}$, $[S]_{0}^{\tril}$, $[j]_{0}^{\triu}$ and $[j]_{0}^{\tril}$.
                \item Invert $\left( \mathbb{I} +  [S]_{0}^{\tril}\Delta t \right)^{-1}$ resp.~$\left( \mathbb{I} -  [S]_{0}^{\triu} \Delta t \right)^{-1}$.
                  \item Construct $[J]$ and take the time derivative numerically to obtain $[K]$.
                  \end{itemize}
Hence, by means of a numerical derivative, a matrix inversion and another numerical derivative, $K(t,\tau)$ can be constructed easily from experimental data. Further details regarding this method can be found in ref.~\cite{meyer2020,meyer2020num}.

Due to the existence of a fluctuation-dissipation relation, \eref{generalizedFDT}, it is straightforward to interpret the linear nsGLE, \eref{nsGLE}, as a stochastic equation of motion, i.e.~we replace the fluctuating force $f_{0,t}$ by a stochastic process $\xi_{0,t}$ which fulfills
\[
 K(t,t')= -\frac{\langle \xi_{0,t}\xi_{0,t'}\rangle}{ \langle \left\vert A_{t}\right\vert^2\rangle} \quad .
  \]
  (For the nonlinear version, \eref{nlGLEcomplete}, the construction of a stochastic alternative to the fluctuating force is not as obvious, because there is no fluctuation-dissipation relation.)
Then we can generate coarse-grained stochastic dynamics by means of the method of additional stochastic variables introduced by Berkowitz \cite{berkowitz1983} (sec.~\ref{sec:NumericalMemory}) in combination with the method to reconstruct non-stationary kernels, which we just described. I.e.~we interpret $K(t,t')$ as a family of functions $K^t(t-t')$ and fit each of them by a series
\[
  K^t(t-t') = \sum_{j}\kappa^t_{j}\exp{\left({-\frac{1}{\tau^t_{j}}}\right)}\cos\left((t-t')\omega^t_j - \phi_{t,j}\right) \quad .
\]
(Note that the superscript $t$ is the label of the fit function and coefficients here. It does not indicate a power of $t$.)
Then we generate Markovian stochastic processes which fullfill \eref{generalizedFDT} for each $K^t$ and use those to construct the coarse-grained dynamics.
The trajectories produced in this way are correct on the level of the ensemble average as well as the fluctuations. Note that they will, in general, be different from the ones generated by \eref{nlGLE}, as \eref{nsGLE} is exact, while \eref{nlGLE} contains approximations.

\subsubsection{Markov State Models}
\label{sec:neMSM}
In this section we focus on the construction of coarse-grained Markovian models for non-equilibrium processes.
As in sec.~\ref{sec:MSM} we follow ref.~\cite{koltai2018} and consider a time-continuous stochastic process $\left\{\vec{x}_t\right\}_{t\ge 0}$ on the space of particle position $X \subset \mathbb{R}^{3N}$ with a probability density $p:X\times X\rightarrow\mathbb{R}_{\ge 0}$, where $p(\vec{y},\vec{x};s,t)$ is the conditional probability density of $N$ particles being at positions $\vec{x}$ at time $t$ given that they were at $\vec{y}$ at time $s$.
Now we need to distinguish between the propagator of a density of states $f(\vec{x})$
\[
\mathcal{P}_{s,t} f(\vec{x}) \coloneqq \int_X f(\vec{x})\,p(\vec{y},\vec{x};s,t) \; \dd \vec{y} \quad ,
  \]
  the corresponding transfer operator $\mathcal{T}_{s,t}: L^2_{\mu_s} \rightarrow  L^2_{\mu_t}$ with respect to an initial density $\mu_s$ and a final density $\mu_t = \mathcal{P}_{s,t}\mu_s$
  \begin{equation}
    \label{eq:transferNE}
\mathcal{T}_{s,t} u(\vec{x}) \coloneqq \frac{1}{\mu_t(\vec{x})} \int_X u(\vec{x})\mu_s(\vec{x})\;p(\vec{y},\vec{x};s,t) \; \dd \vec{y} = \frac{1}{\mu_t(\vec{x})} \mathcal{P}_{s,t} (u\mu_s)
\end{equation}
and the Koopman operator which propagates an observable $\mathbb{A}(\vec{x})$
\[
\mathcal{K}_{s,t} \mathbb{A}(\vec{x}) \coloneqq \int_X p(\vec{x},\vec{y};s,t) \mathbb{A}(\vec{y})\; \dd \vec{y} \quad .
  \]
  $\mathcal{T}_{s,t}$ is still the adjoint of $\mathcal{K}_{s,t}$, but in contrast to the equilibrium case, they are not identical.
  
To construct a non-equilibrium Markov State Model (neMSM), we first need to specify what constitutes a good model. The spectrum of $\mathcal{T}_{s,t}$ will not provide us with a useful set of relaxation times, as the dynamics is not stationary, i.e.~the eigenvalues depend on the initial time $s$ as well as the lag time $t-s$. Hence, in contrast to the equilibrium case, the quality of a neMSM cannot simply be determined by how well it approximates the eigenfunctions to the largest eigenvalues of the propagator. Instead, as we would like to predict the evolution of a system, the minimization of the propagation error is a reasonable objective. In other words, a good neMSM is a rank-$k$ projection of $\mathcal{T}_{s,t}$ (with $k$ preferably small), for which the difference between the evolution of a density according to \eref{transferNE} and the corresponding evolution in the neMSM is minimized.
  
    As in the equilibrium case, we obtain the optimal reduced transfer operator $\mathcal{T}^k_{s,t}$ by means of a variation of the projection operator. However, here we need to take into account that  $L^2_{\mu_t}$ is not identical to $L^2_{\mu_s}$, hence we will need two projection operators $\mathcal{Q}(s) : L^2_{\mu_s} \rightarrow V_s \subset  L^2_{\mu_s}$ and $\mathcal{Q}(t) : L^2_{\mu_t} \rightarrow V_t \subset  L^2_{\mu_t}$ such that\footnote{Note that the symbol $\mathcal{Q}$ does not imply orthogonality here. In this section, the symbol $\mathcal{P}$ is already in use for the propagator. Therefore we needed another symbol for the projector.}
    \[
\mathcal{T}^k_{s,t} = \mathcal{Q}(t)\mathcal{T}_{s,t}\mathcal{Q}(s) \quad .
\]
Expressed as a variational problem, the optimal model is the one for which
\begin{equation}
  \label{eq:OptineMSM}
   \mathcal{T}^k_{s,t} =
    \argmin_{%
      \substack{%
         \mathcal{T}'=\mathcal{Q}'(t)\mathcal{T}_{s,t}\mathcal{Q}'(s)
      }
    }
     \left\vert\left\vert\mathcal{T}_{s,t} -\mathcal{T}^k_{s,t} \right\vert \right\vert \quad ,
  \end{equation}
  where the rank of $\mathcal{Q}'(t)$ and $\mathcal{Q}'(s)$ is $k$ and $\vert\vert \circ \vert\vert$ indicates the induced operator norm of operators mapping  $L^2_{\mu_s}$ to  $L^2_{\mu_t}$.
   \eref[Eq.]{OptineMSM} is solved by a singular value decomposition, i.e.~we pick the $k$ largest singular values $\sigma_1 \ge \ldots \ge   \sigma_k$ of $\mathcal{T}_{s,t}$ and their corresponding right and left singular vectors $\phi_i$ and $\psi_i$ ($\mathcal{T}_{s,t}  \phi_i = \sigma_i\psi_i$) and set
    \[
      \mathcal{Q}'(s)\circ = \sum_{i=1}^k\left(\phi_i,\circ\right)_{\mu_s}\phi_i \quad , \quad
      \mathcal{Q}'(t)\circ = \sum_{i=1}^k\left(\psi_i,\circ\right)_{\mu_t}\psi_i \quad .
    \]
    Note that we did not write out the dependence of $\psi_i,\phi_i$ and $\sigma_i$  on the times $s$ and $t$ to avoid cluttering the equations.  
    Similarly to the equilibrium case, this optimization problem can be tackled numerically \cite{wu2020}.

    The question is, how to interpret the eigenfunctions of the projectors once we have found them. In ref.~\cite{koltai2016,koltai2018} Koltai and co-workers argue, if there are sets on which the eigenfunctions are almost constant, then these sets define meta-stable states. As we need to consider two different types of eigenfunctions, $\psi_i$ and $\phi_i$, in the non-equilibrium case, the initial metastable states will in general be different from the final ones. 
    $\mathcal{T}^k_{s,t}$ then provides us with the transition rates between these states. However, for a general non-equilibrium problem such sets do not necessarily exist. 

    An interesting special case of non-equilibrium Markov State Modelling considers systems under periodic external driving \cite{wang2015,knoch2015,knoch2017,knoch2019}. As such systems eventually enter a non-equilibrium steady state (NESS), the interpretation and analysis of the coarse-grained dynamics is less involved than in the general case. We briefly review a method to tackle this problem, which  Knoch and Speck introduced in a series of articles \cite{knoch2015,knoch2017,knoch2019}. Instead of the continuous state space $X$ we consider a discrete and finite space (note that this is the space of the ``microscopic'' degrees of freedom which we intend to coarse-grain, not the state space of the coarse-grained model). Again, we assume that some degrees of freedom have already been integrated out such that the dynamics is stochastic and governed by a Master equation (see sec.~\ref{sec:MSM}) 
 \[
P_{j}(t+\dd  t) = P_j(t)\left(1-\sum_{j\ne i}k_{ji}(t)\dd  t\right)+ \sum_{j\ne i}P_i(t)k_{ij}(t)\dd  t \quad ,
\]
where $P_i(t)$ is the probability of finding the system in a state $i$ at time $t$ and $k_{ij}(t)$ is the rate of going from state $i$ to cell $j$, for which we have now allowed a time-dependence. We define the rate matrix $[K](t)$ such that $K_{ii}(t)=-\sum_{j\ne i}k_{ij}(t)$ and $K_{i\ne j}(t)= k_{ji}(t)$ to obtain
\begin{equation}
  \label{eq:neMaster}
     \frac{\dd \mathbf{P}(t)}{{\dd }t} = \mathbf{P}(t)\cdot [K](t)\quad .
   \end{equation}
   Note that $[K](t)$ is the equivalent of the conditional probability density $p(\vec{y},\vec{x};s,t)$ used above, and not to $\mathcal{K}$. For a general rate matrix $[K](t)$, \eref{neMaster} cannot be solved in closed form. However, if $[K](t)$ is periodic, i.e.~$[K](t+T)=[K](t)$, one can construct a time-independent rate matrix, which produces the same steady-state dynamics. We define averaged occupation probabilities and fluxes 
   \[
\bar{P}_i \coloneqq \frac{1}{T}\int_t^{t+T}\dd \tau \;P_i(\tau)
     \]
   and 
   \[
\bar{\Phi}_{ij} \coloneqq \frac{1}{T}\int_t^{t+T}\dd \tau\; P_i(\tau)\;K_{ij}(\tau) \quad .
\]
Before we coarse-grain the dynamics, we search for a time-independent matrix $[\tilde{K}]$ which produces the same $\bar{P}_i$ and currents $\bar{J}_{ij}\coloneqq \bar{\Phi}_{ij} -\bar{\Phi}_{ji}$. To solve this task, we make use of the fact that, due to Floquet's theorem \cite{Floquet1883}, the periodicity of the rate matrix is reflected in the solution of \eref{neMaster}
\[
\mathbf{P}(t+nT) = \mathbf{P}(t)\cdot \left(\mathbf{P}(T)\right)^n \quad , \quad n\in \mathbb{N} \quad .
\]
The time-independent part $\mathbf{P}(T)$ can be interpreted as the stationary solution of a Markov process with a time-independent rate $[\tilde{W}]$, $\mathbf{P}(T) = \exp(\tilde{W}T)$. As $\mathbf{P}(T)$ is identical to the stationary solution for $\bar{\mathbf{P}}$, the first part of the task is easily solved, and $[\tilde{W}]$ could be coarse-grained with the methods we described in sec.~\ref{sec:MSM}. However, the coarse-graining scheme also needs produce the correct fluxes.

To address the condition on the fluxes, we note that due to the periodicity of the driving field the system will reach a NESS. However, in the NESS it will not obey detailed balance but rather contain cycles of states $(i \rightarrow j \rightarrow \ldots \rightarrow i)$ along which currents will flow. These currents are associated to entropy production. On average in a transition from a state $i$ to a state $j$ entropy is produced with a rate
\[
\frac{\dd S}{\dd t} = \frac{1}{2}\sum_{ij}J_{ij}\ln{\left(\frac{\Phi_{ij}}{\Phi_{ji}}\right)} \quad ,
\]
where $\Phi_{ij}=\bar{P}_i\tilde{W}_{ij}$ and $J_{ij}\coloneqq \Phi_{ij}- \Phi_{ji}$. 
The algorithm proposed by Knoch and Speck constructs a coarse-grained model by grouping cycles into clusters. The original system is replaced by a set of representatives for these clusters and the transition rates between the representative cycles are determined under the condition that the entropy production rate $\frac{\dd S}{\dd t}$ is preserved. Thus the coarse-grained model is consistent, on average, in the sense of non-equilibrium thermodynamics and it can be used to study e.g.~work relations.

\section{Summary}

To obtain a model of a physical process that can be understood and used by humans, the equations of motion of a large number of microscopic particles need to be reduced to effective equations for a small set of relevant observables. This is the task of {\it coarse-graining}. In this review, we discussed systematic coarse-graining methods, both analytical and numerical, for systems in equilibrium and in non-equilibrium steady-states, systems relaxing into equilibrium and systems fully out of equilibrium, e.g.~due to time-dependent external driving.

In the section on systems in equilibrium we recalled the approach via the BBGKY-hierarchy and the projection operator formalisms of Zwanzig and Mori. We devoted a larger part of this section to a detailed analysis of the nonlinear generalized Langevin equation with a potential of mean force, as this equation is used frequently to analyze simulation data in the context of biomolecular modelling and modelling of soft materials. We showed that, in general, a nonlinear potential of mean force is accompanied by a nonlinear memory term and that there is no fluctuation-dissipation theorem that would relate the memory kernel of the nonlinear GLE to the corresponding fluctuating force. Arguments supporting the contrary, which have been brought forward in the literature, seem to be based on a misunderstanding regarding the exchangeability of various projection operators.
   
After the discussion of analytical equilibrium methods, we presented an overview of numerical methods used in the simulation of soft materials and biological systems such as united-atom models, Dissipative Particle Dynamics, Markov State Models and methods to construct memory kernels. In the case of equilibrium dynamics, many of these methods can be derived systematically from the underlying microscopic dynamics under certain assumptions, such as e.g.~time-scale separation or purely pairwise interactions on the coarse-grained level. In principle, the user of such simulation methods should test whether these assumptions hold before choosing a specific method.

We then moved to coarse-graining methods for systems that relax into equilibrium. The approach via the BBGKY hierarchy can, in principle, be applied straightforwardly in this context, however the equations get so involved that one hardly makes progress beyond the level of the Boltzmann equation and the Vlasov equation. Projection operators seem to be better suited to non-equilibrium dynamics. We briefly reviewed some of the work by Grabert on time-dependent projection operators and recalled how this formalism can be used to derive Dynamic Density Functional Theory, the non-stationary Generalized Langevin Equation and the non-stationary Generalized Fokker-Planck Equation. These equations are already considerably more difficult to handle than their equilibrium counterparts. Therefore care is needed if one intends to use the standard methods described in the equilibrium section to simulate relaxing systems, such as e.g.~materials undergoing phase transitions or molecules undergoing conformational changes.

Finally we reviewed several approaches that have recently been introduced to coarse-grain systems which are fully out of equilibrium, such as active systems or systems under time-dependent external driving. We recalled the concept of power functionals by Schmidt and Brader, which allows the non-equilibrium current density to be derived by means of a variational principle. Then we showed how a time-dependent projection operator of the Mori-type can be used to derive a linear, non-stationary Generalized Langevin equation. Due to the linear projection operator, the memory kernel and the fluctuating force of this equation fullfill a fluctuation-dissipation relation even for systems under external driving. In the last section on analytical methods, we presented a pathway to a nonlinear, non-stationary Generalized Langevin equation via a series expansion of a nonlinear projection operator.

We note that the equations discussed in sec.~\ref{sec:timeDependence} are exact.
 If one wishes to describe a non-equilibrium system by means of a simpler form of the Langevin equation, this constitutes an approximation which needs to be justified. This observation is relevant to several fields of physics that are currently subject to intensive research activity. 
 In the field of stochastic thermodynamics a Markovian version of the Langevin equation is widely used to describe non-equilibrium systems which are in contact with a heat bath \cite{seifert2008,seifert2012}. Fluctuation theorems which are derived in this context, such as e.g.~Jarzynski's equality \cite{jarzynski1997}, are powerful statements regarding non-equilibrium entropy production, but they need to be taken with a grain of salt given that the underlying equation of motion is that of an already simplified, coarse-grained model. 
 In the field of active matter there has recently been a large number of publications, in which the non-equilibrium nature of active systems is discussed, but at the same time the Langevin equation (or the nonlinear Generalized Langevin equation with a linear memory kernel) is used under the assumption that the fluctuation-dissipation theorem holds. Similar observations can be made in the field of biomolecular simulation. It would certainly be interesting to use the equations discussed in sec.~\ref{sec:timeDependence} in these research contexts in order to have a more controlled approach to coarse-grained non-equilibrium dynamics.

 In the final section of this review, we addressed numerical methods to handle coarse-grained models out of equilibrium. There are not many. The author is convinced that it is a worthwhile endeavour to develop such methods and to apply them in soft matter physics, biomolecular modelling and fluid dynamics. 
 
\section*{Acknowledgements}
Sec.~\ref{sec:nlGLE} and sec.~\ref{sec:neZwanzig} are summaries of work by Fabian Glatzel, with whom I had countless stimulating discussions about this review. Sec.~\ref{sec:TimeDependentProjectorMori}, sec.~\ref{sec:neFP} and sec.~\ref{sec:HuguesWork} are summaries of topics covered in Hugues Meyer's PhD thesis. Also Michael Bender, Gerhard Stock, Benjamin Lickert, Matthias Schmid and Thomas Franosch gave me useful feedback on the manuscript. Sture Nordholm kindly scanned his fifty year old PhD thesis for me! Stefan Miller carefully proof-read the manuscript and the references.
 Mark Miller always had a sympathetic ear for my questions regarding English writing.
Last but not least, I acknowledge funding by the Deutsche Forschungsgemeinschaft (DFG, German Research Foundation) grant No.~431945604 (project P4 of FOR 5099).


\bibliography{literature}

\end{document}